\documentclass[useAMS,usenatbib]{mn2e}
\usepackage{amsmath}
\usepackage{amssymb}
\usepackage{color}
\usepackage[dvips]{graphicx}
\usepackage[stable]{footmisc}
\usepackage{multirow}
\usepackage{mathrsfs}
\usepackage{paralist}
\usepackage[active]{srcltx}   % inverse search
\usepackage{url}

\newcommand{\dint}{\mathrm{d}}
\newcommand{\fc}{f\hspace{-0.7mm} c}

\newcommand{\hoM}{\,h\, \mathrm{Mpc}^{-1}}
\newcommand{\Moh}{\,h^{-1}\, \mathrm{Mpc}}
\newcommand{\pw}{p_{\mathrm w}}

\newcommand{\wi}{w_{\mathrm{i}}}

\title[Power spectrum of the SDSS DR7 LRGs: cosmological implications]
{Cosmological implications from the full shape of the large-scale power 
spectrum of the SDSS DR7 luminous red galaxies}
\author[F. Montesano et al.]
{Francesco Montesano$^{1,2}$\thanks{E-mail: montefra@mpe.mpg.de},
Ariel G. S\'anchez$^{2}$ and Stefanie Phleps$^{2}$\\ 
$^{1}$Universit\"atssternwarte M\"unchen, Scheinerstrasse 1, 81679 M\"unchen, Germany\\
$^{2}$Max-Planck-Institut f\"ur Extraterrestrische Physik, Giessenbachstrasse, 85748 Garching, Germany\\
}

\begin{document}

\date{Accepted xxx. Received xxx; in original form xxx}

\pagerange{\pageref{firstpage}--\pageref{lastpage}} \pubyear{0000}

\maketitle

\label{firstpage}

\begin{abstract}

We obtain cosmological constraints from a measurement of the spherically
averaged power spectrum of the distribution of about 90000 luminous red
galaxies (LRGs) across 7646 deg$^{2}$ in the Northern Galactic Cap from the
seventh data release of the Sloan Digital Sky Survey. The errors and mode
correlations are estimated thanks to the 160 LasDamas mock catalogues, created
in order to simulate the same galaxies and to have the same selection as the
data. We apply a model, that can accurately describe the full shape of the
power spectrum with the use of a small number of free parameters. Using the
LRG power spectrum, in combination with the latest measurement of the
temperature and polarisation anisotropy in the cosmic microwave background
(CMB), the luminosity-distance relation from the largest available type 1a
supernovae (SNIa) dataset and a precise determination of the local Hubble
parameter, we obtain cosmological constraints for five different parameter
spaces. When all the four experiments are combined, the flat $\Lambda$CDM model
is characterised by $\Omega_{\mathrm{M}} = 0.259_{-0.015}^{+0.016}$,
$\Omega_{\mathrm{b}} = 0.045 \pm 0.001$, $n_{\mathrm{s}}=0.963\pm0.011$,
$\sigma_8 = 0.802\pm0.021$ and $H_0 =
71.2\pm1.4\,\mathrm{km}\,\mathrm{s^{-1}\,Mpc^{-1}}$. When we consider curvature
as a free parameter, we do not detect deviations from flatness:
$\Omega_{\mathrm{k}} = \left(1.6\pm5.4\right)\times10^{-3}$, when only
CMB and the LRG power spectrum are used; the inclusion of the other two
experiments do not improve substantially this result. We also test for
possible deviations from the cosmological constant paradigm. Considering
the dark energy equation of state parameter $w_{\mathrm{DE}}$ as time
independent, we measure $w_{\mathrm{DE}} = -1.025_{-0.065}^{+0.066}$, if the
geometry is assumed to be flat, $w_{\mathrm{DE}} = -0.981_{-0.084}^{+0.083}$
otherwise. When describing $w_{\mathrm{DE}}$ through a simple linear function
of the scale factor, our results do not evidence any time evolution. In the
next few years new experiments will allow to measure the clustering of galaxies
with a precision much higher than achievable today. Models like the one
used here will be a valuable tool in order to achieve the full potentials of
the observations and obtain unbiased constraints on the cosmological
parameters.

\end{abstract}


\section{Introduction}

The last decade was characterised by a dramatic change in our vision of the
Universe and by an impressive increase in size and quality of available
datasets. At the end of the twentieth century, the analysis of the
luminosity-distance relation in the Type 1a supernovae (SNIa) showed that the
Universe is undergoing a phase of accelerated expansion driven by a new exotic
component, dubbed dark energy, whose energy density is about 70\% of the total
\citep{riess_98,perlmutter_99}. This has been then confirmed by other
observations, such as the cosmic microwave background \citep[CMB,
e.g.,][]{hinshaw_wmpa1_aps, Spergel_03, Spergel_07, komatsu_wmap09,
komatsu_10}, the large scale structure of the Universe \citep[LSS;
e.g.][]{efstathiou02, percival02, tegmark04, Sanchez_06, Sanchez_09,
percival_10_SDSS, reid_10_SDSS, blake_11} and the number density of galaxy
clusters as function of their mass \citep[e.g.][]{vikhlinin_09}. Different
combinations of these probes have been used in the past years to constrain the
dark energy equation of state parameter $w_{\mathrm{DE}}$ with about a
5-10\% error. The analysis presented here aims at constraining cosmological
parameters and shed light on the nature of dark energy.

Many present day and future galaxy redshift surveys, like the Baryonic Oscillation Spectroscopic Survey \citep[BOSS,][]{schlegel_BOSS, eisenstein_11_sdss3}, the Panoramic Survey Telescope \& Rapid Response System \citep[Pan-STARRS,][]{panstarrs}, the Dark Energy Survey \citep[DES,][]{des}, the Hobby Eberly Telescope Dark Energy Experiment \citep[HETDEX,][]{hetdex} and the space based Euclid mission \citep{laureijis_09_euclid}, are designed to constrain the properties of dark energy,  
usually parametrised by its density, the present day value of $w_{\mathrm{DE}}$ and its possible time evolution, with unprecedented precision. This will help to reduce the number of possible models of dark energy that has been proposed in the last decade, that are either based on a time varying field \citep[for a review see, e.g.,][]{Peebles2003}, or on modifications of the equation of general relativity \citep[modified gravity, for a review see, e.g.,][]{tsujikawa_11}.

In this work we concentrate on the distribution of galaxies on large scales as
a means to constrain the history and composition of the Universe, analysing it
statistically through the power spectrum, the Fourier transform of the two
point correlation function. The shape of the galaxy power spectrum and
correlation function contains information about the composition of the Universe
and the non-linear evolution of clustering, bias
\citep[e.g.][]{mcdonald_renorm, mats_LPT2, Jeong_09} and redshift space
distortions \citep[e.g.][]{scoccimarro_04, cabre_09a, cabre_09b, Jennings_2010,
reid_11_rsdist}. Biasing is due to the fact that the objects that we observe
do not trace perfectly the underlying dark matter distribution and redshift
space distortions are introduced when inferring the distance of a galaxy from
the measured redshift, which is a sum of a cosmological component and a Doppler
shift due to the peculiar motion of the emitter. The two-point statistics contain
also a feature, called baryonic acoustic oscillations (BAO), that has been
advocated as a powerful tool to probe the curvature of the Universe and that has
been intensively studied over the past years. BAOs are the relic signature in the
matter distribution of the acoustic oscillations in the baryon-photon plasma in
the hot young Universe. The BAOs show up in the correlation function as a quasi
gaussian bump at scales $r\approx 100-110 \Moh$ \citep{Matsubara_04} and as a
sequence of damped quasi-harmonic oscillations at wave-numbers
$0.01\hoM<k<0.4\hoM$ in the power spectrum \citep{Sugiyama_95, Eisenstein_98,
EH99}. BAOs in the CMB where predicted by \citet{peebles_70} 
and \citet{sunyaev_70} and first detected in the galaxy distribution by \citet{cole_05_2dF}
and \citet{Eisenstein_05}. The BAO scale, as measured from the CMB, depends
only on the plasma physics prior to recombination, which allows in principle to
use it as a standard ruler. Non linear evolution damps and shifts by few
percent the acoustic peaks \citep{crocce_nonlinBAO, Sanchez_08, smith_08}. The full
shape of the power spectrum and of the correlation function, as well as the
BAOs alone, have been used, alone and in
combination with other independent datasets, to constrain cosmological parameters  \citep[e.g.,][]{percival07,
sanchez_cole, cabre_09a, gaztanaga09, Sanchez_09, kazin_10a, percival_10_SDSS,
reid_10_SDSS, blake_11, tinker_11}.

In order to obtain unbiased constraints from the information encoded in the
large scale structure of the Universe, non linear distortions, bias and
redshift space distortions need to be accounted for. Perturbation theory
\citep[PT, see][for a review]{Bernardeau_02} can successfully model the shape
of the power spectrum and of the correlation function at redshifts larger than
$z=1$ or at large scales when including at least third order terms in the
density fluctuations \citep{Jeong_06,Jeong_09}. In the past few years
many groups have proposed different improvements over perturbation theory
\citep[e.g.,][]{crocce_RPT1, crocce_RPT2, mcdonald_renormbias,
matarrese_rengroup1, bernardeau_08,matarrese_rengroup2, mats_LPT1, mats_LPT2,
pietroni_flowtime, taruya_closure, smith09_SZ, taruya_09, Elia_10}. While some
of this approaches, like for instance \citet{mats_LPT2} and \citet{taruya_10},
can account for bias and redshift space distortions, most of them describe only
the clustering of dark matter in real space. For this reason phenomenological
approaches based on a given flavour of PT have been put forward in order to
model the dark matter halo and galaxy distributions in real and in
redshift space \citep[e.g.,][the latter will be thereafter referred to as
M10]{mcdonald_renorm, crocce_nonlinBAO, Sanchez_08, Jeong_09, montesano_10a}.
The model proposed by M10 is inspired by renormalised perturbation theory
\citep[RPT][]{crocce_RPT1, crocce_RPT2} and can describe the full shape of the
mildly non-linear the dark matter and halo-power spectrum from numerical
N-body simulation and allows to obtain unbiased constraints on the dark energy
equation of state parameter.

In this analysis we apply the model developed and tested in M10 to the power spectrum measured from the luminous red galaxies (LRGs), released publicly in the seventh data release of the Sloan Digital Sky Survey (SDSS DR7). Combining this measurement with the latest results from CMB, SNIa and the precise determination of the local Hubble constant, $H_{0}$, we obtain cosmological constraints for five different parameter spaces.
The LRG sample and the mock catalogues used to estimate the covariance matrix of the data are presented in Section \ref{sec:lrg_mocks}. The power spectra and covariance matrix computed from the real and mock datasets are shown in Section \ref{sec:power_spectra}. The CMB and SNIa datasets and the $H_{0}$ measurement are described in Section \ref{sec:extra_exp}. Section \ref{ssec:model} gives a short description of the model that we apply to the LRG power spectrum. In Sections \ref{ssec:parspace} and \ref{ssec:issues} we illustrate the parameter spaces that we explore in our analysis and the technique used to perform the fits. In Section \ref{ssec:test} we test our model against the mean power spectrum of the mock catalogues and of the LRG, in order to determine whether it can describe accurately the full shape of the two-point statistic when a complex geometry is used. Our main results, the cosmological constraints for the five parameter spaces, are discussed in Section \ref{sec:cosmopars} and then compared with some of the most recent works on the topic in Section \ref{sec:comparison}. Finally we summarise our results and draw our conclusions in Section \ref{sec:conclusion}.

\section{The galaxy sample and the mock catalogues.}\label{sec:lrg_mocks}

In this section we describe the galaxy sample (Section \ref{ssec:dr7}) and the mock catalogues (Section \ref{ssec:mocks}) that we use in this work.

\subsection{The luminous red galaxy sample from the 7th data release of SDSS}\label{ssec:dr7}

\begin{figure}
  \includegraphics[width=85mm, keepaspectratio]{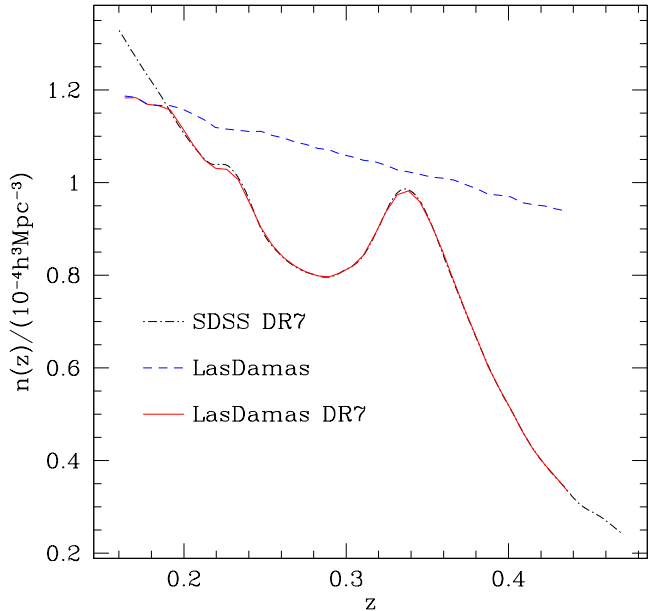}
  \caption{Spline fit to the redshift distribution of the LRGs (dot-dashed line) and the original and modified $n(z)$ of the mock catalogues (dashed and solid lines, respectively)} \label{fig:reddistr} 
\end{figure}

Data release 7 \citep[DR7,][]{abazajian_DR7} is the last data release of the second
phase of SDSS, know as SDSS-II. From the 929,555 galaxies, whose spectra have been
measured, we use the subsample of luminous red galaxies \citep[LRGs,][]{eisenstein_01}
presented in \citet{kazin_10a} and publicly
available\footnote{\url{http://cosmo.nyu.edu/~eak306/SDSS-LRG.html}}. The catalogue
contains 89,791 LRGs, in the redshift range $0.16<z<0.44$ ($\bar{z}=0.314$), from the
large contiguous area of $7646 \,\mathrm{deg^{2}}$ in the Northern Galactic Cap. The full
survey also includes three equatorial stripes, that we do not consider. 
This causes a loss of less than 10\% in galaxy number and volume, but the resulting geometry is simpler. Furthermore, the use of the Northern Galactic Cap only allows us to obtain a more accurate estimate of the statistical errors than for the full survey (see Section \ref{ssec:mocks}).
The dot-dashed line in Figure \ref{fig:reddistr} shows a smooth spline fit to the sample number density
as function of redshift. 
Together with the LRGs
catalogue, we use a random one with about fifty times more objects, designed to
reproduce the geometry and completeness of the galaxy sample and to have a radial number
density proportional to the dot-dashed line in Figure \ref{fig:reddistr}.

\begin{table} 
  \centering 
  \caption{Cosmological parameters and specifications of the LasDamas-Oriana simulations} \label{tab:LasDamas} 
  \begin{tabular}{lcc} 
    \hline 
    matter density & $\Omega_{\mathrm{m}}$ & $0.25$ \\
    cosmological constant density & $\Omega_\Lambda$ & $0.75$ \\
    baryonic density & $\Omega_{\mathrm{b}}$ & $0.04$ \\
    Hubble parameter [$\mathrm{km\,s^{-1}\,Mpc^{-1}}$] & $H$ & $70$ \\
    amplitude of density fluctuations & $\sigma_8$ & $0.8$ \\
    scalar spectral index & $n_{\mathrm{s}}$ & $1.0$ \\
    \hline number of particles & $N_{\mathrm{p}}$ & $1280^{3}$ \\
    box size [$\Moh$] & V & $2400$ \\
    particle mass [$10^{10}\,M_{\odot}$] & $M_{\mathrm{p}}$ & $45.73$ \\
    softening length [$h^{-1} \mathrm{kpc}$] & $\epsilon$ & $53$\\
    \hline 
  \end{tabular} 
\end{table}

\subsection{The mock catalogues}\label{ssec:mocks}

In order to test our analysis technique and to estimate the covariance matrix associated to the LRG power spectrum, we use
the \emph{LasDamas} mock catalogues (McBride et al., in prep.). The mocks have been constructed from a suite of 40 large dark matter N-body simulations, dubbed \emph{Oriana}, that reproduce a part of a universe characterised by a geometrically flat cosmology dominated by a cosmological constant and cold dark matter ($\Lambda$CDM). The
cosmological parameters and specifications of the simulations are listed in Table \ref{tab:LasDamas}. From each
simulation a halo catalogue is extracted using a Friend-of-Friend algorithm \citep[FoF,][]{davis_85} with linking length 0.2
times the mean inter-particle separation. In order to match the LRG clustering signal, the haloes have then been populated with mock
galaxies using a halo occupation distribution \cite[HOD,][]{berlind_02} within the
halo model approach \citep[HM, see][for a review]{cooray_HM}. The HOD parameters have been
chosen in order to reproduce the galaxy number density and the projected correlation
function of the observed SDSS DR7 samples. From each simulation two (four) mock catalogues
of the full SDSS DR7 volume (Northern Galactic Cap only) have been extracted. These mock
catalogues, together with the mocks from two smaller companions of Oriana and the
corresponding randoms, are publicly available \footnote{
\url{http://lss.phy.vanderbilt.edu/lasdamas}}.

In this work we use the 160 mock catalogues of the LRGs in the Northern Galactic Cap
region. We modify the mocks and the corresponding random catalogue, which have the radial number density shown by the dashed line in Figure \ref{fig:reddistr}, in order to reproduce the one of the LRG: the resulting $n(z)$ is indicated by the solid line. The mock catalogues contain on average 91137 galaxies. 

\section{The power spectra}\label{sec:power_spectra}

From the dataset and the mock catalogues just described we compute the power spectra and the covariance matrix that are used in the rest of the analysis. They are presented in this section.

\subsection{The LRG power spectrum}\label{ssec:dr7_ps}

To compute the power spectrum and the window function we need to convert the angular positions and redshifts of the galaxies and of the random points into comoving coordinates. This is done first inferring radial distances from the measured redshifts and then converting the spherical coordinates into cartesian ones. To do the first step we assume as fiducial the cosmology of the LasDamas simulations, shown in Table \ref{tab:LasDamas}.

We compute the power spectrum, as well as the survey window function, using the estimator
introduced by \citet[][thereafter FKP]{fkp}. \citet[][thereafter PVP]{pvp} proposed a modification of the FKP approach to take into account the relative
biases between populations with different luminosities. In Appendix \ref{ap:test_pk} we show that,
thanks to the fact that the LRG sample is almost volume limited and composed by a relatively
homogeneous class of galaxies, the shape of the power spectra recovered with the two
methods are in excellent agreement at linear and mildly non-linear scales. In
Appendix \ref{ap:pk_eqs} we summarise the most important equations of both estimators.

At first we correct the galaxy catalogue for the loss of objects due to fibre collisions \citep[][]{zehavi_02, masjedi_06}. The SDSS spectrographs are fed by optical fibres plugged on plates, which forces the fibres to be separated by at least 55''. It is then impossible, in a single exposure, to obtain spectra of galaxies nearer than this angular distance. The problem is partially alleviated by multiple exposures, but it is not possible to observe all the objects in crowded regions. 
Assuming that in a given region of the $n$ galaxies that satisfy the selection criteria we can measure only $m\leq n$ redshifts due to fibre collision and assuming that the missed galaxies have the same redshift distribution of the observed ones, we assign to the latter a weight $\wi = n/m$. This ensures that the sum of the weights in a given region of the sky is equal to the number of selected galaxies $n$. Secondly to each LRG and random object at position $\mathbf{x}$, where the number density is $n(\mathbf{x})$, we associate a weight $w(\mathbf{x}) = \left(1 + \pw n(\mathbf{x})\right)^{-1}$, with $\pw=40000h^{-3}\mathrm{Mpc}^{3}$. This value has been chosen in order to minimise the variance of the measured power spectrum in the range $0.02\hoM\leq k \leq0.2 \hoM$. In Appendix \ref{ap:test_pk} we show the results of the tests to analyse the impact of different choices of $\pw$ and corrections, namely fibre collision and completeness, on the recovered power spectrum.

To compute the power spectrum we assign the LRGs and the random objects, weighted as
described before, to a cubic grid with $N=1024^{3}$ cells and side $L=2200 \Moh$ using triangular
shaped cloud (TSC) as mass assignment scheme (MAS). For each cell, we compute the
$F(\mathbf{x})$ field of equation (\ref{eq:Fr_fkp}). We then perform the fast Fourier transform (FFT)
using the publicly available software \textsc{fftw}\footnote{\url{http://www.fftw.org/}}
\citep[Fastest Fourier Transform in the West,][]{Frigo_05}. We correct each Fourier mode
by dividing it by $\sum_{\mathbf{n}} |W(\mathbf{k} - 2k_{\mathrm N}\mathbf{n})|^{2}$, where
$W({\mathbf{k}})$ is the Fourier transform of TSC, $k_{\mathrm N}=\pi (N^{1/3}/L)$ is the
Nyquist wavenumber and $\mathbf{n}$ is a 3D integer vector. Finally we spherically
average the Fourier modes and subtract the shot noise (equation \ref{eq:sn_fkp}). 

The window function is evaluated similarly. We assign the objects of the random catalogue
to four cubic grids with $N=1024^{3}$ cells and sides $L=2200, \, 4400, \, 8800\,$ and
$17600\, \Moh$ and compute the field $\bar{G}(\mathbf{x})$ of equation (\ref{eq:Gr_fkp}). We use the four
grids with different dimensions in order to be able to compute the window function up to
very large scales. For each box, we perform the FFT, correct the Fourier modes, spherical
average and subtract the shot noise (equation \ref{eq:Gsn_fkp}). For each window function $G^{2}(k)$, we discard all the modes with wave-number $k>0.65 \,k_{\mathrm N}$
and, when two or more window functions overlap, we consider only the one computed in the larger volume. This choice is motivated by the fact that, for a given band in wavenumber, the larger volume window function has been computed averaging over a larger number of modes than the ones from smaller volumes. Finally we merge the four window functions in order to obtain a single curve.

\begin{figure} 
  \includegraphics[width=79mm, keepaspectratio]{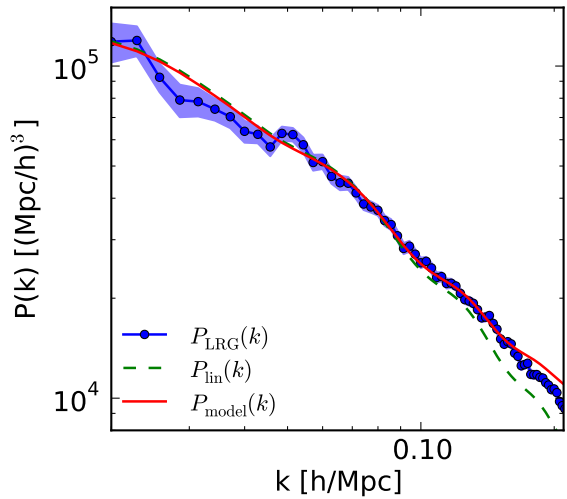} 
  \caption{LRGs power spectrum (blue dots connected with solid line) and corresponding 1-$\sigma$ error bars from the mock catalogues (shaded area). The green dashed and the red solid lines show, respectively, the linear and model power spectra computed using the mean value of the cosmological parameters of the $\Lambda$CDM cosmology show in the last column of Table \ref{tab:1D_LCDM}.} \label{fig:meas_powersp} 
\end{figure}

\begin{figure} 
  \includegraphics[width=79mm, keepaspectratio]{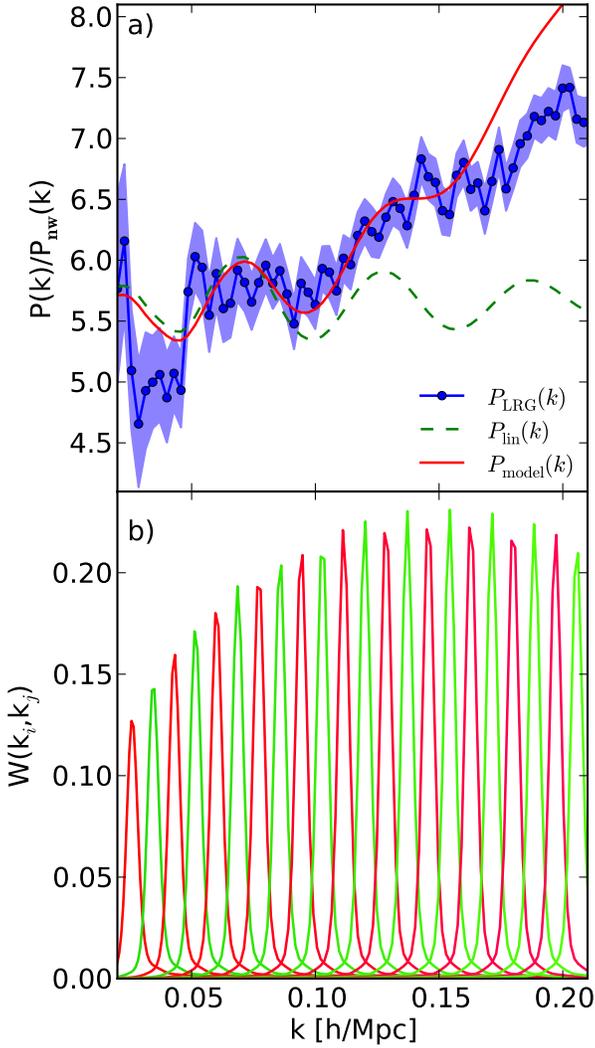}
  \caption{Panel \emph{a)}: power spectra of Figure \ref{fig:meas_powersp} divided by a linear
  power spectrum without BAOs \citep{Eisenstein_98}. Panel \emph{b)}: rows of
  the window matrix corresponding to the $k$-bands of the measured LRG power spectrum. For
  clarity only one every third row is shown.} \label{fig:meas_power} 
\end{figure}

According to equation (\ref{eq:ps_recovered}), the observed power spectrum $P_{\mathrm{o}}(k)$
just described is a convolution of the ``true'' power spectrum $P_{\mathrm{t}}(k)$ with
the window function. This convolution is computationally time consuming, in particular
when it has to be performed repetitively. Therefore we transform this convolution into a matrix
multiplication: 
\begin{equation}\label{eq:pk_winmat} 
  P_{\mathrm{o}}(k_i) = \sum_n W(k_i,k_j) P_{\mathrm{t}}(k_j) - C\,G^{2}(k_{i}).  
\end{equation} 
$W(k_i,k_j) = a_{j}k^{2}_{j} \int_{-1}^{1}\dint \cos(\theta) G^{2}(|\mathbf{k}_{i}-\mathbf{k}_{j}|)$ is the window matrix normalised such that $\sum_{j}W(k_i,k_j)=1$
$\forall i$. The coefficients $a_{j}$ corresponding to the wavenumber $k_{j}$ are derived using the Gauss-Legendre decomposition. The second term in the right hand side arises from the
integral constraint \citep{percival07} where $C$ is a constant determined by
requiring that $P_{\mathrm{o}}(0)=0$. 

The LRG power spectrum is shown in Figure \ref{fig:meas_powersp} with blue dots connected by
a solid line. We also show the linear (green dashed line) and the model (red solid line) power spectra computed from the best fit parameters obtained assuming a flat $\Lambda$CDM cosmology (see section \ref{ssec:parspace1} for more details). Both of the power spectra have
been convolved with the window function as in equation (\ref{eq:pk_winmat}) 
and their amplitude has been boosted in order to match the observed one.  The blue shaded area
represent the variance from the mock catalogues.
Panel a) of Figure \ref{fig:meas_power} shows the same quantities, with the same
colour and line coding, but divided by a smooth linear power spectrum without BAOs
\citep{Eisenstein_98}. 

Panel b) of Figure \ref{fig:meas_power} shows one every third row of the window matrix
$W(k_i,k_j)$.  Because of the relatively simple and uniform geometry of the sample used, the
window function has its maximum at $k=0\hoM$ and decreases very steeply. This translates
into the very sharp peaks at $k_j \sim k_i$ in the window matrix rows, as shown in the
figure.

\subsection{The mock power spectra and covariance matrix}\label{ssec:mocks_ps}

\begin{figure} 
  \begin{minipage}{80mm} 
    \includegraphics[width=79mm, keepaspectratio]{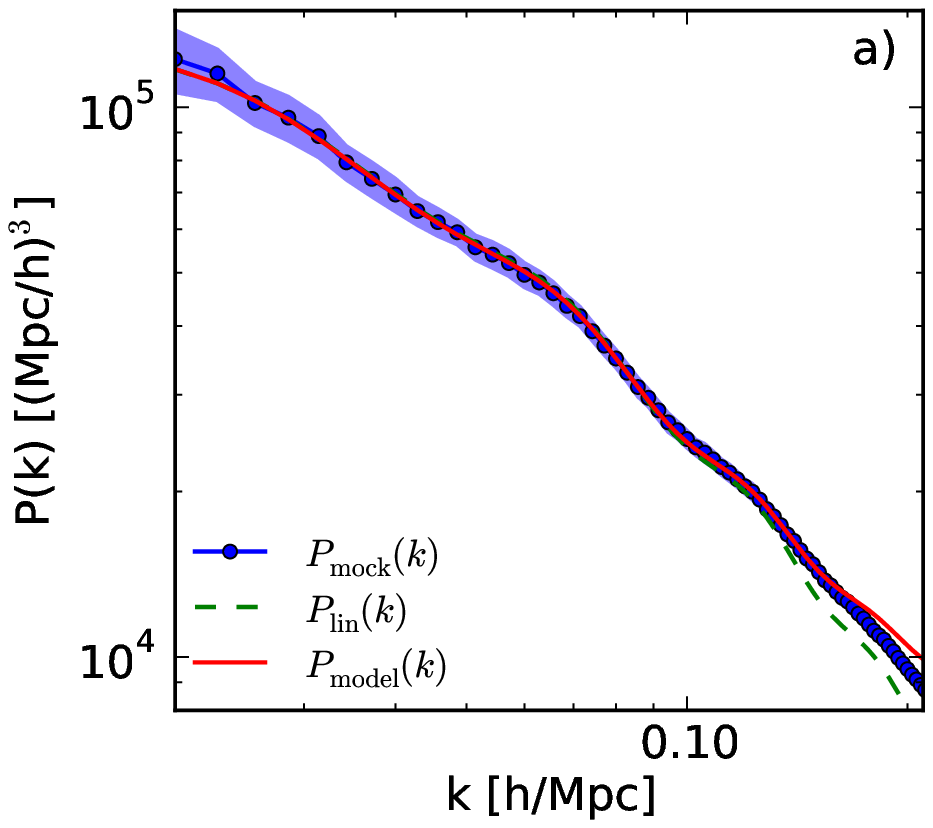} 
  \end{minipage}
  \begin{minipage}{80mm} 
    \includegraphics[width=79mm, keepaspectratio]{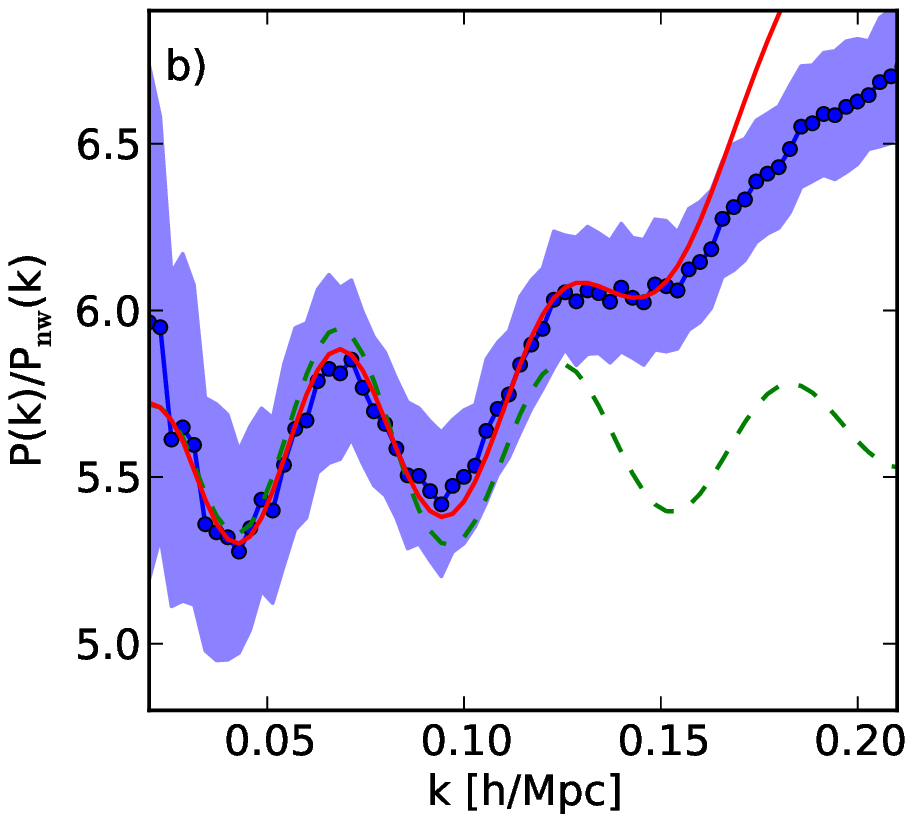}
  \end{minipage}
  \caption{Panel \emph{a)}: mean power spectrum (blue dots connected by solid line) with 1-$\sigma$ variance (blue
  shaded area) from the mock catalogues. The linear and model power spectra, convolved
  with the window function, are shown with green dashed and red solid lines, respectively.
  Panel \emph{b)}: same power spectra divided by a linear power spectrum without BAOs \citep{Eisenstein_98}.}
  \label{fig:meas_powerspmock} 
\end{figure}

We compute the power spectra from the 160 realisations and the window function as we
do for the LRG sample in the previous section. We then compute the mean $\bar{P}(k)$,
the standard deviation and the covariance matrix ${\textsf {\mathbfss C}}$, whose
elements are defined by: 
\begin{equation} 
  C_{mn}=\frac{1}{N_{\mathrm{real}}-1}\sum_{l=1}^{N_{\mathrm{real}}}(P_l(k_{m})-\bar{P}(k_{m}))(P_l(k_{n})-\bar{P}(k_{n})),
\end{equation} 
where $P_l(k_{m})$ corresponds to the measurement of the power spectrum at
the $m$-th $k$-bin in the $l$-th realisation.

\begin{figure} 
  \includegraphics[width=79mm, keepaspectratio]{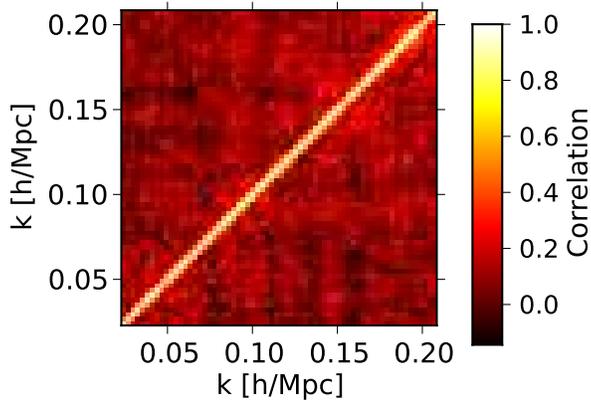}
  \caption{Correlation matrix computed from the LasDamas mock catalogues.} \label{fig:correlation}
\end{figure}

Panel a) of Figure \ref{fig:meas_powerspmock} shows the mean power spectrum and its variance (blue dots with
solid line and shaded area) obtained from the mock catalogues. The green dashed and the red
solid lines show the linear and the best fit model power spectra (see Section \ref{sssec:test_mocks} for more details) convolved with the window
function of the mocks and rescaled by the bias. Panel b) of Figure \ref{fig:meas_powerspmock} shows the same power
spectra of panel a) divided by a power spectrum without oscillations $P_{\mathrm{nw}}(k)$.

The correlation matrix of the mock catalogues, defined as $C_{mn}/\sqrt{C_{mm}C_{nn}}$, is
shown in Figure \ref{fig:correlation}. The mode correlation caused by the convolution with
the window function is visible in particular near the diagonal. Non linear mode coupling
is present at small scales and its strength increases with increasing $k$. Although the
correlation becomes important for $k>0.2 \hoM$, it is not negligible already at $k\approx
0.1 \hoM$. Recently \citet{samushia_11} showed that different methods of constructing the random catalogues can affect the estimated covariance matrix.

The version of the mocks available when writing this article does not contain information
about the luminosity of the galaxies, completeness and fibre collision. We cannot
therefore test the impact of different estimators and corrections on the errors. However
we analyse the impact of different $\pw$ on the power spectra and errors as measured
from the mocks: the results are reported in Appendix \ref{ap:test_err}.

\section{Additional experiments}\label{sec:extra_exp}

Later in this article we use the measurement of the LRG power spectrum described in the previous section to constrain cosmological parameters. In order to obtain tight constraints it is not possible to use this measurements alone, since some parameters have a weak dependence on the shape of the power spectrum and others present strong degeneracies. Using the information coming from independent experiments, like the cosmic microwave background (Section \ref{ssec:cmb}) and the type 1a supernovae (Section \ref{ssec:sn}), or prior knowledge of some of the parameters, like the local Hubble parameter (Section \ref{ssec:h_0}), it is possible to greatly improve the accuracy of the analysis.

\subsection{Cosmic microwave background}\label{ssec:cmb}

Accurate measurements of the CMB temperature and polarisation anisotropies provide a
powerful tool to constrain cosmological parameters. In this work we use the observations
from five different instruments.

The Wilkinson Microwave Anisotropy Probe (WMAP) satellite produced full sky maps of the
CMB with a resolution of $0.2^{\circ}$. We make use of the temperature angular power
spectrum in the multipole range $2 \leq l \leq 1000$ and the temperature-E polarisation
cross power spectrum in the range $2 \leq l \leq 450$ from the 7th year data release
\citep[WMAP7,][]{jarosik_11_WMAP, komatsu_10, larson_11_WMAP}.

We also use measurements of higher multipoles from other four experiments which observe
the CMB temperature anisotropy on small patches of the sky with a much higher resolution.
In order to avoid complex correlations, we use these experiments only for multipoles that
do not overlap with the WMAP measurements. We use the
measurements of the temperature angular power spectrum \begin{inparaenum}[i)] \item in 14
bandpowers in the range $910 \leq l \leq1850$ from the Arcminute Cosmology Bolometer Array
\citep[ACBAR, ][]{kuo_07, reichardt_09}, \item in 6 bandpowers in the range $855 \leq l
\leq1700$ from the Cosmic Background Imager \cite[CBI, for latest results
see][]{sievers_09_cbi}, \item in 7 bandpowers in the range $925 \leq l \leq1400$ from the
2003 flight of the Balloon Observations Of Millimetric Extragalactic Radiation and
Geophysics \citep[BOOMERanG][]{jones_06_Boo, mactavish_06_Boo, montroy_06_Boo,
piacentini_06_Boo} and \item in 11 bandpowers in the range $974 < l < 1864$ from the final
data release of QUEST at DASI \citep[QUAD, for latest results
see][]{brown_09_QUaD}\end{inparaenum}. We also use E and B polarisation (EE and BB) and
the cross temperature-E polarisation (TE) angular power spectra measurements from the
latter three experiment. \begin{inparaenum}[i)] \item CBI: EE in 7 bandpowers in the range
$860 \leq l \leq 1800$, BB in 5 bandpowers in the range $0 \leq l \leq5000$ and TE in 8
bandpowers in the range $860 \leq l \leq1800$; \item BOOMERanG: EE and BB in 3 bandpowers in the
range $600 < l < 1000$ and TE in 6 bandpowers in the range $450 \leq l \leq 950$; \item
QUAD: EE, BB and TE in 17 bandpowers in the range $488 < l < 1864$. \end{inparaenum}

\subsection{Type 1a supernovae}\label{ssec:sn}

Type 1a supernovae (SNIa) provided the first evidences of an accelerating Universe and
the need of a new exotic component, called dark energy, to explain it \citep{riess_98,perlmutter_99}. From the light curve, i.e. the luminosity variation as function of time, of a SNIa it is possible to measure the absolute luminosity, from which the distance to each object is inferred, probing in this way the distance-redshift relation. The first step is performed thanks to models which encode intrinsic variations due to the physics of the SNIa, the effects from galactic and intergalactic medium and selection effects in different ways. Because of this, models produce different results, which impact the accuracy at which cosmological parameters can be measured and can introduce systematic effects \citep[e.g.][]{hicken_09_SN, kessler_09_SN}.

SNIa data have been collected by many surveys designed according to different strategies and carried out using a large variety of telescopes. Each of them observes a relatively small number of events, typically less than a hundred. In order to increase the number of objects, recently collections of SNIa from different surveys have been created. In this work we use one of such samples, the Union2 \citep{amanullah_10_SN}. It consists of 557 SN drawn from 17 datasets in the redshift range  $0.015 \leq z \leq 1.4$ and is the largest available supernovae sample to
date. It extends and improves the Union \citep{kowalski_unionSN1a} and the Constitution
\citep{hicken_09_SN} datasets: all the light curves of the selected SNIa have been fitted
with \textsc{salt2}  \citep{guy_07_salt2} and an improved analysis of systematics is
presented. \citet{amanullah_10_SN} showed that the inclusion of systematic errors when
fitting cosmological parameters, using only SNIa or in combination with independent probes
(BAO, CMB and $H_{0}$), increases the associated errors leaving the best fit value almost
unchanged. In Sections \ref{ssec:parspace1} and \ref{ssec:parspace3} we will further comment on the impact of systematics and of light curve fitters.

\subsection{Hubble parameter}\label{ssec:h_0}

The Supernovae and H0 for the Equation of State Program \citep[SHOES,][]{reiss_09_H} aims
at the direct measurement of the Hubble parameter at present epoch $H_{0}$ to better than
$5\%$ accuracy. The SHOES team identified 6 nearby spectroscopically typical SNIa, that
have been observed before maximum luminosity, that reside in galaxies containing Cepheids
and that are subjects to low reddening. Thanks to 260 Cepheids observed with the
Near-Infrared Camera and Multi-Object Spectrometer (NICMOS) on the Hubble Space Telescope
(HST) in the 6 host galaxies and in the ``maser galaxy'' NGC 4258, the authors could
calibrate directly the peak luminosity of the SNIa. Combining these 6 objects with 240 SNIa
at redshift $z<0.1$, they measure the Hubble parameter to be $H_{0} = 74.2 \pm 3.6
\,\mathrm{km \,s^{-1} Mpc^{-1}}$. The error includes both statistical and systematic
uncertainties. In this work we use this result as a prior knowledge of H$_{0}$, under the assumption that the associated likelihood is a gaussian with the mean and standard deviation measured by the SHOES team.

Recently \citet{moresco_11} measured, using spectral properties of early-type galaxies,
the Hubble parameter to be $H_{0} = 72.6 \pm 2.9 \,\mathrm{km \,s^{-1} Mpc^{-1}}$ at
$68\%$ confidence level. While finishing this work, \citet{reiss_11_H} refined their
analysis obtaining $H_{0} = 73.8 \pm 2.4 \,\mathrm{km \,s^{-1} Mpc^{-1}}$.

\section{Methodology}\label{sec:method}

In this section we describe the model for the full shape of the power spectrum that we use throughout the rest of this article (Section \ref{ssec:model}). In Sections \ref{ssec:parspace} and \ref{ssec:issues}, we present the parameter
spaces explored in Section \ref{sec:cosmopars} and the method used to extract cosmological information. Finally, we test the model against the LasDamas
and the LRG power spectra (Section \ref{ssec:test}).

\subsection{The model}\label{ssec:model}

M10 introduced a model for the full shape of the power spectrum inspired by
renormalized perturbation theory \citep[RPT,][]{crocce_RPT1,crocce_RPT2} and analogous to
the one for the correlation function of \citet{crocce_nonlinBAO} and \citet{Sanchez_08}.
In this model they parametrize the non-linear power spectrum as 
\begin{equation}\label{eq:rpt_mod} 
  P(k,z) = b^{2} \left(e^{-\left(k/k_{\star} \right)^{2}} P_{\mathrm L}(k,z) + A_{\mathrm
  MC}P_{\mathrm{1loop}}(k,z)\right), 
\end{equation} 
where the linear bias $b$, the damping scale of the BAO oscillations $k_\star$ and amplitude of the mode coupling contribution $A_{\mathrm MC}$ are free parameters. $P_{\mathrm L}(k,z)$ is the linear power spectrum and 
\begin{equation}\label{eq:rpt_1loop_app} 
  P_{\mathrm{1loop}}(k) = \frac{1}{4\pi^{3}} \int {\mathrm d}^{3}q |F_{\mathrm2}(\mathbf k-\mathbf q,\mathbf q)|^{2}P_{\mathrm{L}}(|\mathbf k-\mathbf q|)P_{\mathrm L}(q) 
\end{equation} 
is the lowest order term arising from the coupling of two initial modes. The exponential damping and $P_{\mathrm{1loop}}(k)$ are the approximations of the non-linear propagator and of the mode coupling power spectrum described by RPT.

The model of equation (\ref{eq:rpt_mod}) has been successfully tested in M10 against a suite of very
large, median resolution N-body numerical simulations, called L-BASICC II
\citep{Angulo_08,Sanchez_08}. M10 computed power spectra for the dark matter distribution 
and several halo samples, selected according to their mass, both in real and redshift space at z=0, 0.5 and 1. They have shown that in all cases equation
(\ref{eq:rpt_mod}) describes accurately these power spectra for $k\lesssim0.15 \hoM$ and that it allows to obtain unbiased constraints on cosmological parameters. 

\subsection{Parameter spaces}\label{ssec:parspace}

In the description of the datasets presented in Sections \ref{sec:lrg_mocks}-\ref{sec:extra_exp}, a strong assumption was implicitly made: the Universe is, at large scales, statistically homogeneous and
isotropic and that density and velocity fluctuations around their mean values are small. In the following, we further assume that the fluctuations set by the initial conditions were adiabatic, gaussian and almost scale invariant and that they do not present tensor
modes. The WMAP7 data, both alone and in combination with BAO, $H_{0}$ and SNIa,
confirm that those assumptions are correct at the $95\%$ confidence level \citep{komatsu_10,
larson_11_WMAP}.

Within this framework we analyse five different cosmological models, explored using five combinations of experiments: CMB, CMB+$P_{\mathrm{LRG}}(k)$, CMB+$P_{\mathrm{LRG}}(k)$+$H_{0}$, CMB+$P_{\mathrm{LRG}}(k)$+SNIa and CMB+$P_{\mathrm{LRG}}(k)$+$H_{0}$+SNIa.

The ``concordance'' $\Lambda$CDM model is the simplest model able to successfully
describe a large variety of cosmological datasets. It describes a geometrically flat
($\Omega_{\mathrm{k}}=0$) universe with a cosmological constant $\Lambda$,
whose equation of state parameter $w_{\mathrm{DE}}=-1$ is constant in space and time, and
pressureless cold dark matter (CDM) as main components. This cosmology can be characterised by six
parameters: the baryon and dark matter densityies $\omega_{\mathrm{b}}=\Omega_{\mathrm{b}}h^{2}$ and $\omega_{\mathrm{DM}} =
\Omega_{\mathrm{DM}}h^{2}$, the scalar spectral index $n_{\mathrm{s}}$ and the amplitude
$A_{\mathrm{s}}$ of the primordial fluctuations, the optical dept $\tau$ to the last scattering surface, assuming instantaneous reionisation, and the ratio between the horizon
scale at decoupling and the angular diameter distance to the corresponding redshift
$\Theta$: 
\begin{equation}\label{eq:parspace1} 
  \theta_{\Lambda \mathrm{CDM}} = (\omega_{\mathrm{b}}, \omega_{\mathrm{DM}}, n_{\mathrm{s}}, A_{\mathrm{s}}, \tau, \Theta; b, k_{\star}, A_{\mathrm{MC}}, A_{\mathrm{SZ}}).  
\end{equation} 
The four parameters after the semicolon are related to the modelling of the matter power
spectrum ($b, \,k_{\star}$ and $A_{\mathrm{MC}}$ from equation \ref{eq:rpt_mod}) and of the CMB
angular power spectrum ($A_{\mathrm{SZ}}$, amplitude of the contribution to the CMB at
large $l$ from the Sunyaev-Zeldovich effect). These parameters are marginalised over
when showing the cosmological constraints in the section \ref{sec:cosmopars}.

If we then drop one or both assumptions on geometry and on the value of the dark
energy equation of state, we obtain three cosmologies characterised by the following
parameter spaces: 
\begin{itemize} 
  \item variable curvature, $w_{\mathrm{DE}}=-1$:
  \begin{equation}\label{eq:parspace2} 
    \theta_{k\Lambda \mathrm{CDM}} = (\omega_{\mathrm{b}}, \omega_{\mathrm{DM}}, \Omega_{\mathrm{k}}, n_{\mathrm{s}}, A_{\mathrm{s}}, \tau,  \Theta; b, k_{\star}, A_{\mathrm{MC}}, A_{\mathrm{SZ}}); 
  \end{equation} 
  \item zero curvature, $w_{\mathrm{DE}}=\mathrm{const}$: 
  \begin{equation}\label{eq:parspace3}
    \theta_{\mathrm{w CDM}} = (\omega_{\mathrm{b}}, \omega_{\mathrm{DM}}, w_{\mathrm{DE}}, n_{\mathrm{s}}, A_{\mathrm{s}}, \tau, \Theta; b, k_{\star}, A_{\mathrm{MC}}, A_{\mathrm{SZ}}); 
  \end{equation} 
  \item variable curvature, $w_{\mathrm{DE}}=\mathrm{const}$: 
  \begin{equation}\label{eq:parspace4} 
    \theta_{\mathrm{kwCDM}} = (\omega_{\mathrm{b}}, \omega_{\mathrm{DM}}, \Omega_{\mathrm{k}}, w_{\mathrm{DE}}, n_{\mathrm{s}}, A_{\mathrm{s}}, \tau, \Theta; b, k_{\star}, A_{\mathrm{MC}}, A_{\mathrm{SZ}}).  
  \end{equation} 
\end{itemize}

As last case, we consider a flat Universe in which $w_{\mathrm{DE}}(a)$ evolves with time.
We adopt the parametrisation proposed by \citet{chevallier_01} and \citet{linder_03}: 
\begin{equation}\label{eq:wa}
  w_{\mathrm{DE}}(a) = w_{0} + w_{\mathrm{a}}(1-a).  
\end{equation} 
Although not physically motivated, it can describe accurately a big variety of equations of state
derived from scalar fields with the use of only two parameters: its value today,
$w_{0}$, and its first derivative with respect to $a$, $w_{\mathrm{a}}$. The resulting
parameter space is: 
\begin{equation}\label{eq:parspace5} 
  \theta_{\mathrm{waCDM}} = (\omega_{\mathrm{b}}, \omega_{\mathrm{DM}}, w_{\mathrm{0}}, w_{\mathrm{a}}, n_{\mathrm{s}}, A_{\mathrm{s}}, \tau, \Theta; b, k_{\star}, A_{\mathrm{MC}}, A_{\mathrm{SZ}}).  
\end{equation}

Other cosmological quantities can be derived from the ones just listed. In particular we
are interested in: 
\begin{equation}\label{eq:parspaceder} 
  \theta_{\mathrm{der}} = (\Omega_{\mathrm{DE}}, \Omega_{\mathrm{M}},  H, t_{0}, \sigma_8, z_{\mathrm{re}}).
\end{equation} 
The density of dark energy, $\Omega_{\mathrm{DE}}$, is obtained from a combination of $\Omega_{\mathrm{k}}$,
$\omega_{\mathrm{M}}=\omega_{\mathrm{b}}+\omega_{\mathrm{DM}}$ and $\Theta$. From there,
the total matter density $\Omega_{\mathrm{M}} = 1 - \Omega_{\mathrm{k}} -
\Omega_{\mathrm{DE}}$, the Hubble parameter $h = \sqrt{
\omega_{\mathrm{M}}/\Omega_{\mathrm{M}}}$ and the age of the universe $t_{0} =
\int_{0}^{1}\dint a/(aH(a))$ are derived. The present day $rms$ of linear density
fluctuation in a sphere of radius $8\Moh$, $\sigma_{8}$, is computed from $A_{\mathrm{s}}$. From $\tau$, $H$, $\Omega_{\mathrm{b}}$ and
$\Omega_{\mathrm{DM}}$ it is possible to estimate the redshift of reionisation
$z_{\mathrm{re}}$ \citep{tegmark_94}.

\subsection{Practical issues}\label{ssec:issues}

\begin{table} 
  \centering 
  \caption{Prior ranges for the primary cosmological and the model parameters} \label{tab:priors} 
  \begin{tabular}{lrl} 
    \hline 
    Parameter & lower limit & upper limit\\
    \hline $\omega_{\mathrm{b}} = \Omega_{\mathrm{b}} h^{2}$ & 0.005 & 0.1 \\
    $\omega_{\mathrm{DM}} = \Omega_{\mathrm{DM}} h^{2}$ & 0.01 & 0.99 \\
    $\Omega_{\mathrm{k}}$ & -0.3 & 0.3 \\
    $w_{\mathrm{DE}}$ ($w_{0}$) & -2 & 0 \\
    $w_{\mathrm{a}}$ & -2 & 2 \\
    $n_{\mathrm{s}}$ & 0.5 & 1.5 \\
    $\log(10^{10} A_{\mathrm{s}})$ & 2.7 & 4 \\
    $\tau$ & 0.01 & 0.8 \\
    100$\Theta$ & 0.5 & 10 \\
    \hline $k_{\star}$ & 0.01 & 0.35 \\
    $A_{\mathrm{MC}}$ & 0 & 5 \\
    $A_{\mathrm{SZ}}$ & 0 & 2\\
    \hline 
  \end{tabular} 
\end{table}

To constrain the sets of cosmological and model parameters just described, we use the Markov Chain Monte Carlo technique
\citep[MCMC,][]{gilks_MCMC, christensen_00_MCMC} as implemented in the free software
\textsc{cosmomc}\footnote{\url{http://cosmologist.info/cosmomc/}} \citep[Cosmological
MonteCarlo,][]{lewis_02}. The CMB and linear matter power spectra are computed with a
modified version of \textsc{camb} \citep[Code for Anisotropies in the Microwave
Background,][]{lewis_00} that allows to consider time varying $w_{\mathrm{DE}}$
\footnote{\url{http://camb.info/ppf/}}\citep{fang_08}. For each choice of parameter
space and probes we run eight independent chains. Their execution is stopped when the
\citet{gelman_92} criterion $R < 1.02$ is satisfied. The MCMC requires some prior
knowledge of the parameter space that is explored. We assume for all the primary
parameters (equations \ref{eq:parspace1}-\ref{eq:parspace4} and \ref{eq:parspace5}) flat
priors in the ranges listed in Table \ref{tab:priors}. The model
parameter $b$ is analytically marginalised over an infinite flat prior \citep[equation F2
in][]{lewis_02}.

All likelihoods used to compare the results from the cosmological probes described in
sections \ref{ssec:dr7}, \ref{ssec:cmb}, \ref{ssec:sn} and \ref{ssec:h_0} with the
corresponding models are assumed to be Gaussian. In the case of the LRGs
power spectrum, this is 
\begin{equation}\label{eq:likely} 
  \mathcal{L}\propto \exp\left(-\frac{1}{2}\chi^{2}({\mathbf \theta})\right), 
\end{equation} 
where
\begin{equation} 
  \chi^{2}(\theta)=(\mathbf{d}-\mathbf{t}(\theta))^{T} {\textsf
  {\mathbfss C}}^{-1} (\mathbf{d}-\mathbf{t}(\mathbf{\theta})) 
\end{equation} 
is the standard $\chi^{2}$, in which $\mathbf{d}$ is an array containing the band power
$P(k)$, $\mathbf{t}(\mathbf{\theta})$ contains the model computed for the set of
parameters $\theta$ and then convolved with the window function and $\mathbfss{C}$ is the covariance matrix of the measurement
presented in section \ref{ssec:mocks_ps}.

In section \ref{ssec:dr7} we assume a fiducial cosmology in order to convert redshifts and
angles to physical coordinates. Different choices of these parameters result in
modifications of the measured LRG power spectrum. The ideal case would be to
recompute, for each step of the MCMC chain, the power spectrum and the window function
according to the given cosmology, but this is not computationally feasible. Under the assumption that the survey covers a wide solid angle, it is possible to incorporate these distortions in the correlation function
when computed using different cosmologies by rescaling the distance scale by the factor
\citep{Eisenstein_05}: 
\begin{equation}\label{eq:cosmcorr} 
  \alpha = \frac{D_{\mathrm{V}}^{\mathrm{model}}(z_{\mathrm{m}})}{D_{\mathrm{V}}^{\mathrm{fid}}(z_{\mathrm{m}})},
\end{equation} 
where the effective distance $D_{\mathrm{V}}(z_{\mathrm{m}})$ to the mean redshift of the sample $z_{\mathrm{m}}$ is,
\begin{equation}\label{eq:effdist} 
  D_{\mathrm{V}}(z) = \left[ D_{\mathrm{A}}^{2}(z_{\mathrm{m}})\frac{cz_{\mathrm{m}}}{H(z_{\mathrm{m}})} \right]^{1/3}, 
\end{equation} 
with $D_{\mathrm{A}}(z_{\mathrm{m}})$ the comoving angular diameter distance. Since the power spectrum is the Fourier transform of the correlation function, this holds also for the former. Thus,
at each step of the chain we multiply the wave-number of the model power by $\alpha$ and its amplitude by $1/\alpha^{3}$ in order to rescale it to the fiducial cosmology.

\subsection{Testing the model}\label{ssec:test}

In the first part of this section we extend the analysis of M10 and fit our model against the mock catalogues in order to test
wether it provides an accurate description of the measured power spectrum and
unbiased constraints of the dark energy equation of state parameter also when a complex
geometry is involved. We also test the stability of the cosmological parameters
in the wCDM case as the maximum $k$ included in the analysis varies, when CMB
and the $P_{\mathrm{LRG}}(k)$ are combined.

\subsubsection{Mock catalogues}\label{sssec:test_mocks}

In this section we follow M10 and assume all the parameters fixed, except for
$w_{\mathrm{DE}}$. Under this assumption, equation (\ref{eq:cosmcorr}) links
univocally variations of the dark energy equation of states to stretches
$\alpha$ of the model power spectrum. Therefore we consider the latter as a a
free parameter \citep{Huff_07}.

Using a MCMC approach, we explore the parameter space defined by
$\mathbf{\theta}=(k_{\star}, \, b,\,A_{\rm{MC}}, \, \alpha )$. We chose priors with a
constant probability within the following ranges: 
\begin{itemize} 
  \item $0 \,h\,\rm{Mpc^{-1}}$$< k_{\star} < 0.35\,h\,\rm{Mpc^{-1}}$, 
  \item $0\le A_{\rm{MC}} < 10$, 
  \item $0.5 \le \alpha < 1.5$.  
\end{itemize} 
The bias $b$ is marginalised analytically over an infinite flat prior as described in section \ref{ssec:issues}.

\begin{figure} 
  \includegraphics[width=79mm, keepaspectratio]{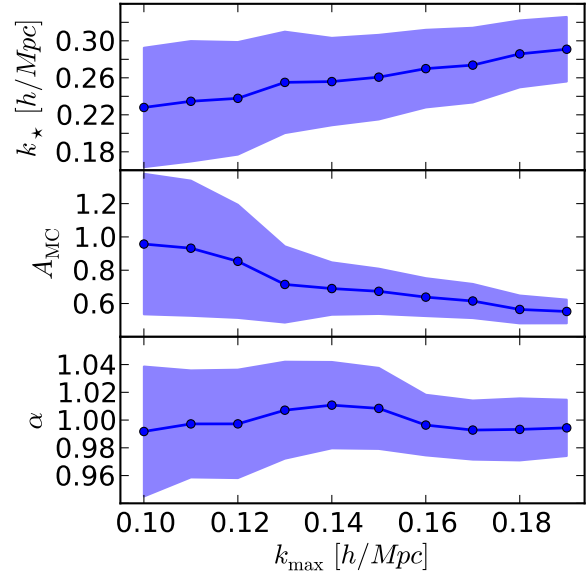} 
  \caption{One-dimensional marginalised constraints on the parameters $k_{\star}$, $A_{\rm{MC}}$ and $\alpha$ as function of the maximum
  value of $k_{i}$ ($k_{\mathrm{max}}$) used to fit the model of equation (\ref{eq:rpt_mod}) to the
  LasDamas mean power spectrum. The mean values and the standard deviation are indicated,
  respectively, by circles connected with solid lines and shaded areas. The maximum value
  of $k$ at which the model is computed before the convolution with the window function is
  $k=0.2 \hoM$.} \label{fig:mocks_parvskmax} 
\end{figure}

Figure \ref{fig:mocks_parvskmax} shows the one-dimensional marginalised constraints on the parameters
$k_{\star}$, $A_{\rm{MC}}$ and $\alpha$ varying the maximum value of the wave number $k_{\mathrm{max}}$ of the measured power spectrum used to perform the fit; we keep the minimum $k$ fixed to $0.02 \hoM$. The blue circles connected by solid lines
and the shaded areas correspond to the mean and the standard deviation of these parameters. For every $k_{\mathrm{max}}$ the model is evaluated for $k\leq0.2 \hoM$ and then convolved with the window function. Since each row of $W(k_i,k_j)$ is sharply peaked at $k_{j} \sim k_{i}$, the main contribution to $P_{\mathrm{o}}(k_i)$ comes from modes near $k_{i}$: therefore the constraints shown in Figure \ref{fig:mocks_parvskmax} depend weakly on the exact wave number range in which the model is computed, as long as it is larger than the range of the measured power spectrum that we fit. As $k_{\mathrm{max}}$ increases, more modes are included in the fit and the errors decrease. The constraints on $\alpha$ are compatible with unity for all
the $k_{\mathrm{max}}$ considered. 
Considering that the volume of the simulations used in M10 is much larger that the one sampled by the LRGs and consequently that the errors in the former case are smaller than in the latter, we decide to further consider scales $0.02 \hoM  \leq k \leq 0.15 \hoM$.  The constraints on $k_{\star}$ and $A_{\mathrm{MC}}$ exhibit, respectively, a monotonic increase and decrease. As explained in M10, the approximate mode coupling power in equation (\ref{eq:rpt_mod}) is about $30\%$ larger than the exact value at $k \approx 0.15-0.2 \hoM$: this forces $A_{\mathrm{MC}}$ to decrease to $\sim 0.7-0.6$ and $k_{\star}$ to increase in
order to maintain the shape of the resulting power spectrum unvaried.

\begin{figure} 
  \includegraphics[width=79mm, keepaspectratio]{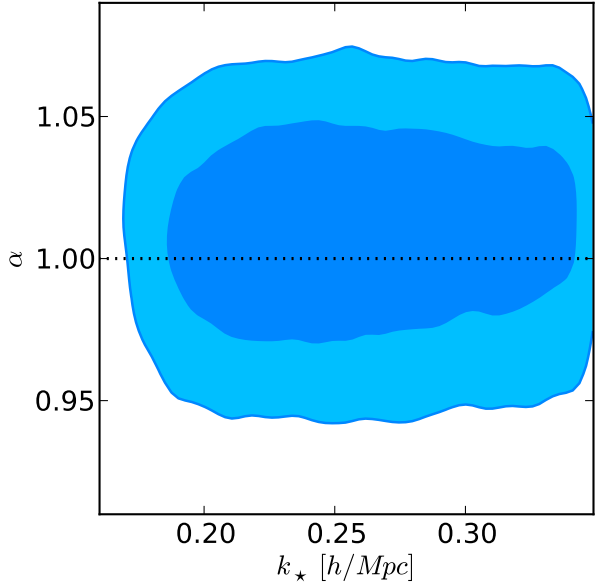} 
  \caption{Two-dimensional marginalised constraints in
  the $k_{\star}-\alpha$ plane obtained applying the model of equation (\ref{eq:rpt_mod}) to the
  LasDamas mean power spectrum. The results are show for $k_{\mathrm{max}}=0.15
  \,h\,\rm{Mpc^{-1}}$. The inner dark and
  outer light areas represent the 68\% and 95\% confidence level, respectively.}
  \label{fig:mocks_alphakstar} 
\end{figure}

Figure \ref{fig:mocks_alphakstar} shows the two-dimensional marginalised constraints in the $k_{\star}-\alpha$
plane as obtained from the mock catalogues for $k_{\mathrm{max}}=0.15 \hoM$. The inner dark and outer light shaded areas represent
regions whose volumes are $68\%$ and $95\%$ of the total likelihood. This representation of the two-dimensional constraints will be used through
all section \ref{sec:cosmopars}. The independence of $\alpha$ from $k_{\star}$ or
$A_{\mathrm{MC}}$ makes the constraints on the former robust. The latter two parameters,
instead, are strongly degenerate, as it is possible to describe accurately the overall
shape of the power spectrum compensating an increase (decrease) of $k_{\star}$ with a
decrease (increase) of $A_{\mathrm{MC}}$.

The model power spectrum indicated by a solid line in the two panes of Figures \ref{fig:meas_powerspmock} has been computed using the best fit values of the parameters as obtained in this section for $k_{\mathrm{max}}=0.15\hoM$: $k_\star=0.26\hoM$, $A_{\mathrm{MC}}=0.61$. The bias has been computed maximising the likelihood of equation (\ref{eq:likely}) with all the other parameters fixed.

\subsubsection{The Luminous red galaxy sample}\label{sssec:test_lrgs}

\begin{figure} 
  \includegraphics[width=79mm, keepaspectratio]{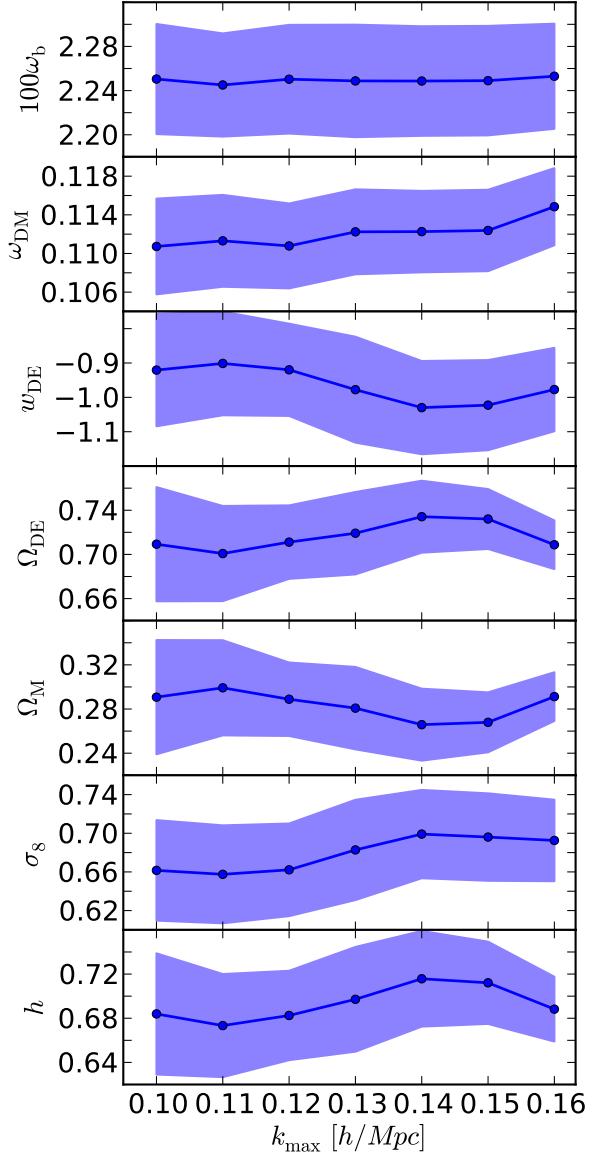}
  \caption{One-dimensional marginalised constraints of the wCDM parameter space on the parameters $\omega_{\mathrm{b}}$,
  $\omega_{\mathrm{DM}}$, $w_{\mathrm{DE}}$,  $\Omega_{\mathrm{DE}}$, $\Omega_{\mathrm{M}}$, $\sigma_{8}$ and
  $h$ as function of the maximum value of $k$ ($k_{\mathrm{max}}$) as obtained
  when combining the LRGs power spectrum with the CMB data. Circles connected by solid
  line and shaded areas are,  respectively, the mean and the 68\% confidence level. The
  model is computed for $k<0.2 \hoM$ and then convolved with the
  window function.} \label{fig:data_parvskmax}
\end{figure}

After testing the robustness of our model at mildly non-linear scales, we test here the
dependence of the cosmological parameters upon $k_{\mathrm{max}}$. We use the wCDM
cosmology defined in equation (\ref{eq:parspace3}) and the combination of the LRG power spectrum and
CMB measurements. Figure \ref{fig:data_parvskmax} shows the one-dimensional marginalised constraints on
the parameters $\omega_{\mathrm{b}}$, $\omega_{\mathrm{DM}}$, $w_{\mathrm{DE}}$,
$\Omega_{\mathrm{DE}}$, $\Omega_{\mathrm{M}}$, $\sigma_{8}$ and $h$ as function of $k_{\mathrm{max}}$. The
circles connected by solid lines and the shaded areas show the mean and the standard
deviation as obtained from the MCMC. The model is computed for $k<0.2 \hoM$ and then convolved with the window function; 
as before, the measured parameters are mostly insensitive to this limit. 
Although some parameters are more stable with respect to changes of 
$k_{\mathrm{max}}$ than others, there are no significant trends or deviations. 
The error-bars of the parameters decrease as $k_{\mathrm{max}}$ increases: 
we interpret this as a sign of new information coming from these scales.

We also test the impact of using CMB measurements from WMAP only. We obtain the
same constraints, but with errors larger by 5-10\%, due to the loss of information at
small angular scales, i.e. large multipoles $l$. Appendix \ref{ap:test_cosm} describes how the results just described depend upon $p_{\mathrm{w}}$ and $w_{\mathrm{i}}$.
A more extensive analysis of the wCDM cosmology is presented in
section \ref{ssec:parspace3}.

\section{Results: the cosmological parameters}\label{sec:cosmopars}

In this section we present the constraints on the cosmological parameters obtained from
the five cosmological scenarios introduced in section \ref{ssec:parspace}.

\subsection{The concordance cosmology}\label{ssec:parspace1}

\begin{table*} 
  \centering 
  \begin{minipage}{160mm} 
    \caption{Marginalised constraints on the cosmological parameters of the $\Lambda$CDM parameter space from
    the combination of probes listed in the header of the table. The quoted values are the
    means and widths of the posterior distribution containing 68\% of the total area.}
    \label{tab:1D_LCDM} 
    \begin{tabular}{ l c c c c c }
      \hline 
      & CMB & CMB+$P(k)$ & CMB+$P(k)$+$H_0$ & CMB+$P(k)$+SNIa & CMB+$P(k)$+$H_0$+SNIa \\
      \hline 
      $100\omega_{\mathrm{b}}$ & $2.254_{-0.052}^{+0.052}$ & $2.258_{-0.048}^{+0.048}$ &
      $2.259_{-0.049}^{+0.050}$ & $2.252_{-0.047}^{+0.047}$ &
      $2.260_{-0.047}^{+0.047}$\\[1.5mm]
      $100\omega_{\mathrm{DM}}$ & $10.96_{-0.52}^{+0.52}$ & $11.23_{-0.36}^{+0.36}$ & $11.10_{-0.35}^{+0.35}$ & $11.22_{-0.33}^{+0.34}$ & $11.13_{-0.32}^{+0.33}$\\[1.5mm]
      100$\Theta$ & $1.0400_{-0.0022}^{+0.0023}$ & $1.0404_{-0.0020}^{+0.0020}$ & $1.0406_{-0.0021}^{+0.0020}$ & $1.0403_{-0.0020}^{+0.0021}$ & $1.0406_{-0.0020}^{+0.0020}$\\[1.5mm]
      $\tau$ & $0.088_{-0.015}^{+0.015}$ & $0.087_{-0.014}^{+0.014}$ &
      $0.086_{-0.014}^{+0.014}$ & $0.085_{-0.015}^{+0.014}$ &
      $0.087_{-0.014}^{+0.014}$\\[1.5mm]
      $n_{\mathrm{s}}$ & $0.963_{-0.013}^{+0.013}$ & $0.963_{-0.012}^{+0.011}$ &
      $0.963_{-0.011}^{+0.011}$ & $0.961_{-0.011}^{+0.012}$ &
      $0.963_{-0.011}^{+0.011}$\\[1.5mm]
      $\log(10^{10}\,A_{\mathrm{s}})$ & $3.065_{-0.033}^{+0.033}$ &
      $3.075_{-0.031}^{+0.031}$ & $3.068_{-0.033}^{+0.033}$ & $3.070_{-0.032}^{+0.032}$ &
      $3.071_{-0.032}^{+0.032}$\\[1.5mm]
      $\Omega_{\mathrm{DE}}$ & $0.741_{-0.026}^{+0.026}$ & $0.730_{-0.018}^{+0.018}$ &
      $0.737_{-0.017}^{+0.017}$ & $0.730_{-0.017}^{+0.017}$ &
      $0.735_{-0.016}^{+0.015}$\\[1.5mm]
      $\mathrm{Age\,[Gyr]}$ & $13.722_{-0.112}^{+0.111}$ & $13.723_{-0.094}^{+0.094}$ &
      $13.711_{-0.093}^{+0.095}$ & $13.735_{-0.092}^{+0.094}$ &
      $13.710_{-0.094}^{+0.091}$\\[1.5mm]
      $\Omega_{\mathrm{M}}$ & $0.259_{-0.026}^{+0.026}$ & $0.270_{-0.018}^{+0.018}$ &
      $0.263_{-0.017}^{+0.017}$ & $0.270_{-0.017}^{+0.017}$ &
      $0.265_{-0.015}^{+0.016}$\\[1.5mm]
      $\sigma_8$ & $0.796_{-0.027}^{+0.027}$ & $0.807_{-0.022}^{+0.022}$ &
      $0.799_{-0.022}^{+0.022}$ & $0.804_{-0.021}^{+0.021}$ &
      $0.802_{-0.021}^{+0.021}$\\[1.5mm]
      $z_{\mathrm{re}}$ & $10.4_{-1.2}^{+1.2}$ & $10.4_{-1.2}^{+1.1}$ & $10.3_{-1.2}^{+1.2}$ & $10.3_{-1.2}^{+1.2}$ & $10.4_{-1.2}^{+1.2}$\\[1.5mm]
      $H_0\,[\mathrm{km}\,\mathrm{s^{-1}\,Mpc^{-1}}]$ & $71.7_{-2.3}^{+2.4}$ & $70.8_{-1.6}^{+1.6}$ & $71.3_{-1.5}^{+1.5}$ & $70.7_{-1.5}^{+1.5}$ & $71.2_{-1.4}^{+1.4}$\\[1.5mm]
      \hline 
    \end{tabular} 
  \end{minipage}
\end{table*}

The flat $\Lambda$CDM cosmology, parametrized by the first six quantities of
equation (\ref{eq:parspace1}), is the minimal model able to describe data coming from
many independent probes. Table \ref{tab:1D_LCDM} summarises the complete list of one-dimensional constraints for the primary
and derived cosmological parameters obtained in this section. We quote the mean
values and width of the posterior distribution
containing 68\% of the total area, which for a Gaussian distribution corresponds to the standard deviation, as obtained for the five different combinations of the four experiments used in this work. This convention will be followed in the
rest of this article.

\begin{figure}
  \includegraphics[width=79mm, keepaspectratio]{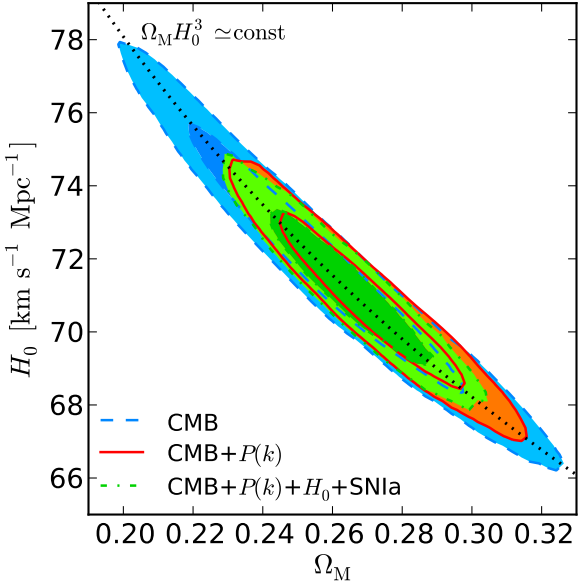}
  \caption{Two-dimensional marginalised constraints of the $\Lambda$CDM parameter space in the $\Omega_{\mathrm{DM}}-H_{0}$
  plane. Blue, red and green shaded areas enclosed in dashed, solid and dot-dashed lines shows
  the constraints from CMB alone, CMB+$P_{\mathrm{LRG}}(k)$ and
  CMB+$P_{\mathrm{LRG}}(k)$+H$_{0}$+SNIa, respectively. The inner darker and the outer
  lighter areas are the 68\% and 95\% confidence level. The black dotted line that runs across the figure shows
  the locus defined by equation $\Omega_{\mathrm{M}}H_{0}^{3}\simeq
  \mathrm{const}$.}\label{fig:LCDM_odm_H0}
\end{figure}

The CMB experiments described in section \ref{ssec:cmb} provide measurements of
the temperature and polarization angular power spectra with very high accuracy.
The blue shaded areas enclosed in dashed lines in Figure \ref{fig:LCDM_odm_H0}
show the 68\% and 95\% confidence level in the $\Omega_{\mathrm{DM}}-H_{0}$ as
obtained when CMB information only is used.  The apparent position of the peaks
in the CMB power spectra is proportional to their physical scale, which depends
on the composition of the early universe (baryons, dark matter and radiation),
and the angular diameter distance to the last scattering surface, a function of
$H_{0}$ and of the density and equation of state parameter of matter, dark energy and of
curvature. Since here we consider a flat geometry and fix $w_{\mathrm{DE}}$ to
-1, a degeneracy between the matter density and the Hubble
parameter appears.  It has been shown by \citet{percival02} that this effect,
together with the preservation of the relative amplitude of the peaks, leads,
in a $\Lambda$CDM universe, to a degeneracy along the curve defined by
$\Omega_{\mathrm{M}}h^{3} \simeq \mathrm{const}$, which is displayed in
Figure \ref{fig:LCDM_odm_H0} by the dotted line. The accurate detection of the
third peak in the temperature power spectrum, whose relative amplitude with
respect to the first two is proportional to the matter-radiation ratio, helps
reducing this degeneracy. In this case we measure
$\Omega_{\mathrm{M}}=0.259\pm0.026$, $H_0=71.7^{+2.4}_{-2.3}\,
\mathrm{km}\,\mathrm{s^{-1}\,Mpc^{-1}}$ and
$\omega_{\mathrm{DM}}=\left(10.96\pm0.52\right)\times10^{-2}$.

The inclusion of information from the large scale structure can break or reduce some of
the degeneracies in the CMB. The shape of the power spectrum depends upon 
$\Omega_{\mathrm{M}}h$ and, more weakly, $\Omega_{\mathrm{b}}/\Omega_{\mathrm{M}}$
\citep[e.g.,][]{efstathiou02,sanchez_cole}. Thanks to this, the errors on the cosmological
parameters decrease up to about 30\%. In particular, we measure a decrease in the allowed region
by about one third in the three parameters considered before:
$\Omega_{\mathrm{M}}=0.27\pm0.018$, $H_0=70.8\pm1.6\,
\mathrm{km\,s^{-1}\,Mpc^{-1}}$ and $\omega_{\mathrm{DM}} =\left(11.23_{-0.36}^{+0.36}\right)\times10^{-2}$. The two-dimensional
constraints in the $\Omega_{\mathrm{M}}-H_0$ plane are shown in Figure \ref{fig:LCDM_odm_H0} with
the red shaded areas within solid lines.

When the SNIa and H$_{0}$ measurements are also used, we obtain
$\Omega_{\mathrm{M}}=0.265_{-0.015}^{+0.016}$, $H_{0}=71.2\pm1.4\,
\mathrm{km\,s^{-1}\,Mpc^{-1}}$ and $\omega_{\mathrm{DM}}=\left(11.13_{-0.32}^{+0.33}\right)\times10^{-2}$, which
means a 10-15\% increase in accuracy. The two dimensional 68\% and 95\% confidence levels for the
former two parameters, when all the four probes are used, are shown in Figure \ref{fig:LCDM_odm_H0}
by the green shaded areas enclosed by dot-dashed lines.
 
\begin{figure}
  \includegraphics[width=79mm, keepaspectratio]{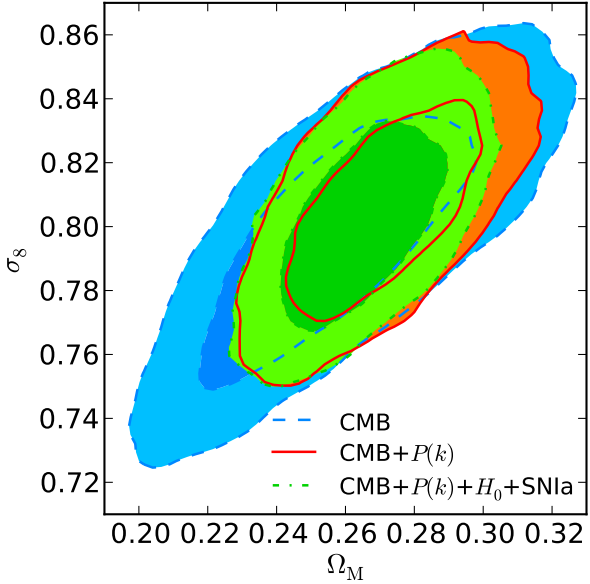}
  \caption{Two-dimensional marginalised constraints of the $\Lambda$CDM parameter space in the
  $\Omega_{\mathrm{m}}-\sigma_{8}$ plane. The colour and line coding is the same as in
  Figure \ref{fig:LCDM_odm_H0}.}\label{fig:LCDM_om_s8}
\end{figure}

Figure \ref{fig:LCDM_om_s8} shows the two-dimensional marginalised constraints in the
$\Omega_{\mathrm{M}}-\sigma_{8}$ plane for the same combination of datasets; colour
and line coding is the same as in Figure \ref{fig:LCDM_odm_H0}. The correlation between the two parameters is
caused by the fact that an increase (decrease) of $\Omega_{\mathrm{M}}$ causes a decrease
(increase) in the amplitude of the power spectrum that can be compensated by a larger
(smaller) value of $\sigma_{8}$. For the three cases shown in the figure we find that
the one-dimensional constraints on the latter parameter are, respectively,
$\sigma_8=0.796\pm0.027$, $\sigma_8=0.807\pm0.022$ and $\sigma_8=0.802\pm0.021$; the
error on the latter two decreases by 20 and 22\% respectively to the CMB only result.

The solid lines in Figures \ref{fig:meas_powersp} and \ref{fig:meas_power} show the model power spectrum computed using the cosmological parameters listed in the last column of Table \ref{tab:1D_LCDM} and the best fit model parameters $k_\star=0.28\hoM$ and $A_{\mathrm{MC}}=0.72$, as obtained from the MCMC for $k_{\mathrm{max}}=0.15\hoM$. The bias has been computed maximising the likelihood of equation (\ref{eq:likely}) with all the other parameters fixed.

\begin{table}
  \centering 
  \begin{minipage}{75mm} 
    \caption{Marginalised constraints on the cosmological parameters of the $\Lambda$CDM parameter space from
    the combination of CMB+$P_{\mathrm{LRG}}(k)$+$H_0$+SNIa when systematic errors are not considered and when
    the \textsc{mlcs2k2} SNIa light curve fitter is used.}
    \label{tab:1D_LCDM_othersn} 
      \begin{tabular}{ l c c } 
      \hline 
      & no systematics & \textsc{mlcs2k2} \\ 
      \hline 
      $100\omega_{\mathrm{b}}$ & $2.258_{-0.048}^{+0.046}$ & $2.244_{-0.046}^{+0.047}$\\[1.5mm] 
      $100\omega_{\mathrm{DM}}$ & $11.16_{-0.29}^{+0.28}$ & $11.61_{-0.31}^{+0.31}$\\[1.5mm]
      100$\Theta$ & $1.0405_{-0.0020}^{+0.0020}$ & $1.0402_{-0.0020}^{+0.0021}$\\[1.5mm]
      $\tau$ & $0.087_{-0.014}^{+0.014}$ & $0.084_{-0.014}^{+0.014}$\\[1.5mm] 
      $n_{\mathrm{s}}$ & $0.963_{-0.011}^{+0.011}$ & $0.957_{-0.011}^{+0.011}$\\[1.5mm] 
      $\log(10^{10}\,A_{\mathrm{s}})$ & $3.072_{-0.033}^{+0.033}$ & $3.081_{-0.032}^{+0.031}$\\[1.5mm] 
      $\Omega_{\mathrm{DE}}$ & $0.734_{-0.013}^{+0.013}$ & $0.709_{-0.016}^{+0.016}$\\[1.5mm] 
      $\mathrm{Age\,[Gyr]}$ & $13.717_{-0.089}^{+0.089}$ & $13.765_{-0.090}^{+0.090}$\\[1.5mm] 
      $\Omega_{\mathrm{M}}$ & $0.266_{-0.013}^{+0.013}$ & $0.291_{-0.016}^{+0.016}$\\[1.5mm] 
      $\sigma_8$ & $0.803_{-0.020}^{+0.020}$ & $0.822_{-0.021}^{+0.021}$\\[1.5mm] 
      $z_{\mathrm{re}}$ & $10.4_{-1.2}^{+1.2}$ & $10.3_{-1.2}^{+1.2}$\\[1.5mm]
      $H_0\,[\mathrm{km}\,\mathrm{s^{-1}\,Mpc^{-1}}]$ & $71.0_{-1.2}^{+1.2}$ & $69.1_{-1.3}^{+1.3}$\\[1.5mm]
      \hline 
    \end{tabular} 
  \end{minipage} 
\end{table}

\subsubsection*{Effects of supernovae systematics and light curve fitters}

\begin{table*} 
  \centering 
  \begin{minipage}{160mm} 
    \caption{Marginalised constraints on the cosmological parameters of the k$\Lambda$CDM parameter space
      from the combination of probes listed in the header of the table. The quoted values
      are the same as in Table \ref{tab:1D_LCDM}.}
    \label{tab:1D_kLCDM} 
    \begin{tabular}{ l c c c c c }
      \hline 
      & CMB & CMB+$P(k)$ & CMB+$P(k)$+$H_0$ & CMB+$P(k)$+SNIa & CMB+$P(k)$+$H_0$+SNIa \\
      \hline 
      $100\omega_{\mathrm{b}}$ & $2.232_{-0.050}^{+0.052}$ & $2.250_{-0.047}^{+0.048}$ &
      $2.252_{-0.049}^{+0.050}$ & $2.245_{-0.048}^{+0.050}$ &
      $2.247_{-0.050}^{+0.049}$\\[1.5mm]
      $100\omega_{\mathrm{DM}}$ & $11.06_{-0.50}^{+0.50}$ & $11.29_{-0.45}^{+0.44}$ & $11.30_{-0.45}^{+0.45}$ & $11.30_{-0.41}^{+0.42}$ & $11.31_{-0.44}^{+0.43}$\\[1.5mm]
      100$\Theta$ & $1.0396_{-0.0022}^{+0.0022}$ & $1.0401_{-0.0021}^{+0.0022}$ & $1.0402_{-0.0022}^{+0.0022}$ & $1.0400_{-0.0021}^{+0.0021}$ & $1.0400_{-0.0021}^{+0.0022}$\\[1.5mm]
      $\tau$ & $0.086_{-0.015}^{+0.014}$ & $0.085_{-0.014}^{+0.014}$ &
      $0.085_{-0.014}^{+0.014}$ & $0.085_{-0.014}^{+0.014}$ &
      $0.085_{-0.014}^{+0.015}$\\[1.5mm]
      $100\Omega_{\mathrm{k}}$ & $-4.90_{-5.10}^{+4.39}$ & $0.16_{-0.54}^{+0.54}$ & $0.30_{-0.49}^{+0.49}$ & $0.14_{-0.53}^{+0.51}$ & $0.30_{-0.47}^{+0.48}$\\[1.5mm]
      $n_{\mathrm{s}}$ & $0.956_{-0.013}^{+0.013}$ & $0.961_{-0.012}^{+0.012}$ &
      $0.961_{-0.012}^{+0.012}$ & $0.960_{-0.012}^{+0.012}$ &
      $0.960_{-0.012}^{+0.012}$\\[1.5mm]
      $\log(10^{10}\,A_{\mathrm{s}})$ & $3.062_{-0.033}^{+0.033}$ &
      $3.072_{-0.031}^{+0.030}$ & $3.073_{-0.032}^{+0.032}$ & $3.072_{-0.032}^{+0.032}$ &
      $3.073_{-0.033}^{+0.033}$\\[1.5mm]
      $\Omega_{\mathrm{DE}}$ & $0.597_{-0.149}^{+0.132}$ & $0.731_{-0.019}^{+0.019}$ &
      $0.736_{-0.016}^{+0.016}$ & $0.730_{-0.016}^{+0.016}$ &
      $0.734_{-0.015}^{+0.015}$\\[1.5mm]
      $\mathrm{Age\,[Gyr]}$ & $15.37_{-1.37}^{+1.46}$ & $13.65_{-0.27}^{+0.27}$ & $13.58_{-0.24}^{+0.23}$ & $13.67_{-0.26}^{+0.26}$ & $13.59_{-0.24}^{+0.24}$\\[1.5mm]
      $\Omega_{\mathrm{M}}$ & $0.452_{-0.175}^{+0.200}$ & $0.267_{-0.021}^{+0.020}$ &
      $0.261_{-0.017}^{+0.017}$ & $0.269_{-0.017}^{+0.018}$ &
      $0.263_{-0.015}^{+0.016}$\\[1.5mm]
      $\sigma_8$ & $0.773_{-0.032}^{+0.032}$ & $0.808_{-0.025}^{+0.024}$ &
      $0.809_{-0.026}^{+0.026}$ & $0.808_{-0.025}^{+0.025}$ &
      $0.810_{-0.026}^{+0.025}$\\[1.5mm]
      $z_{\mathrm{re}}$ & $10.3_{-1.2}^{+1.2}$ & $10.3_{-1.1}^{+1.2}$ & $10.3_{-1.2}^{+1.2}$ & $10.3_{-1.1}^{+1.2}$ & $10.4_{-1.2}^{+1.2}$\\[1.5mm]
      $H_0\,[\mathrm{km}\,\mathrm{s^{-1}\,Mpc^{-1}}]$ & $57.3_{-12.0}^{+11.5}$ & $71.3_{-2.4}^{+2.5}$ & $72.1_{-2.0}^{+2.1}$ & $71.1_{-2.2}^{+2.2}$ & $71.9_{-2.0}^{+1.9}$\\[1.5mm]
      \hline 
    \end{tabular} 
  \end{minipage}
\end{table*}

\begin{figure*} 
  \begin{minipage}{80mm} 
    \includegraphics[width=79mm, keepaspectratio]{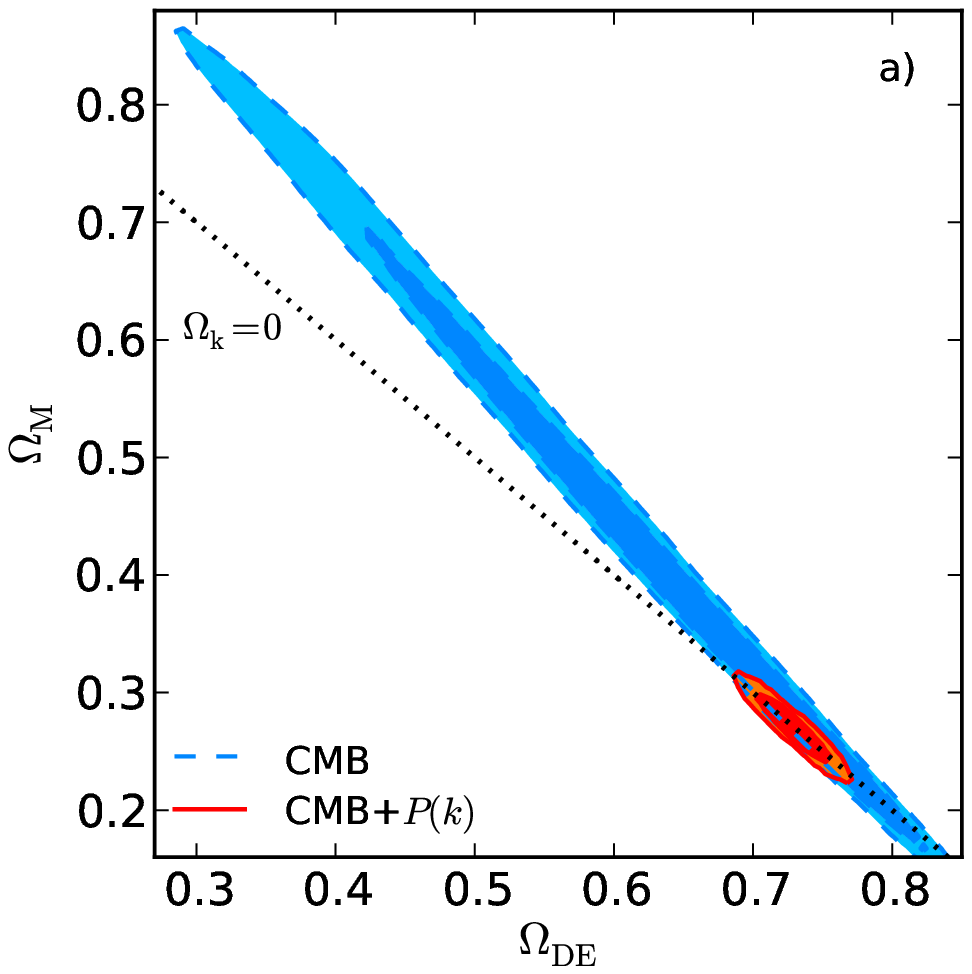} 
  \end{minipage} 
  \begin{minipage}{80mm}
    \includegraphics[width=79mm, keepaspectratio]{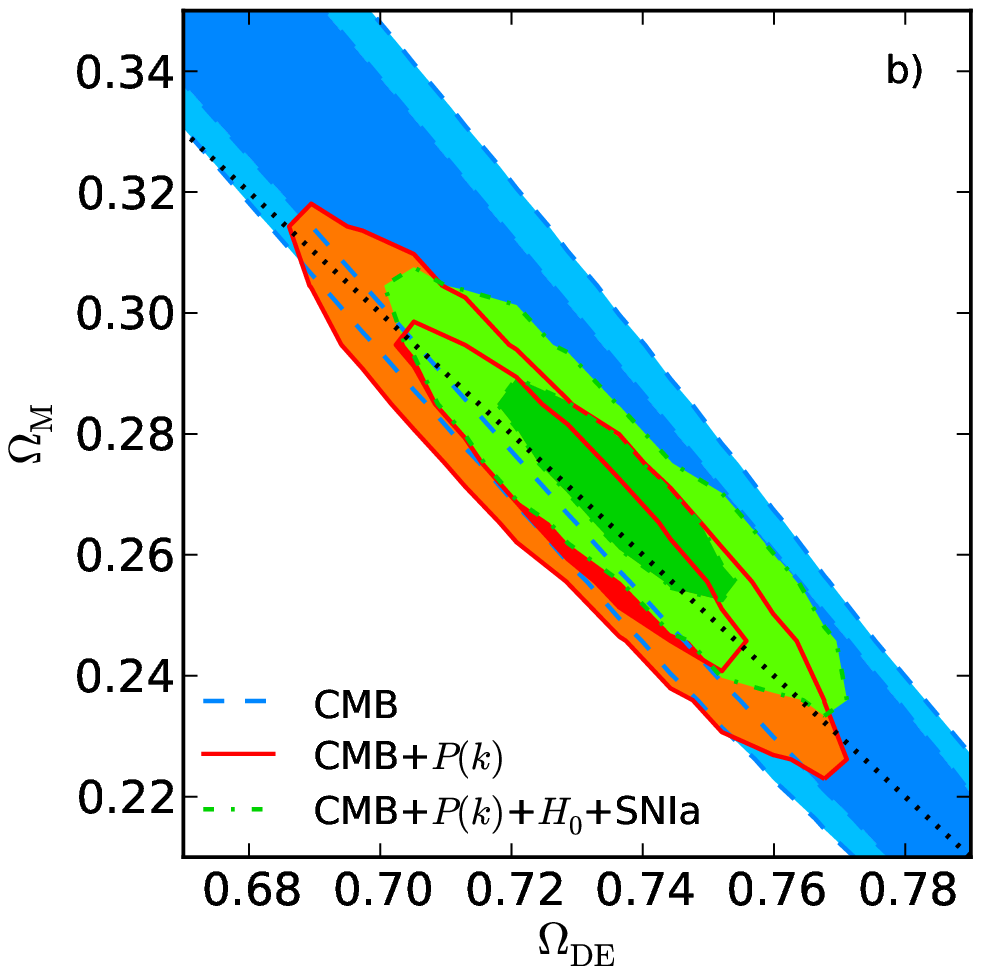} 
  \end{minipage}
  \caption{Two-dimensional marginalised constraints of the k$\Lambda$CDM parameter space in the
  $\Omega_{\mathrm{DE}}-\Omega_{\mathrm{M}}$ plane. Panel \emph{b)} zooms into panel \emph{a)} in order to highlight
  the constraints obtained when combining CMB with the other datasets used in this work.
  The diagonal dotted line is for a flat Universe.  Colour and line coding is the same as in
  Figure \ref{fig:LCDM_odm_H0}.}\label{fig:kLCDM_ode_om} 
\end{figure*}

Uncertainties in the modelling of SNIa can affect the cosmological parameters and
associated errors extracted using a given dataset. To test the impact of systematics and
light curve fitters on the results just presented, we re-analyse the $\Lambda$CDM
cosmology with two different SNIa settings. First we use the same Union2 set, but
neglecting the systematic errors provided with the data; second we substitute this dataset
with the Sloan Digital Sky Survey-II Supernova Survey sample \citep[SDSS
SN,][]{kessler_09_SN}. The sample consists of 288 supernovae from 5 different experiments whose light curves have been fitted both with the \textsc{mlcs2k2} \citep{jha_07_SN} and
\textsc{salt2} models. Since most of the objects in SDSS
SN are also part of the Union2, we do not expect large differences between the two samples if \textsc{salt2} is used. Therefore we show here only the results obtained using the \textsc{mlcs2k2} fitter. 
The one-dimensional constraints that we extract from the combination of all four
probes, when the systematics in the Union2 are ignored and when the SDSS SN with
\textsc{mlcs2k2} are used, are listed in Table \ref{tab:1D_LCDM_othersn}. As expected, and in
agreement with \citet{amanullah_10_SN}, ignoring systematic errors globally reduces the
errors on the recovered values of the parameters without changing sensibly the mean value.
As example we obtain $\Omega_{\mathrm{M}} = 0.266\pm0.013$ (14\% decrease with respect to
the corresponding case in Table \ref{tab:1D_LCDM}), $H_0 = 71\pm1.2
\mathrm{km\,s^{-1}\,Mpc^{-1}}$ (12\%), $\omega_{\mathrm{DM}} = \left(11.16_{-0.29}^{+0.28}\right)\times10^{-2}$ (14\%) and
$\sigma_{8} = 0.803\pm0.02$ (5\%). On the other hand, the use of \textsc{mlcs2k2}
changes the posterior distribution sensibly, without influencing its width. We measure
that $\Omega_{\mathrm{M}}$, $\omega_{\mathrm{DM}}$ and $H_0$ change by more than
1.5-2$\sigma$. This shift agrees with the findings in \citet{kessler_09_SN} and \citet{bengochea_11}
and suggests, in our opinion, that the choice of the model used to fit the light curves can
bias sensibly the cosmological results obtained.

\subsection{Curvature}\label{ssec:parspace2}

In this section we analyse the cosmological constraints obtained when the curvature is
considered as a free parameter. The full list of parameters (mean plus the 68\% confidence
level) is shown in Table \ref{tab:1D_kLCDM} for the combination of experiments listed in the
header of the table.

\begin{figure} 
  \includegraphics[width=79mm, keepaspectratio]{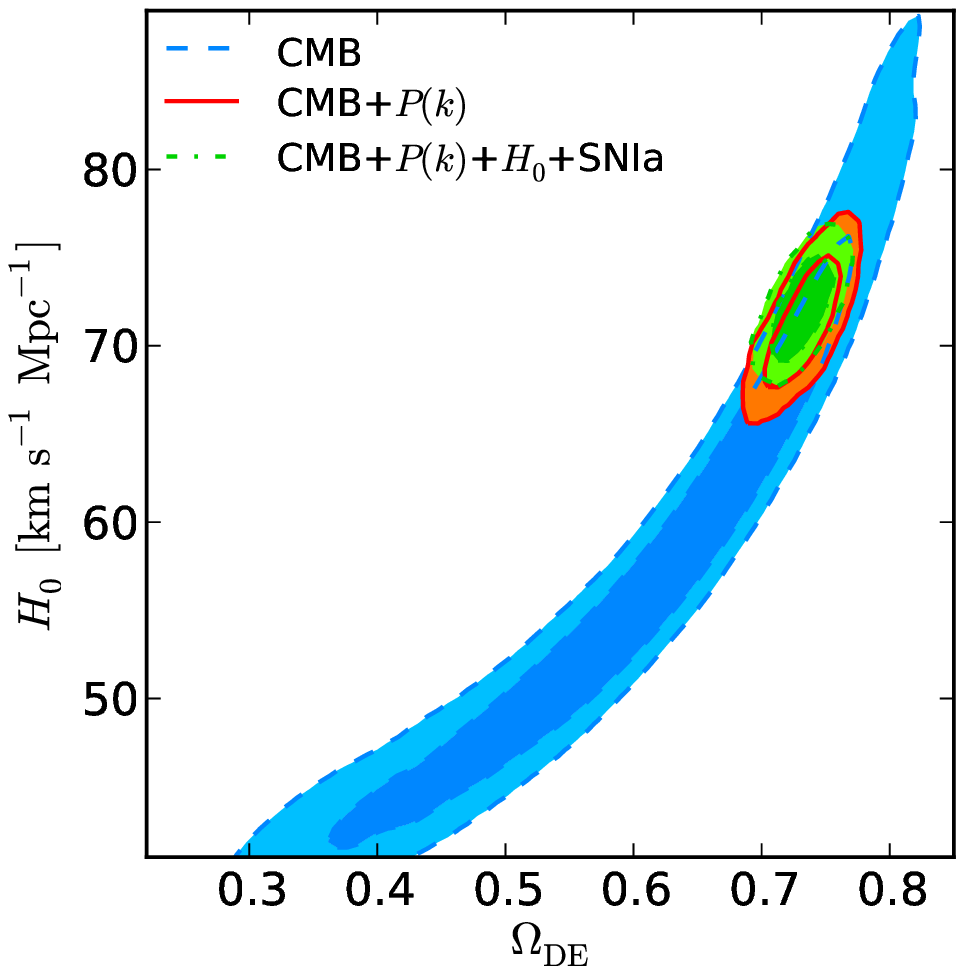}
  \caption{Two-dimensional marginalised constraints of the k$\Lambda$CDM parameter space in the $\Omega_{\mathrm{DE}}-H_{0}$
  plane. Colour and line coding is the same as in
  Figure \ref{fig:LCDM_odm_H0}.}\label{fig:kLCDM_ode_H0} 
\end{figure}

The blue shaded areas within dashed lines in panel a) of Figure \ref{fig:kLCDM_ode_om} and in Figure \ref{fig:kLCDM_ode_H0} show the two-dimensional marginalised constraints in the $\Omega_{\mathrm{DE}}-\Omega_{\mathrm{M}}$ and $\Omega_{\mathrm{DE}}-H_{0}$ planes, respectively, as obtained from CMB data alone. The plots show a very strong degeneracy between these parameters. 
As stated in Section \ref{ssec:parspace1}, the apparent size of the CMB acoustic peaks depends on their physical size and the angular diameter distance. Given that curvature density is small and that it scales as the inverse square of the scale factor, it affects much more $D_{\mathrm{A}}$ than the dynamics of the early universe. It is always possible to find combinations of $\Omega_{\mathrm{k}}$, $\omega_{\mathrm{m}}$ and $\omega_{\mathrm{b}}$ that keeps the angular size of the CMB peaks constant.
Therefore, the constraints on the derived parameters are much weaker than the ones obtained in the previous section:
$\Omega_{\mathrm{DE}} = 0.6_{-0.15}^{+0.13}$, $\Omega_{\mathrm{M}} = 0.45_{-0.17}^{+0.20}$ and $H_{0} =57_{-12}^{+11}\,\mathrm{km\,s^{-1}\,Mpc^{-1}}$. The measured curvature, $\Omega_{\mathrm{k}} = \left(-4.9_{-5.1}^{+4.4}\right) \times 10^{-2}$, is compatible with flatness at about the 1$\sigma$ level. 

Both panels in Figure \ref{fig:kLCDM_ode_om} and Figure \ref{fig:kLCDM_ode_H0} show, with red shaded areas enclosed in solid lines, the same parameter planes as above when explored  combining CMB and large scale structure information. Panel b) of Figure \ref{fig:kLCDM_ode_om} shows a zoom of the constraints of panel a) in the area around $\Omega_{\mathrm{DE}}=0.73$ and $\Omega_{\mathrm{M}}=0.28$ in order to show with more details the constraints obtained when more information is added to the CMB data. The galaxy power spectrum is very sensitive to the matter density. Additionally the BAOs allow to measure the angular diameter distance to the mean redshift of the LRG sample, $\bar{z}=0.313$. The degeneracies in the CMB alone are therefore strongly reduced when LSS measurements are included. The four variables discussed in the previous paragraphs become $\Omega_{\mathrm{DE}} = 0.731\pm0.019$, $\Omega_{\mathrm{M}} = 0.267_{-0.021}^{-0.020}$, $H_{0} = 71.3\pm2.5$ and $\Omega_{\mathrm{k}} = \left(1.6\pm5.4\right) \times 10^{-3}$: their errors are almost one order of magnitude smaller than in the CMB only case. With respect to the corresponding $\Lambda$CDM case, the uncertainties on the first three quantities increase up to 50\%.

The inclusion of SNIa and $H_{0}$ measurement, decreases the errors on curvature by about 10\% ($\Omega_{\mathrm{k}} =  \left[3^{+4.8}_{-4.7}\right] \times 10^{-3}$) and on the other three parameters considered before by circa 20\% ($\Omega_{\mathrm{DE}} =0.734_{-0.015}^{+0.015}$, $\Omega_{\mathrm{M}} =0.263_{-0.015}^{+0.016}$ and $H_{0} = 71.9_{-2.0}^{+1.9}$). When comparing with the results from the previous section, the errors on dark energy and matter density are unchanged, while increasing by less than 40\% for the Hubble parameters. 
The two-dimensional marginalised constraints in the $\Omega_{\mathrm{DE}}-\Omega_{\mathrm{M}}$ and $\Omega_{\mathrm{DE}}-H_{0}$ planes are shown in panel b) of Figure \ref{fig:kLCDM_ode_om} and in Figure \ref{fig:kLCDM_ode_H0} with green shaded lines enclosed in dot-dashed lines.

\subsection{Beyond the cosmological constant}\label{ssec:parspace3}

\begin{table*} 
  \centering 
  \begin{minipage}{160mm} 
    \caption{Marginalised constraints on the cosmological parameters of the wCDM parameter space
      from the combination of probes listed in the header of the table. The quoted values
      are the same as in Table \ref{tab:1D_LCDM}.}
    \label{tab:1D_wCDM} 
    \begin{tabular}{ l c c c c c } 
      \hline 
      & CMB & CMB+$P(k)$ & CMB+$P(k)$+$H_0$ & CMB+$P(k)$+SNIa & CMB+$P(k)$+$H_0$+SNIa \\
      \hline 
      $100\omega_{\mathrm{b}}$ & $2.242_{-0.054}^{+0.053}$ &
      $2.251_{-0.050}^{+0.049}$ & $2.256_{-0.049}^{+0.049}$ & $2.250_{-0.052}^{+0.051}$ &
      $2.257_{-0.050}^{+0.048}$\\[1.5mm]
      $100\omega_{\mathrm{DM}}$ & $11.01_{-0.53}^{+0.53}$ & $11.25_{-0.43}^{+0.43}$ & $11.28_{-0.44}^{+0.44}$ & $11.26_{-0.41}^{+0.40}$ & $11.25_{-0.40}^{+0.40}$\\[1.5mm]
      100$\Theta$ & $1.0397_{-0.0022}^{+0.0023}$ & $1.0402_{-0.0022}^{+0.0022}$ & $1.0403_{-0.0021}^{+0.0021}$ & $1.0403_{-0.0021}^{+0.0021}$ & $1.0404_{-0.0020}^{+0.0020}$\\[1.5mm]
      $\tau$ & $0.088_{-0.015}^{+0.015}$ &
      $0.086_{-0.015}^{+0.015}$ & $0.087_{-0.014}^{+0.014}$ & $0.086_{-0.014}^{+0.015}$ &
      $0.086_{-0.014}^{+0.014}$\\[1.5mm]
      $w_{\mathrm{DE}}$ & $-0.742_{-0.303}^{+0.324}$ & $-1.022_{-0.128}^{+0.129}$ & $-1.069_{-0.106}^{+0.107}$ & $-1.009_{-0.069}^{+0.069}$ & $-1.025_{-0.065}^{+0.066}$\\[1.5mm]
      $n_{\mathrm{s}}$ & $0.959_{-0.014}^{+0.014}$ &
      $0.962_{-0.012}^{+0.012}$ & $0.962_{-0.012}^{+0.012}$ & $0.961_{-0.012}^{+0.012}$ &
      $0.963_{-0.012}^{+0.012}$\\[1.5mm]
      $\log(10^{10}\,A_{\mathrm{s}})$ & $3.064_{-0.033}^{+0.032}$ &
      $3.073_{-0.033}^{+0.034}$ & $3.076_{-0.032}^{+0.033}$ & $3.073_{-0.032}^{+0.032}$ &
      $3.074_{-0.031}^{+0.032}$\\[1.5mm]
      $\Omega_{\mathrm{DE}}$ & $0.637_{-0.133}^{+0.123}$ & $0.732_{-0.028}^{+0.028}$ &
      $0.743_{-0.020}^{+0.020}$ & $0.729_{-0.018}^{+0.018}$ &
      $0.735_{-0.016}^{+0.015}$\\[1.5mm]
      $\mathrm{Age\,[Gyr]}$ &
      $14.062_{-0.396}^{+0.431}$ & $13.733_{-0.117}^{+0.118}$ & $13.692_{-0.099}^{+0.095}$
      & $13.736_{-0.102}^{+0.102}$ & $13.713_{-0.089}^{+0.090}$\\[1.5mm]
      $\Omega_{\mathrm{M}}$ & $0.363_{-0.123}^{+0.133}$ & $0.268_{-0.028}^{+0.028}$ &
      $0.257_{-0.020}^{+0.020}$ & $0.271_{-0.018}^{+0.018}$ &
      $0.265_{-0.015}^{+0.016}$\\[1.5mm]
      $\sigma_8$ & $0.726_{-0.090}^{+0.085}$ &
      $0.818_{-0.064}^{+0.065}$ & $0.838_{-0.060}^{+0.059}$ & $0.810_{-0.043}^{+0.042}$ &
      $0.817_{-0.042}^{+0.043}$\\[1.5mm]
      $z_{\mathrm{re}}$ & $10.6_{-1.2}^{+1.2}$ & $10.4_{-1.2}^{+1.2}$ & $10.4_{-1.1}^{+1.1}$ & $10.4_{-1.2}^{+1.2}$ & $10.3_{-1.2}^{+1.2}$\\[1.5mm]
      $H_0\,[\mathrm{km}\,\mathrm{s^{-1}\,Mpc^{-1}}]$ & $63.1_{-11.2}^{+10.6}$ & $71.2_{-3.8}^{+3.8}$ & $72.8_{-2.8}^{+2.8}$ & $70.7_{-2.0}^{+2.0}$ & $71.4_{-1.7}^{+1.7}$\\[1.5mm]
      \hline    
    \end{tabular} 
  \end{minipage}
\end{table*}

In sections \ref{ssec:parspace1} and \ref{ssec:parspace2} we assume that dark
energy is modelled as a cosmological constant. Despite its simplicity and its
success in describing simultaneously many independent observations, the present
day value of the density of dark energy does not have a solid physical
explanation; this led, in the past decade, to the exploration of a large number
of alternative model, most of which present a time dependent equation of
state\footnote{All theories of modified gravity can be also represented through
an effective $w_{\mathrm{DE}}(z)$.}. Ideally, one would like to be able to
constrain the full time, or redshift, dependence of $w_{\mathrm{DE}}(z)$ in
order to restrict the range of possible models. Usually, parametric forms for
the dark energy equation of state are assumed, which allow to measure time
dependencies, but do not necessarily reproduce the correct
$w_{\mathrm{DE}}(z)$. With non-parametric approaches and principal component
analysis it is possible to overcome some of these limitations
\citep[e.g.,][]{huterer_03, serra_09, holsclaw_10}. In this section we assume
the simplest parametric form possible: $w_{\mathrm{DE}}$ is a constant,
independent of time, with a flat prior in the range $[-2,0]$. Deviations from
the value $w_{\mathrm{DE}}=-1$ would suggest that the cosmological
constant is not a viable model of dark energy. If this is the case, the results
shown in the previous two sections could be biased as consequence of wrongly
assuming that dark energy is described by $\Lambda$. In section
\ref{ssec:parspace5} we analyse an alternative scenario in which
$w_{\mathrm{DE}}$ is parametrized as a linear function of the scale factor.

Table \ref{tab:1D_wCDM} lists the constraints on the cosmological parameters as obtained for the different combinations of datasets analysed in this work. They are in perfect agreement with the ones presented with the previous two parameter spaces.

\begin{figure*} 
  \begin{minipage}{80mm} 
    \includegraphics[width=79mm, keepaspectratio]{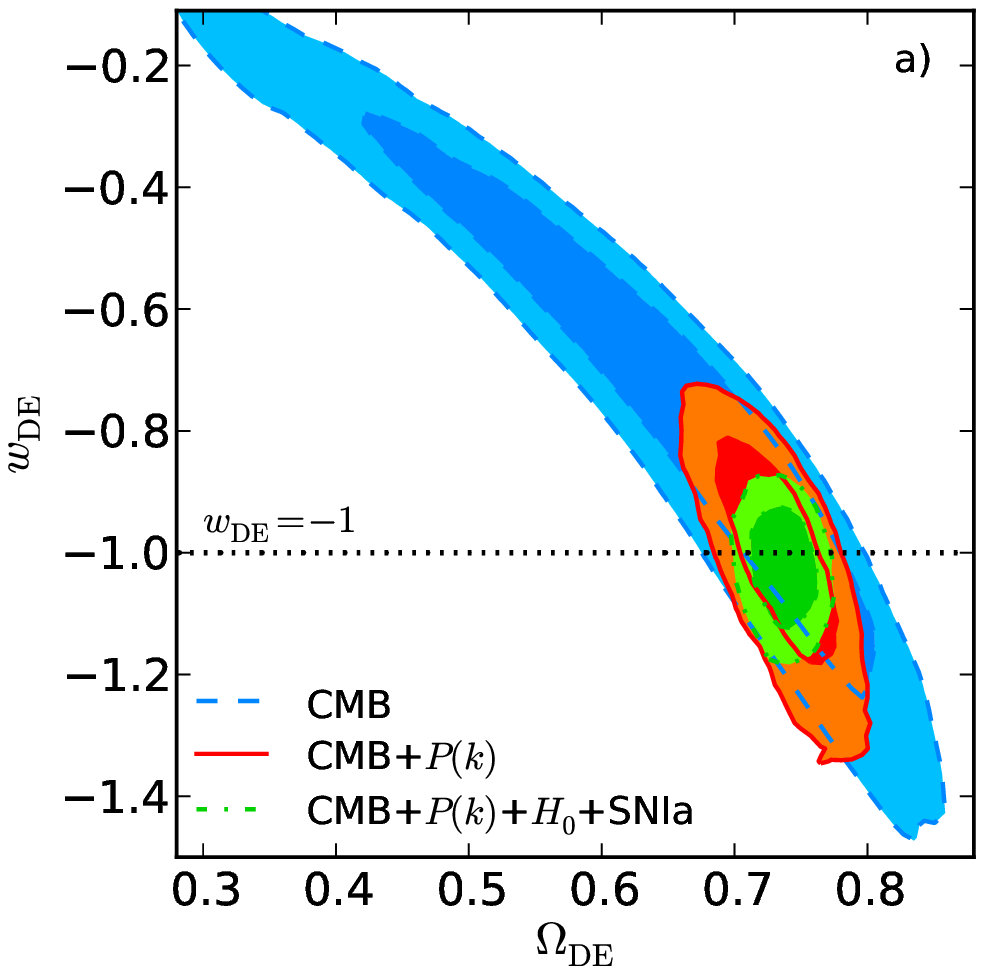} 
  \end{minipage} 
  \begin{minipage}{80mm}
    \includegraphics[width=79mm, keepaspectratio]{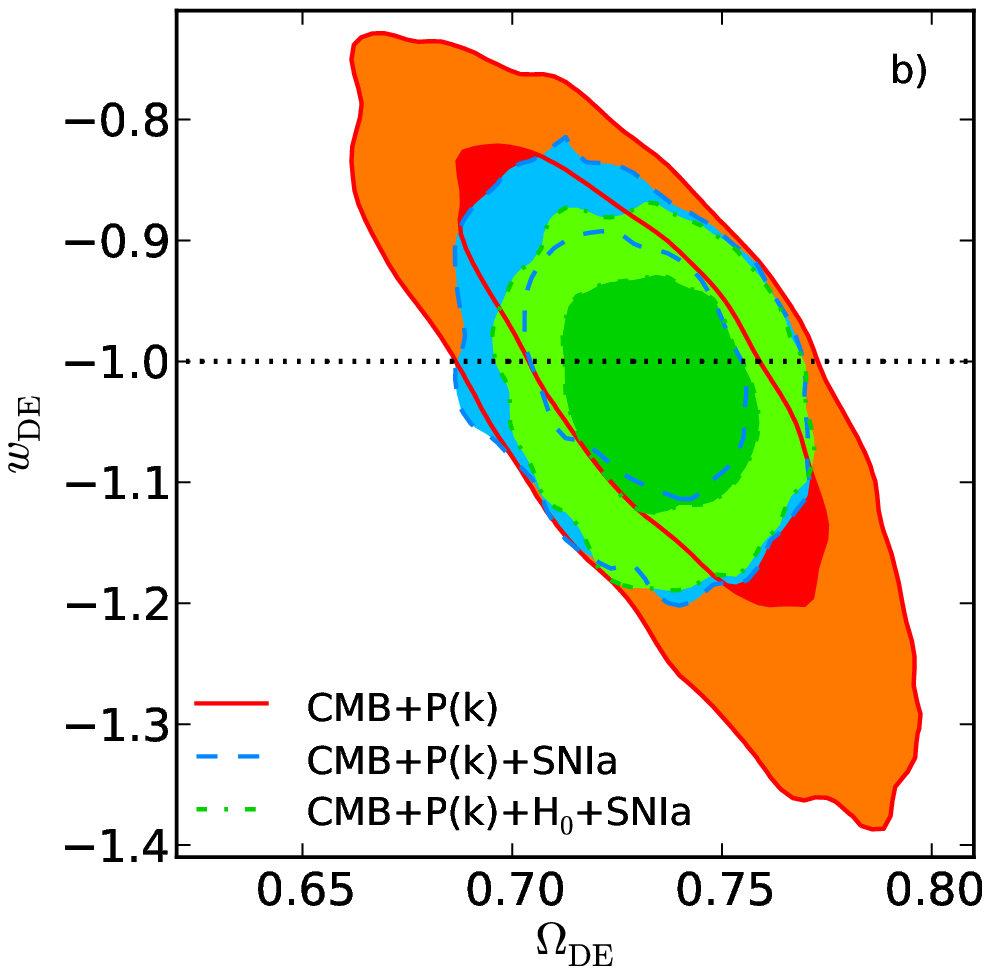} 
  \end{minipage}
    \caption{Panel \emph{a)}: two-dimensional marginalised constraints of the wCDM parameter space in the $w_{\mathrm{DE}}-\Omega_{\mathrm{DE}}$ plane. Colour and line coding is the same as in Figure \ref{fig:LCDM_odm_H0}. Panel \emph{b)}: same as the upper panel, but for CMB+$P_{\mathrm{LRG}}(k)$, red areas within solid lines, CMB+$P_{\mathrm{LRG}}(k)$+SNIa, blue areas within dashed lines, and CMB+$P_{\mathrm{LRG}}(k)$+$H_{0}$+SNIa, green areas within dot-dashed lines. Note that the lower panel is a zoom of the upper one in the region occupied by the CMB+$P_{\mathrm{LRG}}(k)$ contours. The dotted horizontal line shows $w_{\mathrm{DE}}=-1$}\label{fig:wCDM_wde_ode} 
\end{figure*}

Panel a) of Figure \ref{fig:wCDM_wde_ode} shows the constraints in the
$\Omega_{\mathrm{DE}}-w_{\mathrm{DE}}$ plane. The horizontal dotted line
corresponds to $w_{\mathrm{DE}}=-1$. If the CMB alone is considered a strong
degeneracy between these two parameters is present, as shown by blue shaded
areas enclosed by dashed lines. The reason for this degeneracy is analogous to
the case analysed in the previous section when curvature is left as free
parameter. In fact, while dark energy is subdominant at early times and does
not influence the physical scale of the acoustic peaks, its equation of state parameter
and density have a strong effect on the angular diameter distance to the
last scattering surface.  Thus different combinations of
$\omega_{\mathrm{DM}}$, $\omega_{\mathrm{b}}$ and $w_{\mathrm{DE}}$ can result
in the same apparent position of the acoustic oscillations in the CMB
temperature power spectrum. The constraints on the dark energy density and
equation of state parameter are therefore very weak: $\Omega_{\mathrm{DE}} =
0.64_{-0.13}^{+0.12}$ and $w_{\mathrm{DE}} = -0.74_{-0.30}^{+0.32}$. 

With the inclusion of the LRGs data, the degeneracies in the CMB are broken, as
shown in the panel a) of Figure \ref{fig:wCDM_wde_ode} (solid line enclosing
red areas). From the same figure it is clear that the impact of supernovae and
$H_{0}$ (dot-dashed line enclosing green areas) information is much larger in
this scenario than in the previous two. In order to understand what causes such
a difference, we show in panel b) of Figure \ref{fig:wCDM_wde_ode} also the two
dimensional constraints obtained from the combination of
CMB+$P_{\mathrm{LRG}}(k)$+SNIa (dashed lines enclosing blue areas). The
horizontal line corresponds to $w_{\mathrm{DE}}=-1$. The plot shows clearly
that supernovae have a large constraining power for this parameter space. In
fact the supernovae luminosity-distance relation traces the expansion history
of the Universe and clearly identifies the transition between the deceleration
and acceleration phases \citep{riess_04}. This transition depends on the
densities of the cosmic components and is very sensitive to the dark energy
equation of state. The inclusion of a precise measurement of $H_{0}$ increases
the precision on the parameters by a small factor. The constraints on the dark
energy equation of state are \begin{inparaenum}[i)] \item $w_{\mathrm{DE}} =
-1.02\pm0.13$, from CMB+$P_{\mathrm{LRG}}(k)$, \item $w_{\mathrm{DE}} =
-1.07\pm0.11$, from CMB+$P_{\mathrm{LRG}}(k)$+$H_{0}$, \item $w_{\mathrm{DE}} =
-1.009\pm0.069$, CMB+$P_{\mathrm{LRG}}(k)$+SNIa and \item $w_{\mathrm{DE}} =
-1.025_{-0.065}^{+0.066}$, from CMB+$P_{\mathrm{LRG}}(k)$+$H_{0}$+SNIa.
\end{inparaenum} Therefore the inclusion of the large scale structure
measurements decreases the error on $w_{\mathrm{DE}}$ by about a factor 3 and
SNIa further halve it. 

The main result of this section is that, when combining CMB, LSS, SNIa
information with the prior on $H_{0}$, the equation of state parameter of dark energy is
constrained to be -1 with about 6.5\% accuracy. This, although perfectly
compatible with the cosmological constant models, does not exclude other dark
energy scenarios. Excluding CMB, which has very little constraining power, the
other three experiments give us access only to redshift $z\lesssim1$.
Furthermore many dark energy models can be tuned in order to mimic $\Lambda$ at
low redshifts and deviate from it at earlier epochs. In order to narrow the
range of possible models, more precise measurements spanning a larger range of
redshifts, together with the inclusion of tests of the growth of structures,
will be necessary. 

\subsubsection*{Effects of supernovae systematics and light curve fitters}

\begin{table} 
  \centering 
  \begin{minipage}{80mm} 
    \caption{Marginalised constraints on the cosmological parameters of the wCDM parameter space from
    the combination of CMB+$P_{\mathrm{LRG}}(k)$+$H_0$+SNIa when systematic errors are not considered and when
    the \textsc{mlcs2k2} SNIa light curve fitter is used.}
    \label{tab:1D_wCDM_othersn} 
    \begin{tabular}{ l c c } 
      \hline 
      & no systematics & \textsc{mlcs2k2} \\ 
      \hline 
      $100\omega_{\mathrm{b}}$ & $2.257_{-0.048}^{+0.047}$ &
      $2.264_{-0.049}^{+0.048}$\\[1.5mm]
      $100\omega_{\mathrm{DM}}$ & $11.21_{-0.40}^{+0.40}$ & $11.05_{-0.41}^{+0.41}$\\[1.5mm]
      100$\Theta$ & $1.0405_{-0.0021}^{+0.0021}$ & $1.0411_{-0.0022}^{+0.0021}$\\[1.5mm]
      $\tau$ & $0.087_{-0.014}^{+0.014}$ &
      $0.089_{-0.014}^{+0.014}$\\[1.5mm]
      $w_{\mathrm{DE}}$ & $-1.007_{-0.046}^{+0.046}$ &
      $-0.875_{-0.055}^{+0.054}$\\[1.5mm]
      $n_{\mathrm{s}}$ & $0.962_{-0.012}^{+0.012}$ &
      $0.964_{-0.012}^{+0.012}$\\[1.5mm]
      $\log(10^{10}\,A_{\mathrm{s}})$ & $3.074_{-0.032}^{+0.032}$ &
      $3.072_{-0.033}^{+0.032}$\\[1.5mm]
      $\Omega_{\mathrm{DE}}$ & $0.732_{-0.014}^{+0.013}$ &
      $0.704_{-0.017}^{+0.017}$\\[1.5mm]
      $\mathrm{Age\,[Gyr]}$ &
      $13.719_{-0.090}^{+0.091}$ & $13.795_{-0.093}^{+0.095}$\\[1.5mm]
      $\Omega_{\mathrm{M}}$ & $0.268_{-0.013}^{+0.014}$ &
      $0.296_{-0.017}^{+0.017}$\\[1.5mm]
      $\sigma_8$ & $0.809_{-0.036}^{+0.036}$ &
      $0.752_{-0.037}^{+0.037}$\\[1.5mm]
      $z_{\mathrm{re}}$ & $10.4_{-1.1}^{+1.2}$ & $10.5_{-1.2}^{+1.1}$\\[1.5mm]
      $H_0\,[\mathrm{km}\,\mathrm{s^{-1}\,Mpc^{-1}}]$ & $71.0_{-1.2}^{+1.2}$ & $67.2_{-1.5}^{+1.5}$\\[1.5mm]
      \hline
    \end{tabular} 
  \end{minipage} 
\end{table}

Similarly to what has been done in section \ref{ssec:parspace1}, we test the
impact of SNIa systematic effects and different light curve fitters on the
measured cosmological parameters. The constraints for these two cases are
listed in Table \ref{tab:1D_wCDM_othersn}. As before, we find that neglecting
systematics does not change the mean values of the parameters, but generally
reduces the associated errors. In particular the constraints on the dark energy
equation of state are reduced by almost 30\%, $w_{\mathrm{DE}} =
-1.007\pm0.046$. The use of the data from the SDSS SN project, with the light
curves fitted with \textsc{mlcs2k2}, changes some of the parameters, as for
instance $H_{0}$, $w_{\mathrm{DE}}$ and $\Omega_{\mathrm{M}}$, by more than
2-$\sigma$ with respect to the value in our standard case. In particular we
measure $w_{\mathrm{DE}}$ to be $-0.875_{-0.055}^{+0.054}$. This highlights,
better than for the $\Lambda$CDM case, the importance of SNIa modelling in
improving cosmological constraints from future generation experiments.

\subsection{Curvature and dark energy equation of state as free parameters}\label{ssec:parspace4}

\begin{table*} 
  \centering 
  \begin{minipage}{160mm} 
    \caption{Marginalised constraints on the cosmological parameters of the kwCDM parameter space
      from the combination of probes listed in the header of the table. The quoted values
      are the same as in Table \ref{tab:1D_LCDM}.}
    \label{tab:1D_kwCDM} 
    \begin{tabular}{ l c c c c c } 
      \hline 
      & CMB & CMB+$P(k)$ & CMB+$P(k)$+$H_0$ & CMB+$P(k)$+SNIa & CMB+$P(k)$+$H_0$+SNIa \\
      \hline 
      $100\omega_{\mathrm{b}}$ & $2.236_{-0.052}^{+0.052}$ &
      $2.241_{-0.047}^{+0.047}$ & $2.255_{-0.048}^{+0.048}$ & $2.246_{-0.047}^{+0.047}$ &
      $2.251_{-0.049}^{+0.050}$\\[1.5mm]
      $100\omega_{\mathrm{DM}}$ & $11.04_{-0.52}^{+0.52}$ & $11.29_{-0.44}^{+0.43}$ & $11.32_{-0.45}^{+0.46}$ & $11.27_{-0.43}^{+0.43}$ & $11.30_{-0.46}^{+0.45}$\\[1.5mm]
      100$\Theta$ & $1.0397_{-0.0022}^{+0.0022}$ & $1.0400_{-0.0021}^{+0.0020}$ & $1.0401_{-0.0021}^{+0.0021}$ & $1.0401_{-0.0022}^{+0.0022}$ & $1.0402_{-0.0022}^{+0.0021}$\\[1.5mm]
      $\tau$ & $0.087_{-0.014}^{+0.014}$ &
      $0.086_{-0.014}^{+0.014}$ & $0.087_{-0.015}^{+0.015}$ & $0.086_{-0.014}^{+0.014}$ &
      $0.085_{-0.014}^{+0.014}$\\[1.5mm]
      $100\Omega_{\mathrm{k}}$ & $-2.72_{-6.51}^{+6.28}$ & $2.72_{-2.13}^{+2.18}$ & $1.69_{-1.92}^{+2.11}$ & $0.33_{-0.72}^{+0.71}$ & $0.45_{-0.65}^{+0.65}$\\[1.5mm]
      $w_{\mathrm{DE}}$ & $-0.907_{-0.607}^{+0.524}$ &
      $-0.685_{-0.207}^{+0.200}$ & $-0.856_{-0.272}^{+0.262}$ & $-0.973_{-0.088}^{+0.091}$ &
      $-0.981_{-0.084}^{+0.083}$\\[1.5mm]
      $n_{\mathrm{s}}$ & $0.956_{-0.013}^{+0.013}$ &
      $0.959_{-0.012}^{+0.012}$ & $0.962_{-0.012}^{+0.012}$ & $0.960_{-0.012}^{+0.012}$ &
      $0.961_{-0.012}^{+0.012}$\\[1.5mm]
      $\log(10^{10}\,A_{\mathrm{s}})$ & $3.063_{-0.032}^{+0.033}$ &
      $3.074_{-0.032}^{+0.032}$ & $3.077_{-0.033}^{+0.032}$ & $3.072_{-0.032}^{+0.032}$ &
      $3.073_{-0.032}^{+0.031}$\\[1.5mm]
      $\Omega_{\mathrm{DE}}$ & $0.566_{-0.153}^{+0.139}$ & $0.671_{-0.045}^{+0.047}$ &
      $0.710_{-0.043}^{+0.042}$ & $0.728_{-0.017}^{+0.017}$ &
      $0.733_{-0.017}^{+0.016}$\\[1.5mm]
      $\mathrm{Age\,[Gyr]}$ & $15.01_{-1.88}^{+1.78}$ & $13.11_{-0.45}^{+0.47}$ & $13.24_{-0.58}^{+0.57}$ & $13.61_{-0.32}^{+0.33}$ & $13.53_{-0.29}^{+0.29}$\\[1.5mm]
      $\Omega_{\mathrm{M}}$ & $0.461_{-0.176}^{+0.197}$ & $0.302_{-0.031}^{+0.030}$ &
      $0.273_{-0.026}^{+0.026}$ & $0.268_{-0.017}^{+0.017}$ &
      $0.262_{-0.016}^{+0.016}$\\[1.5mm]
      $\sigma_8$ & $0.725_{-0.117}^{+0.109}$ &
      $0.692_{-0.077}^{+0.080}$ & $0.757_{-0.102}^{+0.106}$ & $0.797_{-0.045}^{+0.044}$ &
      $0.802_{-0.043}^{+0.044}$\\[1.5mm]
      $z_{\mathrm{re}}$ & $10.6_{-1.3}^{+1.3}$ & $10.7_{-1.2}^{+1.2}$ & $10.6_{-1.2}^{+1.2}$ & $10.4_{-1.1}^{+1.2}$ & $10.4_{-1.2}^{+1.2}$\\[1.5mm]
      $H_0\,[\mathrm{km}\,\mathrm{s^{-1}\,Mpc^{-1}}]$ & $56.7_{-11.5}^{+11.2}$ & $67.2_{-3.3}^{+3.4}$ & $70.8_{-3.2}^{+3.2}$ & $71.0_{-2.2}^{+2.2}$ & $71.9_{-2.0}^{+2.0}$\\[1.5mm]
      \hline 
    \end{tabular} 
  \end{minipage}
\end{table*}

We now analyse the accuracy that can be achieved when both $w_{\mathrm{DE}}$ and $\Omega_{\mathrm{k}}$ are considered as free parameters. The full list of constraints on the cosmological parameters for the kwCDM cosmology is summarised in Table \ref{tab:1D_kwCDM}.

\begin{figure} 
  \includegraphics[width=79mm, keepaspectratio]{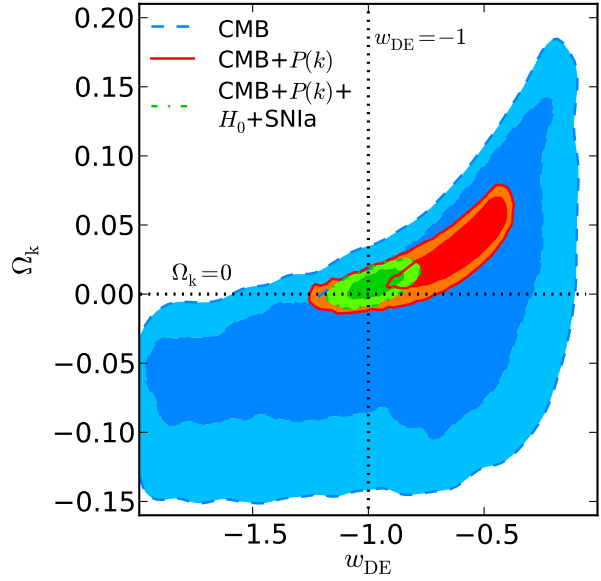}
  \caption{Two-dimensional marginalised constraints of the kwCDM parameter space in the $w_{\mathrm{DE}}-\Omega_{\mathrm{k}}$ plane. Colour and line coding is the same as in Figure \ref{fig:LCDM_odm_H0}. The  horizontal and vertical dotted lines show the values of these parameters for the flat $\Lambda$CDM case.}\label{fig:kwCDM_wde_ok} 
\end{figure}

\begin{figure} 
  \includegraphics[width=79mm, keepaspectratio]{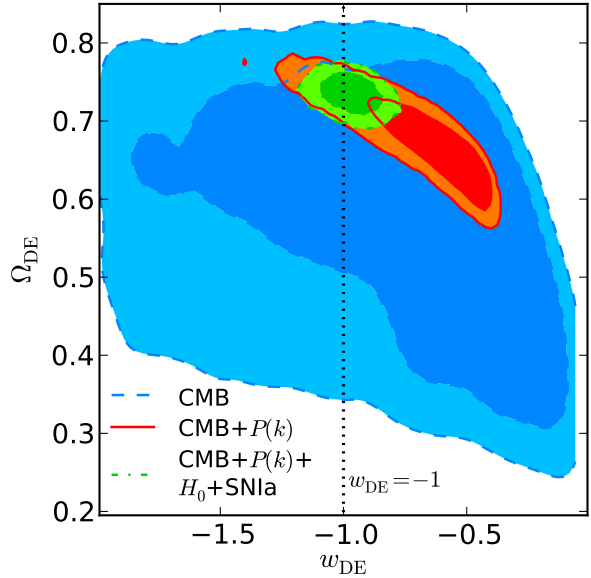}
  \caption{Two-dimensional marginalised constraints of the kwCDM parameter space in the
  $w_{\mathrm{DE}}-\Omega_{\mathrm{DE}}$ plane. Colour and line coding is the same as in Figure \ref{fig:LCDM_odm_H0}. The vertical dotted line shows $w_{\mathrm{DE}}=-1$}\label{fig:kwCDM_wde_ode} 
\end{figure} 

The introduction of an extra degree of freedom with respect to the previous two sections affects the CMB degeneracy already discussed. The two-dimensional marginalised constraints in the $w_{\mathrm{DE}}-\Omega_{\mathrm{k}}$ and $w_{\mathrm{DE}}-\Omega_{\mathrm{DE}}$ planes from CMB data only are shown in Figures \ref{fig:kwCDM_wde_ok} and \ref{fig:kwCDM_wde_ode} with the blue shaded areas within dashed lines. By comparing Figure \ref{fig:kwCDM_wde_ode} with panel a) of Figure \ref{fig:wCDM_wde_ode} it becomes clear that the allowed area in the parameter space increases because of the larger degeneracy. This causes the errors on $\Omega_{\mathrm{k}}$ and  $w_{\mathrm{DE}}$ to increase by 50-60\% with respect to the corresponding cases in sections \ref{ssec:parspace2} and \ref{ssec:parspace3}.

The red contours within solid lines in Figures \ref{fig:kwCDM_wde_ok} and
\ref{fig:kwCDM_wde_ode} show the constraints in the
$w_{\mathrm{DE}}-\Omega_{\mathrm{k}}$ and
$w_{\mathrm{DE}}-\Omega_{\mathrm{DE}}$ planes from the combination of CMB with
the LRG data.  The inclusion of the power spectrum reduces the region allowed
by CMB alone to a one-dimensional degeneracy, that can be broken using the information on
the amplitude of $P_{\mathrm{LRG}}(k)$. Therefore a degeneracy arises
between the bias and the shape of the power spectrum when both the curvature
and the dark energy equation of state are treated as free parameters. When
considering the range of scales $0.02 \hoM  \leq k \leq 0.15 \hoM$ larger
values of $w_{\mathrm{DE}}$, $\Omega_{\mathrm{M}}$ and $\Omega_{\mathrm{k}}$
and lower values of $\sigma_{8}$ can be compensated by unphysically large
biases $b$ and larger values of $A_{\mathrm{MC}}$ in order to obtain a
comparable $\chi^{2}$. If we used a finite flat prior on $b$ we would decrease
the large $\Omega_{\mathrm{k}}$ and $w_{\mathrm{DE}}$ tail and reduce the
degeneracy.  Alternatively, the detection of the turnover in the power
spectrum, not possible nowadays because of the still too small volumes
probed by galaxy surveys, might help constraining better its overall
shape and break the degeneracy just described.

The addition of $H_{0}$ and supernovae measurements breaks the bias-shape degeneracy, returning parameters perfectly consistent with the $\Lambda$CDM cosmology. As for the previous section, the biggest change in precision is due to the SN, which improves the accuracy of the three parameters previously discussed, by a factor 2-3. The final constraints, when all four probes are used, are shown in Figures \ref{fig:kwCDM_wde_ok} and \ref{fig:kwCDM_wde_ode} by the green shaded areas enclosed within dot-dashed lines. For the quantities shown in the plot we obtain  $\Omega_{\mathrm{k}} = \left(4.5\pm6.5\right)\times10^{-3}$ (36\% increase in the errors with respect to k$\Lambda$CDM case), $w_{\mathrm{DE}} = -0.981_{-0.084}^{+0.083}$ (29\% increase in the errors with respect to the wCDM case) and $\Omega_{\mathrm{DE}} =0.733_{-0.017}^{+0.016}$ (8\% increase with respect to both). 

\subsection{Time varying dark energy equation of state parameter}\label{ssec:parspace5}

\begin{table*} 
  \centering 
  \begin{minipage}{160mm} 
    \caption{Marginalised constraints on the cosmological parameters of the waCDM parameter space
      from the combination of probes listed in the header of the table. The quoted values
      are the same as in Table \ref{tab:1D_LCDM}.}
    \label{tab:1D_waCDM} 
    \begin{tabular}{ l c c c c c }
      \hline 
      & CMB & CMB+$P(k)$ & CMB+$P(k)$+$H_0$ & CMB+$P(k)$+SNIa & CMB+$P(k)$+$H_0$+SNIa \\
      \hline 
      $100\omega_{\mathrm{b}}$ & $2.249_{-0.053}^{+0.052}$ &
      $2.253_{-0.049}^{+0.050}$ & $2.255_{-0.049}^{+0.049}$ & $2.251_{-0.050}^{+0.050}$ &
      $2.257_{-0.048}^{+0.049}$\\[1.5mm]
      $100\omega_{\mathrm{DM}}$ & $10.99_{-0.52}^{+0.52}$ & $11.27_{-0.43}^{+0.43}$ & $11.31_{-0.43}^{+0.43}$ & $11.30_{-0.44}^{+0.43}$ & $11.29_{-0.45}^{+0.45}$\\[1.5mm]
      100$\Theta$ & $1.0399_{-0.0022}^{+0.0022}$ & $1.0402_{-0.0022}^{+0.0022}$ & $1.0402_{-0.0021}^{+0.0021}$ & $1.0401_{-0.0022}^{+0.0022}$ & $1.0403_{-0.0021}^{+0.0021}$\\[1.5mm]
      $\tau$ & $0.088_{-0.015}^{+0.015}$ &
      $0.086_{-0.014}^{+0.014}$ & $0.086_{-0.014}^{+0.014}$ & $0.086_{-0.015}^{+0.015}$ &
      $0.086_{-0.014}^{+0.014}$\\[1.5mm]
      $w_{\mathrm{0}}$ & $-0.71_{-0.49}^{+0.47}$ & $-0.75_{-0.48}^{+0.44}$ & $-1.04_{-0.31}^{+0.31}$ & $-0.98_{-0.14}^{+0.14}$ & $-1.00_{-0.14}^{+0.14}$\\[1.5mm]
      $w_{\mathrm{a}}$ & $-0.32_{-1.01}^{+1.00}$ & $-0.63_{-0.99}^{+1.07}$ & $-0.11_{-0.86}^{+0.85}$ & $-0.15_{-0.52}^{+0.52}$ & $-0.13_{-0.53}^{+0.53}$\\[1.5mm]
      $n_{\mathrm{s}}$ & $0.961_{-0.013}^{+0.014}$ &
      $0.961_{-0.012}^{+0.012}$ & $0.962_{-0.012}^{+0.012}$ & $0.960_{-0.012}^{+0.012}$ &
      $0.962_{-0.012}^{+0.012}$\\[1.5mm]
      $\log(10^{10}\,A_{\mathrm{s}})$ & $3.065_{-0.033}^{+0.033}$ &
      $3.074_{-0.032}^{+0.031}$ & $3.075_{-0.033}^{+0.032}$ & $3.074_{-0.033}^{+0.032}$ &
      $3.075_{-0.032}^{+0.032}$\\[1.5mm]
      $\Omega_{\mathrm{DE}}$ & $0.665_{-0.102}^{+0.102}$ & $0.702_{-0.054}^{+0.055}$ &
      $0.741_{-0.028}^{+0.027}$ & $0.728_{-0.018}^{+0.018}$ &
      $0.734_{-0.016}^{+0.016}$\\[1.5mm]
      $\mathrm{Age\,[Gyr]}$ &
      $13.921_{-0.295}^{+0.295}$ & $13.741_{-0.115}^{+0.118}$ & $13.699_{-0.104}^{+0.102}$
      & $13.733_{-0.105}^{+0.109}$ & $13.710_{-0.104}^{+0.104}$\\[1.5mm]
      $\Omega_{\mathrm{M}}$ & $0.335_{-0.102}^{+0.102}$ & $0.298_{-0.055}^{+0.054}$ &
      $0.259_{-0.027}^{+0.028}$ & $0.272_{-0.018}^{+0.018}$ &
      $0.266_{-0.016}^{+0.016}$\\[1.5mm]
      $\sigma_8$ & $0.742_{-0.080}^{+0.082}$ &
      $0.770_{-0.104}^{+0.108}$ & $0.837_{-0.076}^{+0.077}$ & $0.810_{-0.044}^{+0.043}$ &
      $0.818_{-0.045}^{+0.045}$\\[1.5mm]
      $z_{\mathrm{re}}$ & $10.5_{-1.2}^{+1.2}$ & $10.4_{-1.2}^{+1.2}$ & $10.4_{-1.2}^{+1.2}$ & $10.4_{-1.2}^{+1.2}$ & $10.4_{-1.1}^{+1.1}$\\[1.5mm]
      $H_0\,[\mathrm{km}\,\mathrm{s^{-1}\,Mpc^{-1}}]$ & $65.0_{-9.7}^{+9.9}$ & $68.2_{-6.2}^{+6.4}$ & $72.6_{-3.7}^{+3.7}$ & $70.6_{-2.0}^{+2.0}$ & $71.4_{-1.7}^{+1.7}$\\[1.5mm]
      \hline 
    \end{tabular} 
  \end{minipage}
\end{table*}
 
In section \ref{ssec:parspace3} we encode possible deviations from the
cosmological constant model in a constant effective dark energy equation of
state. In this section we include explicitly a time dependency on
$w_{\mathrm{DE}}$ through the simple parameterization of equation
(\ref{eq:wa}). We treat the two parameters of this model, $w_{\mathrm{0}}$ and
$w_{\mathrm{a}}$ as free and we consider the curvature to be fixed 0.
The constraints on the cosmological parameters for the cosmology analysed in
this section from the different combinations of probes are listed in Table
\ref{tab:1D_waCDM}.

\begin{figure} 
  \includegraphics[width=79mm, keepaspectratio]{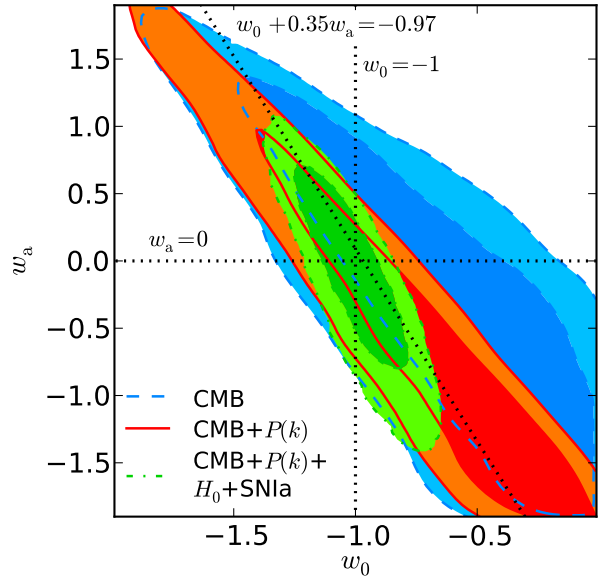}
  \caption{Two-dimensional marginalised constraints of the waCDM parameter space in the $w_{\mathrm{0}}-w_{\mathrm{a}}$ plane. Colour and line coding is the same as in Figure \ref{fig:LCDM_odm_H0}. The vertical and horizontal dotted lines show the values of these parameters for the flat $\Lambda$CDM case ($w_{\mathrm{0}}=-1$ and $w_{\mathrm{a}}=0$). The dotted diagonal line shows the equation $w_{\mathrm{DE}}(a_{\mathrm{p}}) = -0.97 = w_{\mathrm{0}} + (1-a_{\mathrm{p}})w_{\mathrm{a}}$, where $a_{\mathrm{p}}=0.65$ is the pivot scale factor from the combination of CMB and P$_{\mathrm{LRG}}$(k) (see text).}\label{fig:waCDM_wde_wa} 
\end{figure}

Figure \ref{fig:waCDM_wde_wa} shows the two-dimensional marginalised constraints in the
$w_{\mathrm{0}}-w_{\mathrm{a}}$ plane from CMB alone (blue shaded areas within dashed
lines), CMB plus LRG power spectrum (red shaded areas within solid lines) and
the combination of CMB, $P_{\mathrm{LRG}}(k)$, $H_{0}$ and SNIa (green shaded
areas within dot-dashed lines). The vertical and horizontal dotted lines at
$w_{\mathrm{0}}=-1$ and $w_{\mathrm{a}}=0$ show the values of the two
parameters for the $\Lambda$CDM case. In the three cases a degeneracy 
is visible, which is reduced as more independent data are included. From CMB
alone we obtain $w_{\mathrm{0}} = -0.71_{-0.49}^{+0.47}$ and $w_{\mathrm{a}} =
-0.32_{-1.01}^{+1.00}$; the inclusion of the LRGs power spectrum information
does not change substantially the one-dimensional constraints ($w_{\mathrm{0}}
= -0.75_{-0.48}^{+0.44}$ and $w_{\mathrm{a}} = -0.63_{-0.99}^{+1.07}$) but
reduces by almost a factor 2 the figure of merit, i.e. the area of the 95\%
confidence level \citep{albrecht_06_DETF}. As shown in the previous two
sections, supernovae have an important role in constraining dark energy
properties and the inclusion of their infomation reduces the errors in the two
parameters by about a factor 3 and 2, respectively. The one-dimensional
constraints from the combination of all the datasets are $w_{\mathrm{0}} =
-1.00\pm0.14$ and $w_{\mathrm{a}} = -0.13\pm0.53$. The attempt to constrain
also the time evolution of the dark energy equation of state results in a
degradation of its present value.

\begin{figure} 
  \includegraphics[width=79mm, keepaspectratio]{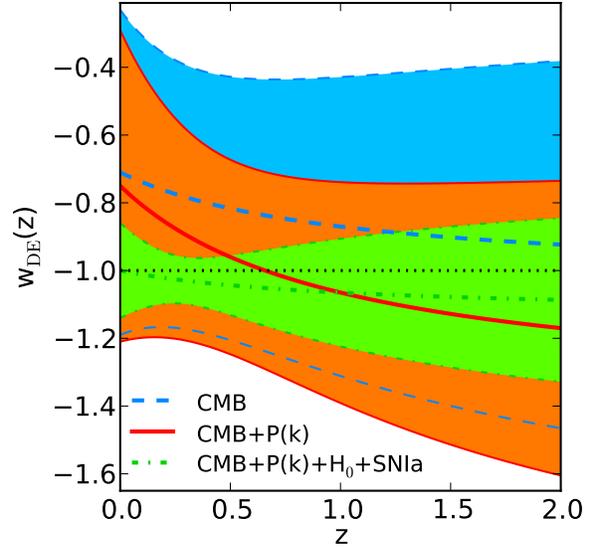} 
  \caption{$w_{\mathrm{DE}}(z)$ as function of the redshift $z$ as obtained from the constraints
  shown in Figure \ref{fig:waCDM_wde_wa}. The thick dashed blue, solid red and green dot-dashed lines are the mean $w_{\mathrm{DE}}(z)$ for CMB, CMB+$P_{\mathrm{LRG}}(k)$ and CMB+$P_{\mathrm{LRG}}(k)$+$H_{0}$+SNIa. The blue, red and green dashed lines enclosed within the outermost dashed, solid and dot-dashed lines are the 68\% confidence level for the same three combinations as computed from equation (\ref{eq:err_wde_a}).}\label{fig:waCDM_wde_a} 
\end{figure}

From these results it is possible to reconstruct the time dependence of the
dark energy equation of state parameter. The thick dashed, solid and
dot-dashed lines in Figure \ref{fig:waCDM_wde_a} show the value of
$w_{\mathrm{DE}}(z)$ from CMB, CMB plus $P_{\mathrm{LRG}}(k)$ and the
combination of the four probes, respectively. The corresponding 1-$\sigma$
errors, which vary with redshift, are indicated by the shaded areas within the
thin lines. They are computed according to \citet{albrecht_06_DETF}:
\begin{equation}\label{eq:err_wde_a} 
  \langle \delta w_{\mathrm{DE}}^{2}(a)\rangle = \langle\left(\delta w_{\mathrm{0}} + (1-a) \delta w_{\mathrm{a}}\right)^{2}\rangle.
\end{equation}
In the redshift range shown in the plot, $w_{\mathrm{DE}}$ is always compatible
with a cosmological constant at the 1$\sigma$ level. Furthermore the errors
show a minimum at a redshift called ``pivot''
\citep{huterer_01,hu_04,albrecht_06_DETF}. For the three cases shown in Figure
\ref{fig:waCDM_wde_a}, we obtain a pivot redshift of $z_{\mathrm{p}}= 0.4$
(CMB), $z_{\mathrm{p}} = 0.54$ (CMB+$P_{\mathrm{LRG}}(k))$ and $z_{\mathrm{p}}
= 0.3$ (CMB+$P_{\mathrm{LRG}}(k)$+$H_{0}$+SNIa) and an equation of state
parameter $w_{\mathrm{DE}}(z_{\mathrm{p}}) = -0.80\pm0.37$,
$w_{\mathrm{DE}}(z_{\mathrm{p}}) = -0.97\pm0.29$ and
$w_{\mathrm{DE}}(z_{\mathrm{p}}) = -1.03\pm0.069$. It is interesting to note
that in the last case the precision is comparable to the one presented in
Section \ref{ssec:parspace3}. The diagonal dotted line in Figure
\ref{fig:waCDM_wde_wa} shows that the degeneracy between $w_{\mathrm{0}}$ and
$w_{\mathrm{a}}$ is very close to $w_{\mathrm{0}} + (1-a_{\mathrm{p}})
w_{\mathrm{a}}=-0.97$, where $a_{\mathrm{p}}=0.65$ is the pivot scale factor
from the CMB+$P_{\mathrm{LRG}}(k)$ constraints.

\begin{figure} 
  \includegraphics[width=79mm, keepaspectratio]{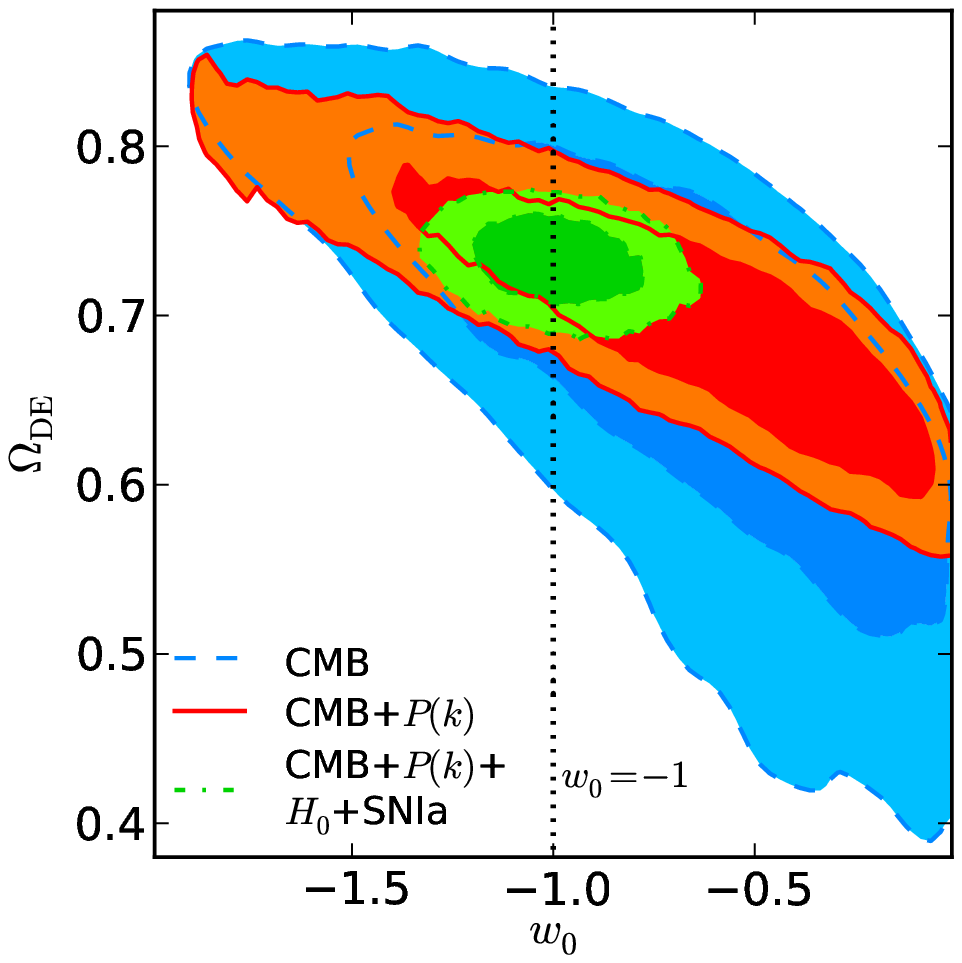}
  \caption{Two-dimensional marginalised constraints of the waCDM parameter space in the $w_{\mathrm{0}}-\Omega_{\mathrm{DE}}$
  plane. Colour and line coding is the same as in Figure \ref{fig:LCDM_odm_H0}. The vertical line is for $w_{\mathrm{0}}=-1$.}\label{fig:waCDM_wde_ode} 
\end{figure}

Figure \ref{fig:waCDM_wde_ode} shows the two-dimensional marginalised constraints in the
$w_{\mathrm{0}}-\Omega_{\mathrm{DE}}$ plane. The colour and line code is the same as in
Figure \ref{fig:waCDM_wde_wa}. This figure illustrates that the addition of large scale structure
information to the CMB data halves the errors on the dark energy density from $\Omega_{\mathrm{DE}}= 0.66\pm0.10$ to $\Omega_{\mathrm{DE}}=0.702_{-0.054}^{+0.055}$. The inclusion of SNIa and H$_{0}$
measurements decreases the errors further by 70\% to $\Omega_{\mathrm{DE}} = 0.734\pm0.016$.

\section{Comparison with previous studies} \label{sec:comparison}

In this section we compare our results with recent work focused on the analysis of the large scale structure of the Universe.
In Section \ref{ssec:R10} we perform a quantitative comparison of our results 
with those of \citet[][thereafter R10]{reid_10_SDSS}, whose 
measurement and model are publicly available\footnote{\url{http://lambda.gsfc.nasa.gov/toolbox/lrgdr/}}.
Section \ref{ssec:others}, instead, lists a number of similar analyses, 
for which this is not the case, and we can only discuss the differences qualitatively.

\subsection{\citet{reid_10_SDSS}}\label{ssec:R10}

The LRG distribution from the SDSS DR7 has already been used to
extract cosmological parameters by R10. The analysis in that
work differs from the one we perform mostly in five important details:
\begin{inparaenum}[i)] \item all the LRGs from the Northern Galactic Cap and
the 3 southern stripes were used (110576 galaxies in 7931 deg$^{2}$), \item
thanks to the count-in-cylinders technique \citep[CiC,][]{reid_09a,reid_09b},
the authors extracted a halo catalogue from the LRG distribution and then
computed the halo power spectrum using the PVP estimator, 
\item they computed the covariance matrix
from 10000 lognormal catalogues \citep[LN,][]{coles_91}, \item they used a
model based on \textsc{halofit} \citep{smith_03_halofit}, that required
additional calibration against numerical simulations and \item they analysed
the power spectrum in the range $0.02\hoM < k < 0.2\hoM$. \end{inparaenum} The
advantage of using the reconstructed underlying density field, instead of the
galaxies, is that the intra-halo peculiar motions, which cause the so called
fingers-of-gods, are erased, leaving weaker small scale redshift-space
distortions. 
They combine their halo power spectrum with the 5th year data from
the WMAP satellite \citep[WMAP5,][]{komatsu_wmap09} and the Union Supernovae
dataset \citep{kowalski_unionSN1a}. The comparison between the results in table
3 of R10 and the corresponding ones in this work (third column
in Tables \ref{tab:1D_LCDM}, \ref{tab:1D_kLCDM}, \ref{tab:1D_wCDM} and third
and fifth columns in Table \ref{tab:1D_kwCDM}) shows that, with the exception
of the $\Lambda$CDM case and despite the smaller $k$ modes used by us, the
errors that we obtain here are slightly smaller and that there are significant
offsets in the preferred values of some parameters. For example, the constraints on $w_{\mathrm{DE}}$ 
in the wCDM parameter space, when combining CMB and $P(k)$, are 
$w_{\mathrm{DE}}=-1.02\pm0.13$ in our analysis and $w_{\mathrm{DE}}=-0.79\pm0.15$ in R10.

\begin{table*} 
  \centering 
  \begin{minipage}{160mm} 
    \caption{Comparison between this work and R10: marginalised
    constraints on the cosmological parameters of the wCDM parameter space
    from the combination of CMB and $P(k)$. The central four columns are for 
    $0.02 \hoM \leq k_{i} \leq 0.15 \hoM$ and $k_{j}\leq 0.2 \hoM$ while the last one for 
    $0.02 \hoM \leq k_{i} \leq 0.20 \hoM$ and $k_{j}\leq 0.5 \hoM$. The quoted values are the same as in Table \ref{tab:1D_LCDM}.}
    \label{tab:1D_wCDM_comp} 
    \begin{tabular}{ l c c c c c }
    \hline
      & \phantom{R}this work\phantom{R} & \phantom{R}$P_{\mathrm{LRG}}(k)$ vs. \phantom{R}& \phantom{R}$P_{\mathrm{R10}}(k)$ vs.\phantom{R} & \phantom{R} $P_{\mathrm{R10}}(k)$ vs. \phantom{R}& $P_{\mathrm{R10}}(k)$ vs.\\
      & & R10 model & our model & R10 model & R10 model, extended \\
      \hline
      $100\omega_{\mathrm{b}}$ & $2.251_{-0.050}^{+0.049}$ & $2.251_{-0.049}^{+0.049}$ & $2.246_{-0.053}^{+0.050}$ & $2.257_{-0.053}^{+0.055}$ & $2.259_{-0.053}^{+0.053}$\\[1.5mm]
      $100\omega_{\mathrm{DM}}$ & $11.25_{-0.43}^{+0.43}$ & $11.92_{-0.42}^{+0.43}$ & $11.04_{-0.46}^{+0.46}$ & $11.22_{-0.51}^{+0.50}$ & $11.18_{-0.48}^{+0.46}$\\[1.5mm]
      100$\Theta$ & $1.0402_{-0.0022}^{+0.0022}$ & $1.0401_{-0.0021}^{+0.0021}$ & $1.0399_{-0.0021}^{+0.0021}$ & $1.0398_{-0.0022}^{+0.0022}$ & $1.0399_{-0.0022}^{+0.0022}$\\[1.5mm]
      $\tau$ & $0.086_{-0.015}^{+0.015}$ & $0.085_{-0.015}^{+0.015}$ & $0.087_{-0.015}^{+0.015}$ & $0.087_{-0.015}^{+0.014}$ & $0.086_{-0.014}^{+0.014}$\\[1.5mm]
      $w_{\mathrm{DE}}$ & $-1.022_{-0.128}^{+0.129}$ & $-1.056_{-0.191}^{+0.184}$ & $-0.798_{-0.127}^{+0.127}$ & $-0.847_{-0.169}^{+0.172}$ & $-0.817_{-0.152}^{+0.154}$\\[1.5mm]
      $n_{\mathrm{s}}$ & $0.962_{-0.012}^{+0.012}$ & $0.963_{-0.013}^{+0.013}$ & $0.959_{-0.012}^{+0.013}$ & $0.966_{-0.014}^{+0.014}$ & $0.966_{-0.014}^{+0.014}$\\[1.5mm]
      $\log[10^{10}\,A_{\mathrm{s}}]$ & $3.073_{-0.033}^{+0.034}$ & $3.102_{-0.032}^{+0.032}$ & $3.066_{-0.034}^{+0.034}$ & $3.078_{-0.032}^{+0.032}$ & $3.075_{-0.030}^{+0.030}$\\[1.5mm]
      $\Omega_{\mathrm{DE}}$ & $0.732_{-0.028}^{+0.028}$ & $0.704_{-0.032}^{+0.033}$ & $0.673_{-0.041}^{+0.041}$ & $0.679_{-0.041}^{+0.042}$ & $0.673_{-0.036}^{+0.037}$\\[1.5mm]
      $\mathrm{Age\,[Gyr]}$ & $13.733_{-0.117}^{+0.118}$ & $13.755_{-0.115}^{+0.118}$ & $13.932_{-0.156}^{+0.148}$ & $13.889_{-0.156}^{+0.154}$ & $13.905_{-0.137}^{+0.141}$\\[1.5mm]
      $\Omega_{\mathrm{M}}$ & $0.268_{-0.028}^{+0.028}$ & $0.296_{-0.033}^{+0.032}$ & $0.327_{-0.041}^{+0.041}$ & $0.321_{-0.042}^{+0.041}$ & $0.327_{-0.037}^{+0.036}$\\[1.5mm]
      $\sigma_8$ & $0.818_{-0.064}^{+0.065}$ & $0.857_{-0.069}^{+0.072}$ & $0.725_{-0.057}^{+0.057}$ & $0.758_{-0.071}^{+0.072}$ & $0.745_{-0.066}^{+0.065}$\\[1.5mm]
      $z_{\mathrm{re}}$ & $10.4_{-1.2}^{+1.2}$ & $10.4_{-1.2}^{+1.2}$ & $10.5_{-1.2}^{+1.2}$ & $10.5_{-1.2}^{+1.2}$ & $10.4_{-1.1}^{+1.2}$\\[1.5mm]
      $H_0\,[\mathrm{km}\,\mathrm{s^{-1}\,Mpc^{-1}}]$ & $71.2_{-3.8}^{+3.8}$ & $69.6_{-4.3}^{+4.5}$ & $64.2_{-3.9}^{+4.0}$ & $65.2_{-4.7}^{+4.6}$ & $64.4_{-4.1}^{+4.0}$\\[1.5mm]
      \hline
    \end{tabular} 
  \end{minipage}
\end{table*}

As the full measurement from R10 and a \textsc{cosmomc} module which includes the model is publicly 
available, we can 
understand the origin of the differences between the two studies. 
We concentrate here on the wCDM parameter space.
Table \ref{tab:1D_wCDM_comp} compares the results shown in the 
third column of Table \ref{tab:1D_wCDM} (second column) with the constraints that we obtain applying \begin{inparaenum}[i)] 
\item R10 model to our measurement (third column), \item our model to R10 measurement (fourth column) and 
\item R10 model to R10 measurement (last two columns). \end{inparaenum} To do it, 
we combine the large scale structure information with the CMB measurements 
presented in Section \ref{ssec:cmb}, we consider the scales $0.02 \hoM \leq k_{i} \leq 0.15 \hoM$ and
we compute the model for $k_{j}\leq 0.2 \hoM$ before the convolution with the window function. 
In the last column we use $0.02 \hoM \leq k_{i} \leq 0.20 \hoM$ and $k_{j}\leq 0.50 \hoM$, 
as originally done in R10. The similarity between the constraints obtained in 
this last case and those of R10 shows that the effect of using WMAP7 
plus small angular scale CMB data instead of WMAP5 is very small.
If we consider the parameters constrained mainly by the CMB ($\omega_{\mathrm{b}}$, 
$\omega_{\mathrm{DM}}$, $\Theta$, $\tau$, $A_{\mathrm{s}}$, $z_{\mathrm{ze}}$
and $n_{s}$) we find that they  
are, as expected, mostly independent of the power spectrum measurement and model used in the analysis. 
Excluding the last column, for the other parameters we find
\begin{inparaenum}[i)] \item that, given a measurement, the errors that are obtained using the model of equation (\ref{eq:rpt_mod}) 
are generally smaller than the ones from R10's 
model and \item that, given a model, 
the parameters obtained from our measurement  agree better with the bulk of the 
literature than the ones recovered from the R10 power spectrum.  \end{inparaenum}
In particular, these data prefer larger values of $w_{\mathrm{DE}}$ and of 
$\Omega_{\mathrm{M}}$, which lead to smaller values of $\sigma_{8}$ and $H_{0}$. Using 
$P_{\mathrm{LRG}}(k)$
presented in Section \ref{ssec:dr7_ps} we measure the dark energy equation of state 
parameter to be $w_{\mathrm{DE}=-1.022_{-0.128}^{+0.129}}$ 
when using our model and $-1.056_{-0.191}^{+0.184}$ when using R10 model. This is more 
than a 50\% increase in the uncertainty. If we consider 
the R10 measurement, we obtain, for the two models, $w_{\mathrm{DE}}=-0.798_{-0.127}^{+0.127}$ 
and $-0.847_{-0.169}^{+0.172}$, with $w_{\mathrm{DE}}>-1$ 
at 1.5 and 0.9 $\sigma$ level, respectively. Including scales up to $0.2\hoM$, 
we measure $w_{\mathbf{DE}}=-0.817_{-0.152}^{+0.154}$: the error 
decreases, but remains larger than when using our model, and disfavours the cosmological 
constant scenario at 1.2$\sigma$. 
These results clearly point towards differences in both the modelling and the measurement of the power spectrum.

\begin{figure} 
  \includegraphics[width=79mm, keepaspectratio]{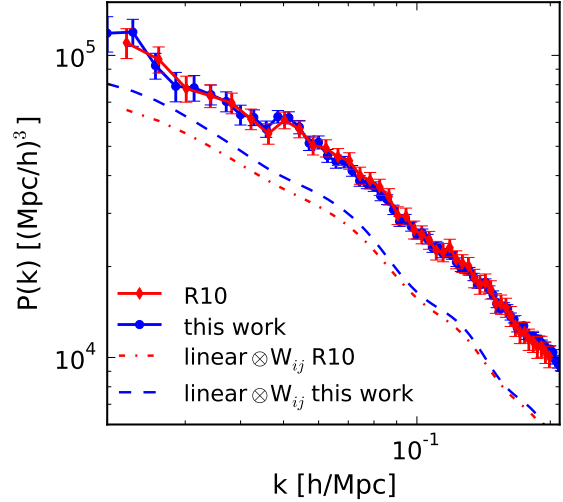}
  \caption{Comparison of the LRG power spectrum described in Section \ref{ssec:dr7_ps} (blue points) and the halo one from R10 (red diamonds) with the corresponding errors. The amplitude of the power spectrum form R10 has been increased by a factor 4 in order to make the comparison easier. 
The best fit linear power spectrum of Figure \ref{fig:meas_powersp} convolved with the window matrix presented in Section \ref{ssec:dr7_ps} and from R10 are shown with a blue dashed and a red dot-dashed lines, respectively. In order to make the plot more readable, the amplitude of these two 
power spectra is not matched to the one of $P_{\mathrm{LRG}}$.  }\label{fig:compare_ps} 
\end{figure}

\begin{figure} 
  \includegraphics[width=79mm, keepaspectratio]{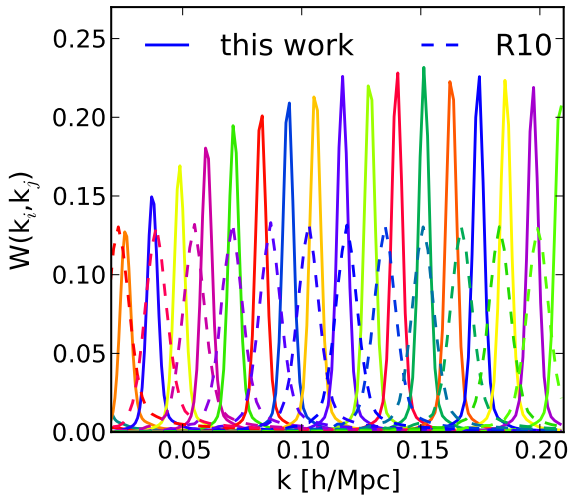}
  \caption{Comparison of the rows of the window matrix corresponding to the $k$-bands of the measured LRG 
  power spectrum analysed in this work (solid lines) and from R10 (dashed lines). For
  clarity only one every fourth row is shown.}
  \label{fig:compare_win} 
\end{figure}

While equation (\ref{eq:rpt_mod}) makes use of only three free parameters and 
does not require any calibration, the model in R10 relies largely on numerical simulations 
in order to fix some of the parameters and to correct for the imprecision of \textsc{halofit}, 
which at $k\approx 0.1-0.2 \hoM$ can describe a $\Lambda$CDM power spectrum with
about a 5\% accuracy \citep[see e.g.][]{heitmann_10}. Furthermore the calibration is 
performed only against simulations that reproduce the clustering of a flat $\Lambda$CDM
universe, which could potentially introduce biases for more general parameter spaces, 
as the comparison between the second and third column in Table \ref{tab:1D_wCDM_comp} might suggest.
The parameters shown in Tables \ref{tab:1D_wCDM}-\ref{tab:1D_waCDM} are almost insensitive 
to the cosmological model assumed, whilst the ones listed in table 3 of R10 exhibit larger variations.
This might indicate that our model is better suited to analyse a large variety of 
cosmological parameter spaces.

The tension between the constraints obtained when using different measurements, instead, points
in the direction of differences in the shape of the power spectrum or of the window matrix. 
Figure \ref{fig:compare_ps} shows the LRG power spectrum described in Section 
\ref{ssec:dr7_ps} (with blue points) and the halo one from R10 (red diamonds). 
The error-bars show the corresponding variances. 
Besides a factor of about four in the amplitude, due to the different samples and estimators 
used, which we include in the figure, the shapes of the two power spectra are almost identical,
as we expect. We would also expect the window
functions to be similar, but Figure \ref{fig:compare_win} shows 
that the rows of $W(k_{i}, k_{j})$ corresponding to the $k$-bands of the measured power 
spectra are substantially different. This is intriguing, given that the bulk of our and R10 samples the same. This difference is even clearer when convolving a model $P(k)$
with the window function. The dashed and dot-dashed lines in Figure \ref{fig:compare_ps}
show the best fit linear power spectrum from Figure \ref{fig:meas_powersp} after the 
convolution with our and R10 $W(k_{i},k_{j})$, respectively. In order to avoid clutter the amplitude of these two power spectra is not matched to $P_{\mathrm{LRG}}(k)$. 
The differences in the window matrix are transformed into different shapes of the convolved
$P_{\mathrm{lin}}(k)$, which implies that these measurements cannot be described by the same set of cosmological parameters.
This difference in the window matrix is responsible for the main differences shown in Table \ref{tab:1D_wCDM_comp}.

\subsection{Other works}\label{ssec:others}

\citet{percival_10_SDSS} analysed almost 900,000 galaxies from the combination
of the full SDSS DR7 galaxy sample and the two-degree Field Galaxy Redshift
Survey \citep[2dFGRS,][]{colless_03_2dF} and extracted the BAO feature from the
power spectrum in 7 redshift bins. The main cosmological results are shown in
their table 5 and can be compared with the third column in Tables
\ref{tab:1D_LCDM}, \ref{tab:1D_kLCDM} and \ref{tab:1D_wCDM} and with Table
\ref{tab:1D_kwCDM}. Their constraints are compatible with ours and with the
findings of R10. Significantly, despite the much smaller
sample that we use, the constraints that we obtain are comparable or tighter
than the ones of \citet{percival_10_SDSS}. As an example, for the wCDM case we
obtain from CMB+$P_{\mathrm{LRG}}(k)$ $w_{\mathrm{DE}}=-1.02\pm0.13$, while
they report the value $w_{\mathrm{DE}}=-0.97\pm0.17$, which corresponds to an
improvement of about 25\% in our results. This comparison suggests the
importance of the use of the full information content in the galaxy power
spectrum, as already noticed by different authors \citep[e.g.,][]{Sanchez_08,
shoji_09, blake_11}.

The correlation function has also been intensively used with similar goals.
\citet{Sanchez_09} applied a model equivalent to the one presented in section
\ref{ssec:model} \citep{crocce_nonlinBAO,Sanchez_08} to the correlation
function of the LRG sample from the SDSS DR6, as measured by \citet{cabre_09a}.
Combining this with Union SNIa sample and the CMB measurements from WMAP5, as
well as with the position of the BAO peak along the line of sight in the
two-dimensional correlation function from \citet{gaztanaga09}, they extracted
cosmological parameters for the same parameter spaces presented in this work.
The overall results of \citet{Sanchez_09} are consistent with the ones
presented in section \ref{sec:cosmopars}. Interestingly our errors on the cosmological parameters are
systematically smaller when curvature is kept as free variable and larger for
the wCDM and waCDM cases. This shows that the models and the measurements of
the power spectrum and the correlation function, although generally coherent,
show some small differences in the sensitivity to cosmological parameters for
different parameter spaces. In the kwCDM case, the constraints of
\citet{Sanchez_09} show the same bias-shape degeneracy that shifts the values
of $\Omega_{\mathrm{k}}$ and $w_{\mathrm{DE}}$ upwards with respect to the flat
$\Lambda$CDM paradigm (compare Figure \ref{fig:kwCDM_wde_ok} with figure 14 in
\citet{Sanchez_09}). The inclusion of SNIa and of H$_{0}$ (radial BAO) in our
(their) analysis leads to the tightest constraints in both works. For flat and
non flat $\Lambda$CDM, the differences are negligible, while for the other
three cases, where SNIa measurements play an important role, the errors that we
obtain are larger: this is due to the inclusion of the supernovae systematic
errors, which where neglected in \citet{Sanchez_09}, in our analysis. When we
do not use the systematics, we obtain tighter constraints, shown by the middle
column of Table \ref{tab:1D_wCDM_othersn}. In this case we measure the dark
energy equation of state parameter to be $w_{\mathrm{DE}}=-1.007\pm0.046$, about 11\%
tighter than the corresponding value measured by
\citet[][$w_{\mathrm{DE}}=-0.969\pm0.052$]{Sanchez_09}.

A more recent analysis of the correlation function from the DR7 LRG sample has
been made by \citet{chuang_10}, using a simplified version of the model
presented by \citet{reid_10_SDSS}. They combined this measurement with WMAP7
and Union2 data in order to extract cosmological parameters. Some of their
constraints are offset by 1-2$\sigma$ with respect to our findings and their
measurements of the dark energy equation of state are compatible with ours,
once the SNIa systematics are neglected. \citet{carnero_11} measured the
angular correlation function and detected the BAO feature from the LRGs in the
SDSS DR7 photometric catalogue, which consists of a sample of about 1.5 million
galaxies. Fixing all the other parameters, with the exception of $H_{0}$,
to the best fit from WMAP7, they measure $w_{\mathrm{DE}}$ to be
$-1.03\pm0.16$. 

The small scale clustering has Also been used to perform cosmological analyses.
From the small scale projected correlation function and mass-number ratio in
clusters, modelled in the halo model framework, \citet{tinker_11} extracted
cosmological and model parameters, obtaining, when combining with WMAP7,
$\Omega_{\mathrm{M}} = 0.290 \pm 0.016$ and $\sigma_{8} = 0.826 \pm 0.02$. The
difference between these results and the values of Table \ref{tab:1D_LCDM}
might be due to the smaller number of cosmological degrees of freedom and the
higher number of model parameters. 

In a recent article \citet{blake_11} presented the first cosmological results
from the WiggleZ survey \citep{drinkwater_10_wigglez}. They used a sample of
about 130,000 emission line galaxies across 1000 deg$^{2}$ in the redshift
range $0.3<z<0.9$. From the correlation function, the power spectrum, the BAOs
and the band-filtered correlation function \citep{xu_10} they extracted BAO
parameters, like the effective distance $D_{\mathrm{V}}(z)$ of equation
(\ref{eq:effdist}) and the ``acoustic parameter'', defined by $A(z) \equiv
D_{\mathrm{V}} \sqrt{\Omega_{\mathbf{M}} H^{2}_{0}}/cz$ \citep{Eisenstein_05}.
From those, they measured cosmological parameters for the k$\Lambda$CMD and the
wCDM models. When using LSS information only, the uncertainties are large but
the results show a strong preference for the presence of dark energy, with
$w_{\mathrm{DE}}=-1.6^{+0.6}_{-0.7}$. Combining the BAO parameters with WMAP7
distance priors they measured the dark energy equation of state parameter to be
$w_{\mathrm{DE}}=-0.982^{+0.154}_{-0.189}$, which became
$w_{\mathrm{DE}}=-1.026\pm0.081$ if also Union2 SN are used. Those results are
in agreement with ours, although their uncertainties are about 20-30\% larger
than the ones that we show in Table \ref{tab:1D_wCDM}. The agreement between
our results and the ones in \citet{blake_11} show that, despite
the differences in the galaxies selected, the survey volume and geometry and
the procedure used to extract cosmological information, both analyses are robust enough for
the precision achievable today.

\section{Summary and conclusions} \label{sec:conclusion}

We have computed the power spectrum of the distribution of about 90,000 luminous red galaxies, extracted from the spectroscopic part of the seventh data release of the Sloan Digital Sky survey. To compute the covariance matrix for the power spectrum we make use of the 160 LasDamas mock catalogues, which have the same angular and redshift distribution as the observed LRGs. We describe the measured power spectrum with a model inspired by renormalised perturbation theory and modified in order to describe biased objects in redshift space (section \ref{ssec:model}). This model has been successfully tested against numerical simulations (M10) and mock catalogues (section \ref{ssec:test}) for $k\lesssim0.15\hoM$ at z=0-0.5.

Combining the large scale structure information with measurements of the CMB
temperature and polarisation power spectrum from the seven year data release of
WMAP, ACBAR, BOOMERanG, CBI and QUAD and with the luminosity-distance relation
from the Union2 supernovae sample and using a precise determination of the
local Hubble parameter as a gaussian prior, we explore the constraints in five
different cosmological parameter spaces described in section
\ref{ssec:parspace}. They are the flat $\Lambda$CDM concordance model, a
similar one where curvature is a free parameter (k$\Lambda$CDM) and three
models in which the dark energy equation of state has a parametric form: in two
cases it is assumed to be constant, with and without the assumption of a flat
geometry (wCDM and kwCDM), and in the last case (waCDM) we model
$w_{\mathrm{DE}}$ with the simple parametric formula of equation (\ref{eq:wa}).

Overall, we obtain tight constraints on the cosmological parameters for all the
five cases and we do not detect deviations from the flat $\Lambda$CDM paradigm.
The different combinations of the four experiments used in this analysis do not
show any evidence of tensions between cosmological probes. We find that the
curvature is null at 1-$\sigma$ level with errors of the order of
$10^{-2}-10^{-3}$. We measure the dark energy equation of state parameter to be
consistent with a cosmological constant with 13\% uncertainty for
CMB+$P_{\mathrm{LRG}}(k)$ and 6.5\% uncertainty for
CMB+$P_{\mathrm{LRG}}(k)$+$H_{0}$+SNIa in the flat wCDM case. If we discard the
systematic errors in the SNIa, the precision increases to 4.6\%. In the kwCDM,
because of the added degree of freedom, we have a degradation of the
constraints to 8.4\%, when combining all the four samples. The constraints
obtained with the CMB and large scale structure together are shifted by
1.5-2$\sigma$ with respect to the best fit $\Lambda$CDM because of a bias-shape
degeneracy in the power spectrum that allows very large, and unphysical, values
of the bias when $w_{\mathrm{DE}}$, $\Omega_{\mathrm{M}}$,
$\Omega_{\mathrm{k}}$ increase and $\sigma_{8}$ decreases. If we assume the
parametric form of equation (\ref{eq:wa}) for the dark energy equation of
state, we obtain $w_{\mathrm{DE}}(z=0.54) = -0.97\pm0.29$
(CMB+$P_{\mathrm{LRG}}(k)$) and $w_{\mathrm{DE}}(z=0.3) = -1.03\pm0.069$ (all
four experiments combined). The latter is only slightly worse than the flat
wCDM result.

In the near future new and larger galaxy redshift catalogues both spectroscopic, like BOSS and HETDEX, and photometric, like Pan-STARRS and DES, will become available, together with the new measurements of the CMB anisotropies from the Planck satellite \citep{plank_mission}. These datasets will enable us to improve the constraints presented in this work even further. Some of these experiments are explicitly designed in order to extract the maximum amount of information regarding dark energy. This would allow to exclude many classes of dark energy models or of modifications of general relativity. In order to use the full information from the power of these new large scale observations and to avoid introducing systematic effects, accurate models of the large scale distribution, scale dependent bias and redshift space distortions are necessary. In M10 we showed that the model used in this work is accurate enough to describe the power spectrum shape also for surveys with volumes larger than available nowadays. The explicit inclusion of bias and redshift space distortions in the model would allow however to use a larger range of scales than now possible, helping to improve the quality of the cosmological constrains even further.

\section*{Acknowledgments}

We thank the LasDamas project for releasing publicly the mock catalogues. We thank Shaun Cole and Andr\'es Balaguera-Antol\'inez for discussions about the power spectrum computation and Friedrich R\"opke and Sandra Benitez for the clarifications about SNIa. FM acknowledges support by the Trans-regional Collaborative Research Centre TRR33 ÔThe Dark UniverseÕ of the German Research Foundation (DFG).

Funding for the SDSS and SDSS-II has been provided by the Alfred P. Sloan Foundation, the Participating Institutions, the National Science Foundation, the U.S. Department of Energy, the National Aeronautics and Space Administration, the Japanese Monbukagakusho, the Max Planck Society, and the Higher Education Funding Council for England. The SDSS Web Site is \url{http://www.sdss.org/}.

    The SDSS is managed by the Astrophysical Research Consortium for the Participating
    Institutions. The Participating Institutions are the American Museum of Natural
    History, Astrophysical Institute Potsdam, University of Basel, University of
    Cambridge, Case Western Reserve University, University of Chicago, Drexel University,
    Fermilab, the Institute for Advanced Study, the Japan Participation Group, Johns
    Hopkins University, the Joint Institute for Nuclear Astrophysics, the Kavli Institute
    for Particle Astrophysics and Cosmology, the Korean Scientist Group, the Chinese
    Academy of Sciences (LAMOST), Los Alamos National Laboratory, the Max-Planck-Institute
    for Astronomy (MPIA), the Max-Planck-Institute for Astrophysics (MPA), New Mexico
    State University, Ohio State University, University of Pittsburgh, University of
    Portsmouth, Princeton University, the United States Naval Observatory, and the
    University of Washington.

\appendix

\section{Basic equations to compute the power spectrum}\label{ap:pk_eqs}

In this section we summarise the basic equations of the FKP and PVP estimators used to
compute the power spectrum and the window function from galaxy surveys. PVP is a
generalisation of FKP and takes in account the change of the galaxy bias with the
luminosity of the galaxies \citep[see e.g.,][]{davis_76,norberg_01,norberg_02,
zehavi_02, phleps_06}. 

The observed power spectrum $P_{\mathrm{o}}(k)$ is obtained from the squared average
Fourier transform of the weighted density field defined by
\begin{subequations}\label{eqs:Fr} 
  \begin{equation}\label{eq:Fr_fkp} 
    \text{FKP:} \,F(\mathbf{x}) = \frac{1}{N} w(\mathbf{x})\left[n_{\mathrm{g}}(\mathbf{x}) -\alpha n_{\mathrm{r}}(\mathbf{x})\right], 
  \end{equation} 
  \begin{equation}\label{eq:Fr_pvp} 
    \text{PVP:}\, F(\mathbf{x}) = \frac{1}{N} \int \dint L \frac{w(\mathbf{x},L)}{b(\mathbf{x},L)}\left[n_{\mathrm{g}}(\mathbf{x},L) -\alpha n_{\mathrm{r}}(\mathbf{x},L)\right], 
  \end{equation} 
\end{subequations}
where $n_{\mathrm{g}}(\mathbf{x},L)$ and $n_{\mathrm{r}}(\mathbf{x},L)$ are the number density
of galaxies and randoms of luminosity $L$ at position $\mathbf{x}$. The corresponding quantities of equation (\ref{eq:Fr_fkp}) can be obtained integrating over the luminosity.
$w(\mathbf{x})$, $w(\mathbf{x},L)$ are weighting functions and $b(\mathbf{x}, L)$, only PVP, is the
bias relative to a specific galaxy population with luminosity $L_{\star}$. The
normalisation $N$ is defined by 
\begin{subequations} \label{eqs:N}
  \begin{equation}\label{eq:N_fkp} 
    \text{FKP:} \,N^{2} = \int \dint^{3}x\,\bar{n}^{2}(\mathbf{x})w^{2}(\mathbf{x}), 
  \end{equation}
  \begin{equation} \label{eq:N_pvp} 
    \text{PVP:}\, N^{2} = \int \dint^{3}x \left[\int \dint L \bar{n}(\mathbf{x},L)w(\mathbf{x},L)\right]^{2}, 
  \end{equation}
\end{subequations} 
where $\bar{n}(\mathbf{x}, L)$ and $\bar{n}(\mathbf{x})$ are, respectively, the
mean expected number density, i.e. in absence of clustering, of galaxies of
luminosity $L$ at position $\mathbf{x}$ and its integral over $L$. Finally $\alpha$ is a constant
introduced to match the two catalogues and is chosen requiring that\footnote{In
FKP, the authors use a definition of $\alpha$ such that $N=1$.} $\langle F(\mathbf{x}) \rangle = 0$:
\begin{subequations} \label{eqs:alpha} 
  \begin{equation} \label{eq:alpha_fkp}
    \text{FKP:} \,\alpha = \frac{\int \dint^{3}x\, w(\mathbf{x})n_{\mathrm{g}}(\mathbf{x})}{\int \dint^{3}x\,w(\mathbf{x})n_{\mathrm{r}}(\mathbf{x})}, 
  \end{equation}
  \begin{equation} \label{eq:alpha_pvp} 
    \text{PVP:}\, \alpha = \frac{\int \dint^{3}x\, \dint L\left[w(\mathbf{x},L)/b(\mathbf{x},L)\right]n_{\mathrm{g}}(\mathbf{x},L)}{\int \dint^{3}x\, \dint L\left[w(\mathbf{x},L)/b(\mathbf{x},L)\right]n_{\mathrm{r}}(\mathbf{x},L)}.
  \end{equation} 
\end{subequations}

The observed power spectrum can be then written for both estimators as:
\begin{equation} \label{eq:ps_recovered} 
  P_{\mathrm{o}}(\mathbf{k}) = \int \frac{\dint k'^{3}}{(2\pi)^{3}} P_{\mathrm{t}}(\mathbf{k}')G^{2}(\mathbf{k}-\mathbf{k}') = \langle |F(\mathbf{k})|^{2} \rangle - P_{\mathrm{sn}}, 
\end{equation} 
where $P_{\mathrm{t}}(\mathbf{k}')$ is the ``true'' underlying power spectrum, the shot noise $P_{\mathrm{sn}}$ is given by 
\begin{subequations} \label{eqs:sn}
  \begin{equation} \label{eq:sn_fkp} 
    \text{FKP:} \,P_{\mathrm{sn}} = \frac{1+\alpha}{N^{2}}\int \dint^{3}x\, \bar{n}(\mathbf{x})w^{2}(\mathbf{x}),
  \end{equation} 
  \begin{equation} \label{eq:sn_pvp} 
    \text{PVP:}\, P_{\mathrm{sn}} = \frac{1+\alpha}{N^{2}}\int \dint^{3}x\, \dint L\, \bar{n}(\mathbf{x},L)\frac{w^{2}(\mathbf{x},L)}{b^{2}(\mathbf{x},L)} 
  \end{equation}
\end{subequations} 
and $G^{2}(k)$ is the window function, which encodes information about the survey geometry.

It can be shown that the window function is computed from the spherical averaged Fourier transform of the field
\begin{subequations} \label{eqs:Gr} 
  \begin{equation} \label{eq:Gr_fkp} 
    \text{FKP:} \, \bar{G}(\mathbf{x}) = \frac{1}{N} \bar{n}(\mathbf{x})w(\mathbf{x}), 
  \end{equation}
  \begin{equation} \label{eq:Gr_pvp} 
    \text{PVP:}\, \bar{G}(\mathbf{x}) = \frac{1}{N}\int \dint L\, \bar{n}(\mathbf{x},L)w(\mathbf{x},L),
  \end{equation} 
\end{subequations}
as $G^{2}(k) = \langle|\bar{G}(k)|^{2}\rangle - G_{\mathrm{sn}}$. $G_{\mathrm{sn}}$ is the shot noise defined by
\begin{subequations} \label{eqs:Gsn}
  \begin{equation} \label{eq:Gsn_fkp} 
    \text{FKP:} \, G_{\mathrm{sn}} = \frac{1}{N^{2}} \int \dint^{3}x\, \bar{n}(\mathbf{x})w^{2}(\mathbf{x}), 
  \end{equation} 
  \begin{equation} \label{eq:Gsn_pvp} 
    \text{PVP:}\, G_{\mathrm{sn}} = \frac{1}{N^{2}} \int \dint^{3}x\, \dint L\, \bar{n}(\mathbf{x},L)w^{2}(\mathbf{x},L).  
  \end{equation} 
\end{subequations}

In this work, we use the weighting functions, designed to minimise the variance, proposed
in FKP and \citet{cole_05_2dF}: 
\begin{subequations} \label{eqs:weights}
  \begin{equation} \label{eq:weight_fkp} 
    \text{FKP:}\,w(\mathbf{x}) = \frac{w_{\mathrm{i}}}{1 + P(k) n(\mathbf{x})} 
  \end{equation}
  \begin{equation} \label{eq:weight_pvp} 
    \text{PVP:}\,w(\mathbf{x}, L) = \frac{ w_{\mathrm{i}} \,b^{2}(\mathbf{x}, L)}{1 + P(k) \int \dint L\, b^{2}(\mathbf{x}, L)n(\mathbf{x},L)}.  
  \end{equation} 
\end{subequations} 
The intrinsic weight of the objects, $w_{\mathrm{i}}$, can contain information about completeness
and/or fibre collision \citep[][]{zehavi_02,masjedi_06}. $P(k)$ is an estimate
of the recovered power spectrum and it is usually substituted with a constant
$p_{\mathrm w}$, chosen in order to minimise the variance around the wave-number
$\bar{k}$ for which $P(\bar{k})\sim p_{\mathrm w}$.

\section{Impact of weights on power spectra and cosmological constraints}

In the following we test the impact of $p_{\mathrm{w}}$, $w_{\mathrm{i}}$ and estimator on the LRG  and mock power spectra, on the errors and on the cosmological parameters.

\subsection{Testing the luminous red galaxies power spectrum}\label{ap:test_pk}

In this appendix we test the 
impact of different choices of the estimator, $\pw$ and $\wi$ on
the LRG power spectrum and window function. For both estimators, FKP and PVP, we test four values of $p_{\mathrm w} = 40000,\,10000,\,4000,\,0$. We also consider four
different intrinsic weights: \begin{inparaenum} [i)] \item $w_{\mathrm{i}}=1$ (all the
objects have equal weight), \item $w_{\mathrm{i}}=c$ (areas with low completeness have less weight than areas with higher one), \item $w_{\mathrm{i}}=\fc$
(the loss of galaxies due to fibre collisions is compensated as described in Section \ref{ssec:dr7_ps}) and \item $w_{\mathrm{i}}=c\times\fc$ (both completeness
and fibre collision corrections applied).\end{inparaenum}

\begin{figure}
  \includegraphics[width=79mm, keepaspectratio]{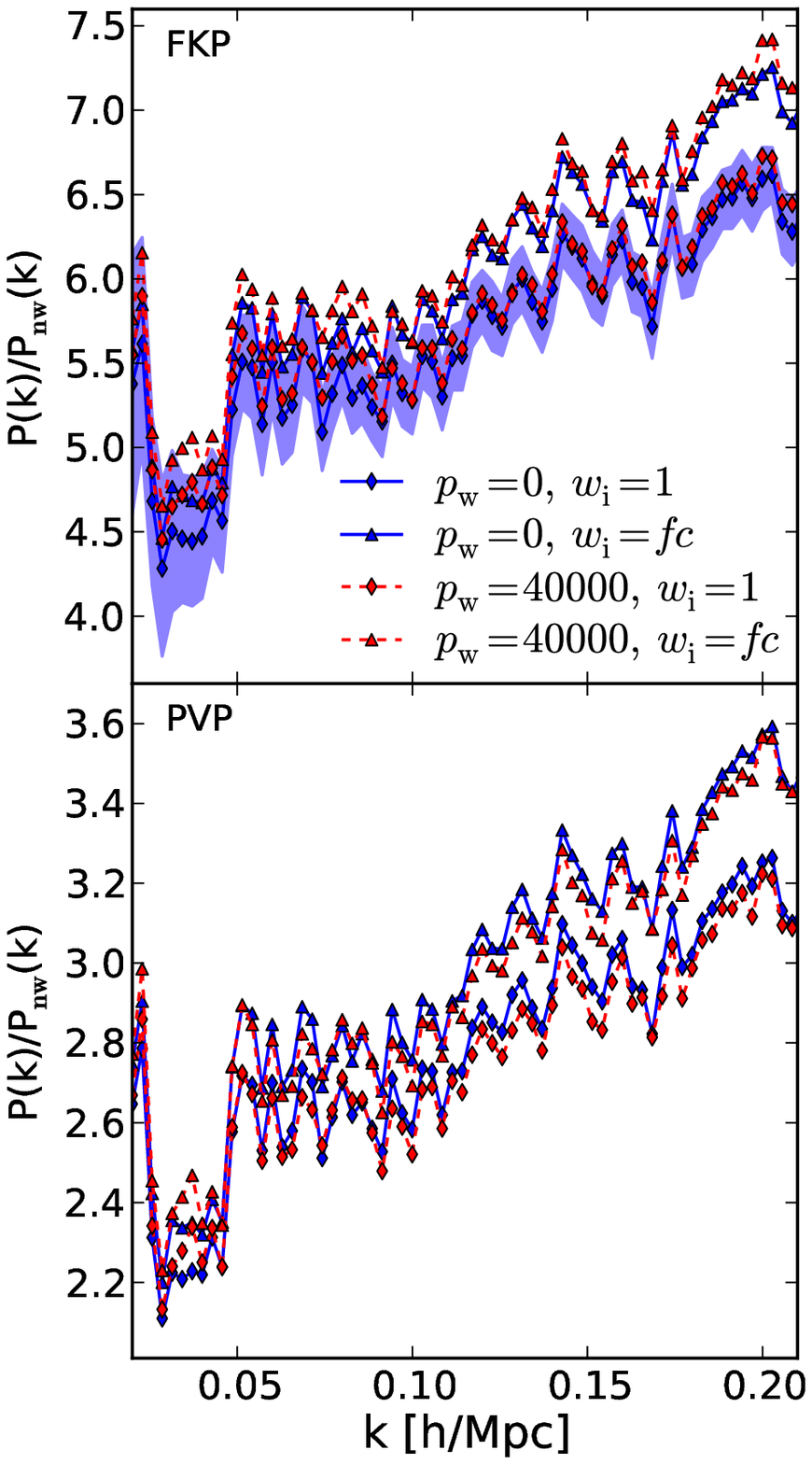}
  \caption{Comparison between power spectra computed with $p_{\mathrm w}=40000, 0$ (red dashed and blue solid lines) and
  $w_{\mathrm{i}}=1, \fc$ (diamonds and up triangles) divided by a linear
  power spectrum without BAOs. The upper panel is for the FKP, the lower for PVP. The
  shaded area denotes the standard deviation computed from the mock catalogues using
  $\pw=40000$.} \label{fig:test_wpw} 
\end{figure}

Figure \ref{fig:test_wpw} shows the differences in the power spectra measured with the FKP (upper
panel) and PVP (lower panel) estimators for the different choices of $\pw$ and $\wi$. For clarity, in the figure we show only the combination of
$p_{\mathrm w}=40000, 0$ (red dashed and blue solid lines respectively) and
$w_{\mathrm{i}}=1, \fc$ (diamonds and up triangles, respectively). The shaded area shows
the standard deviation as measured from the mock catalogues for $\pw=40000$. All the power
spectra have been divided by a non-wiggle one with the same cosmological parameters as the
mock catalogues. Different choices of $\pw$ change marginally the shape of the power
spectrum. The results for $\pw = 10000, 4000$ fall in between the two extreme cases shown in Figure \ref{fig:test_wpw}.

Instead, the correction for fibre collision has a significant effect. When obtaining spectra of crowded fields, not all the objects of interest
can be targeted with a limited number of pointings. Because of this loss of objects, the
amplitude of the highest peak in the density fields decreases, while the low density
regions are unaffected. This causes the amplitude of the fluctuations, and consequently of
the power spectrum, to be lower. We have indeed measured a few percent scale 
independent decrease in the amplitude of $\langle |F(\mathbf{k})|^{2} \rangle$ (i.e. the power
spectrum before subtracting the shot noise) in the case when fibre collision correction is not
applied. This difference is visible at the large scale, low $k$, limit in
Figure \ref{fig:test_wpw} where the power spectrum with the fibre collision (triangles) is always
larger than the one without (diamonds). On the other side the shot noise (equations
\ref{eqs:sn}) depends on the expected non-clustered number density ($\bar{n}$) and the
associated weights, which are not influenced (or influenced in a uniform way) by fibre
collisions. This causes the shot noise to be the same in both cases, changing the shape
of the final power spectrum $P_{\mathrm{o}}(k)$. This effect is clearly visible in the
small scale, large $k$, limit in Figure \ref{fig:test_wpw}, where the fibre collision corrected
power spectra become increasingly larger than the non corrected ones. We do not show the
results when completeness correction is included, since the measured power spectra overlap
almost exactly the non corrected ones. This is because both the galaxy and the random
catalogues are weighted in the same way, therefore the density fluctuation field $F(\mathbf{k})$ is unchanged.

\begin{figure}
  \includegraphics[width=79mm, keepaspectratio]{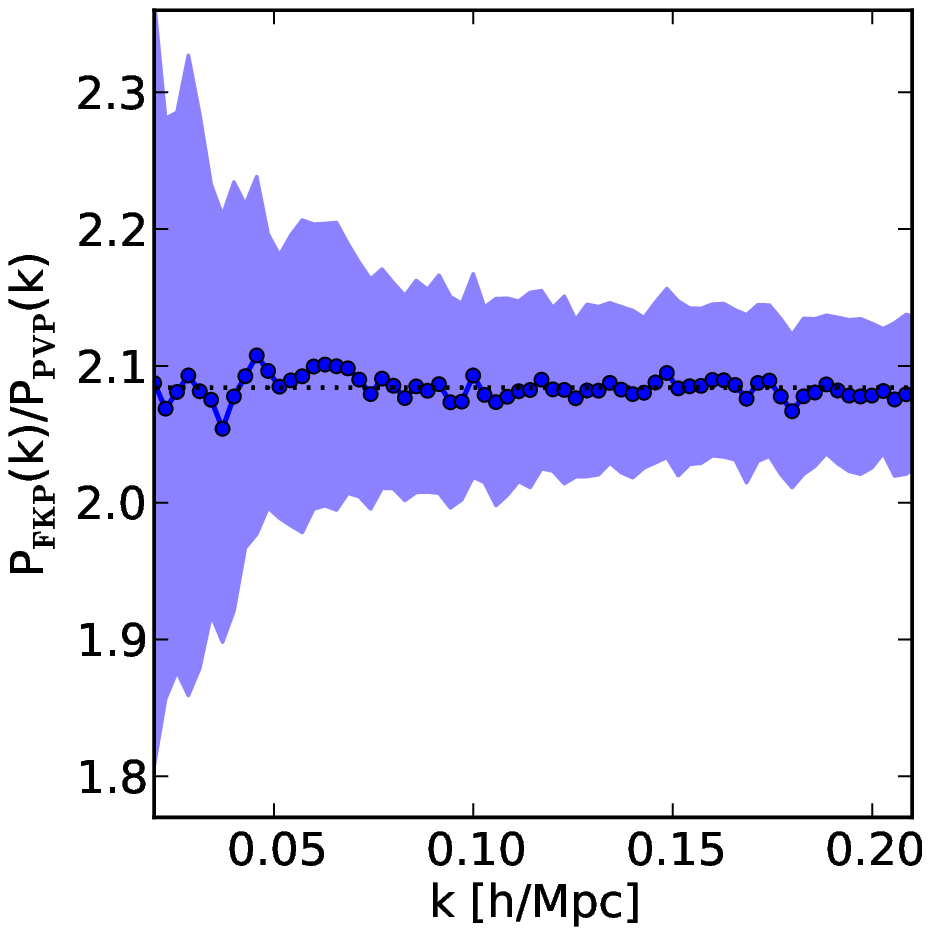}
  \caption{Ratio between power spectra computed the FKP and PVP estimators for
  $\pw=40000$, $\wi=\fc$. The shaded area shows the standard deviation computed from the
  mock catalogues for $\pw=40000$. The dotted horizontal line is the mean of the ratio in
  the plotted interval. The other choices of $\pw$ and $\wi$ show similar behaviour.}
  \label{fig:test_fkpvspvp} 
\end{figure}

\begin{figure}
  \includegraphics[width=79mm, keepaspectratio]{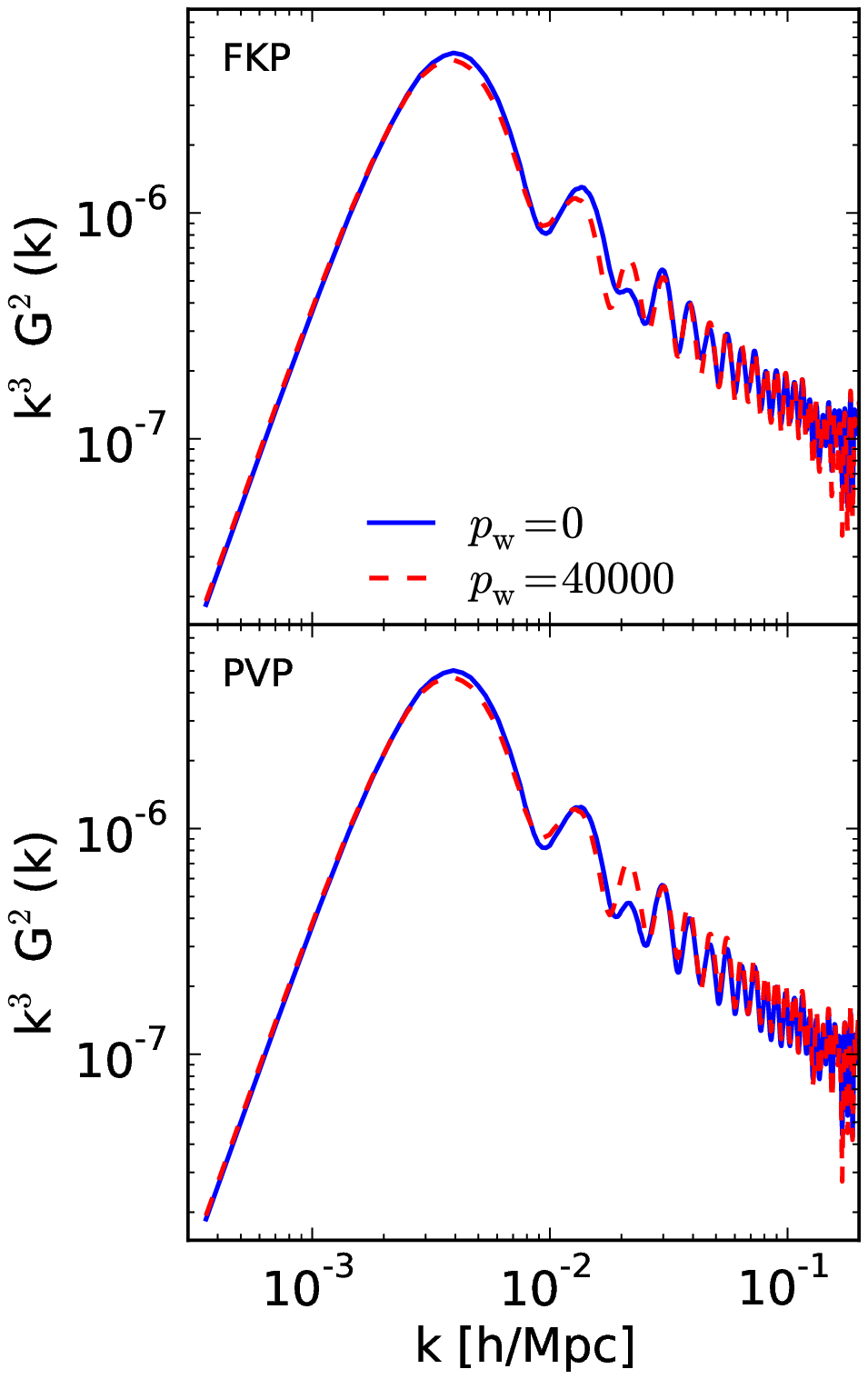}
  \caption{Window functions $G^{2}(k)$ multiplied by the cube of the wave-number $k$ as
  function of $k$ for $\pw=0,40000$ (solid and dashed lines, respectively) for $\wi=1$.
  The upper panel is for FKP, the lower for PVP.} \label{fig:test_winfunc} 
\end{figure}

Comparing the two panels of Figure \ref{fig:test_wpw}, is it clear that the choice of the
estimator has a strong impact on the recovered power spectra: in the PVP case the
amplitude is systematically lower than for FKP. Figure \ref{fig:test_fkpvspvp} shows the ratio of
the power spectrum computed with the latter estimator with respect to the one computed
with the former for the case $\pw=40000$, $\wi=\fc$. As before, the shaded area corresponds to the
standard deviation computed from the LasDamas catalogues. Although the amplitude is
different, the relative bias between the two estimators is scale invariant in the range $0.02 \hoM  \leq
k \leq 0.2 \hoM$. This is expected from the fact that the galaxy sample that
we use is almost volume limited and contains a very uniform population of galaxies. 

Recently, \citet{balaguera_10} have shown that, in volume limited mock catalogues of the REFLEX II cluster survey, the power spectrum computed with the PVP estimator has higher correlations than the FKP one already at $k>0.15 \hoM$. Because of this and of the scale independent relative bias, we can safely use the power spectra as estimated with FKP in order to constrain cosmological parameters.

The window function, describing only the radial and angular selection function
of the survey, is not affected by fibre collision effect; on the other hand different choices of $\pw$
and the completeness correction change the weights of equations \eqref{eqs:weights}, which
influence the effective survey volume. As for the power spectrum, we do not measure
differences when the completeness weighting is applied. Figure \ref{fig:test_winfunc} shows the
``dimensionless'' window function $k^{3}G^{2}(k)$ computed with FKP (upper panel) and
PVP (lower panel) for $\pw = 40000, 0$ (dashed and solid line respectively). Although the
overall shape is similar, there are small differences in the oscillations due to different
effective volumes in the two cases. The results for $\pw = 10000, 4000$ fall in between
the two extreme cases. The difference between the FKP and PVP window functions are
negligible.

When convolving these window functions with a linear power spectrum as in
equation (\ref{eq:pk_winmat}), the differences in the range
$0.02 \hoM  \leq k \leq 0.2 \hoM$ are negligible.

\subsection{Testing the power spectra of the mock catalogues}\label{ap:test_err}

\begin{figure}
  \includegraphics[width=79mm, keepaspectratio]{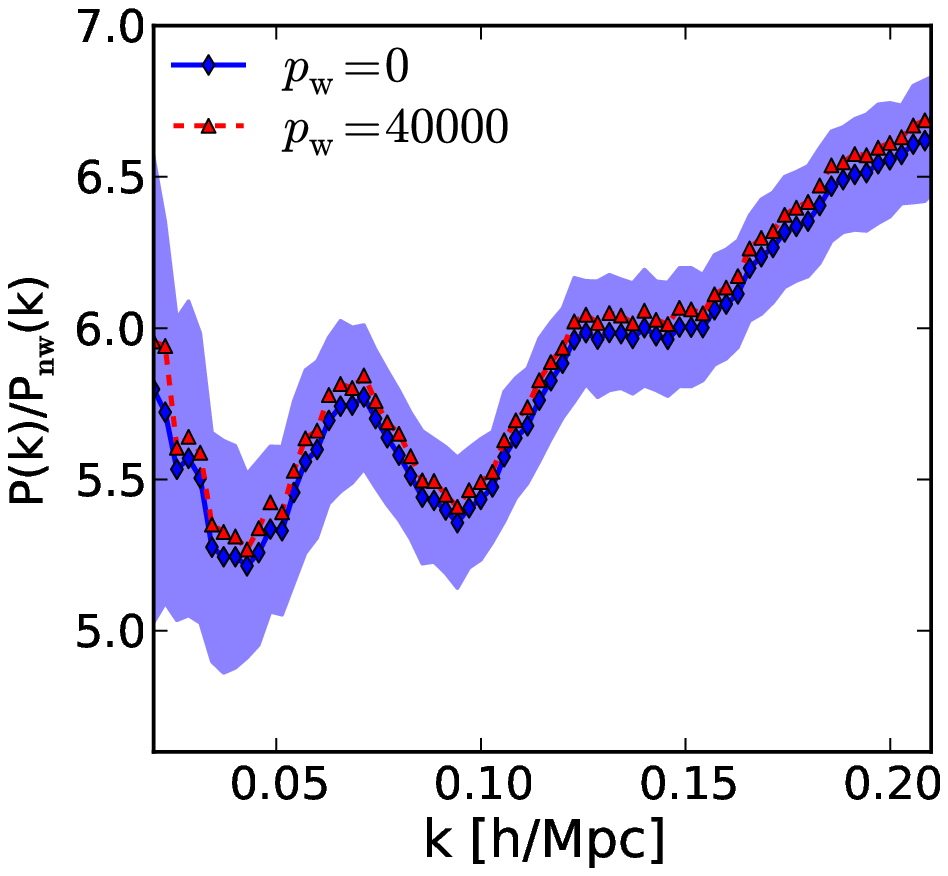}
  \caption{Comparison between mean mock power spectra computed for $p_{\mathrm w}=40000,
  0$ (red dashed with triangles and blue solid lines with diamonds respectively) divided
  by a power spectrum without oscillations. The shaded area denotes the standard deviation
  for $\pw=40000$.} \label{fig:pkmockd} 
\end{figure}

In order to test the impact of $\pw$ on the results from the LasDamas, we compute the power spectra of the 160 mocks, their mean, standard deviation and covariance matrix for $\pw=40000, 10000, 4000, 0$. Figure \ref{fig:pkmockd} shows the mean power
spectra computed for the two extreme cases: the only difference is a small change in amplitude, which confirms the
findings of the previous appendix, i.e. that the impact of different $\pw$ on the computed power
spectrum is negligible.  The shaded area denotes the standard deviation for $\pw=40000$. 

\begin{figure}
  \includegraphics[width=79mm, keepaspectratio]{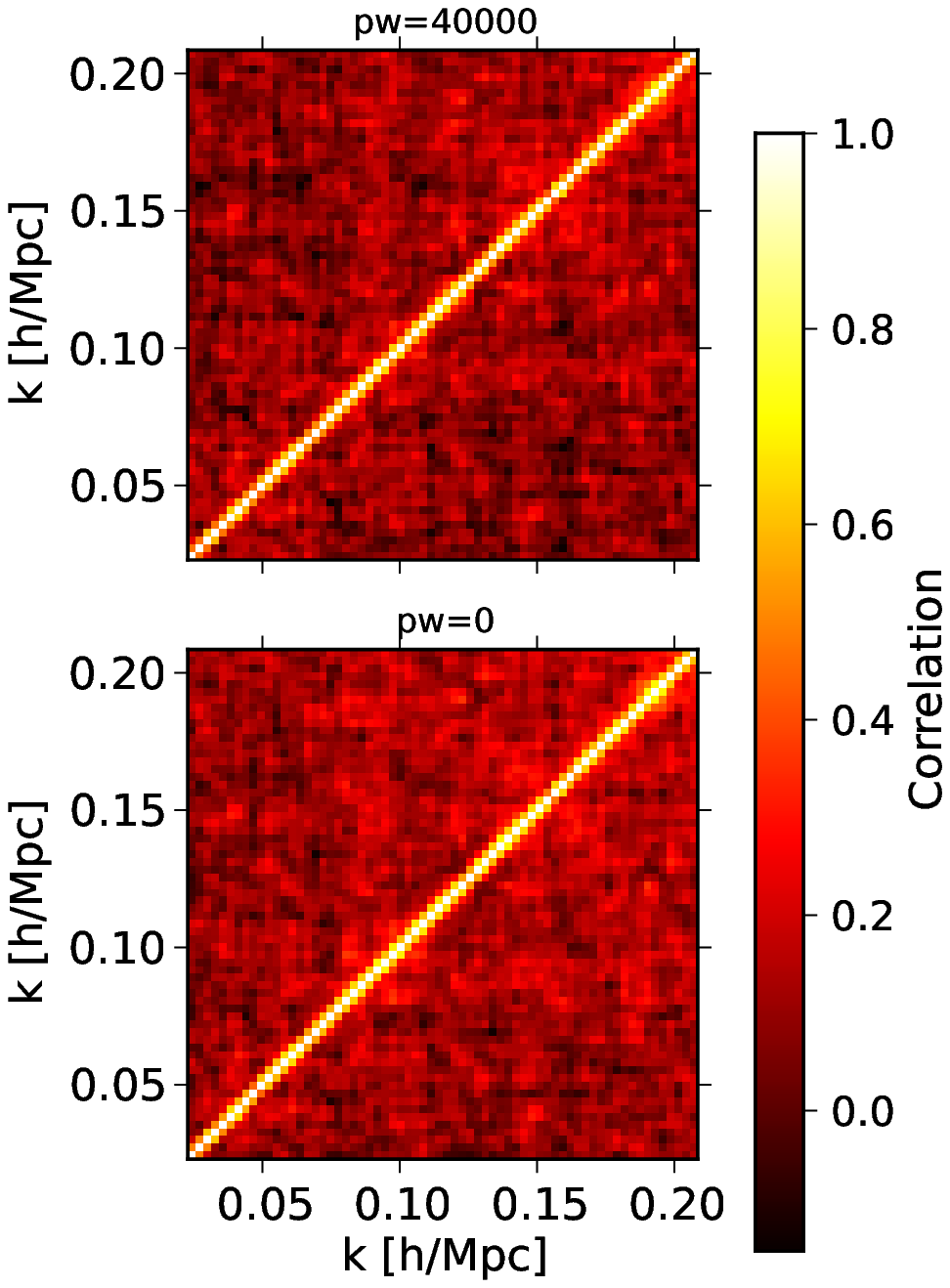}
  \caption{Correlation matrices for $\pw=40000,0$ (upper and lower panels respectively).}
  \label{fig:cor_mock} 
\end{figure}

\begin{figure}
  \includegraphics[width=79mm, keepaspectratio]{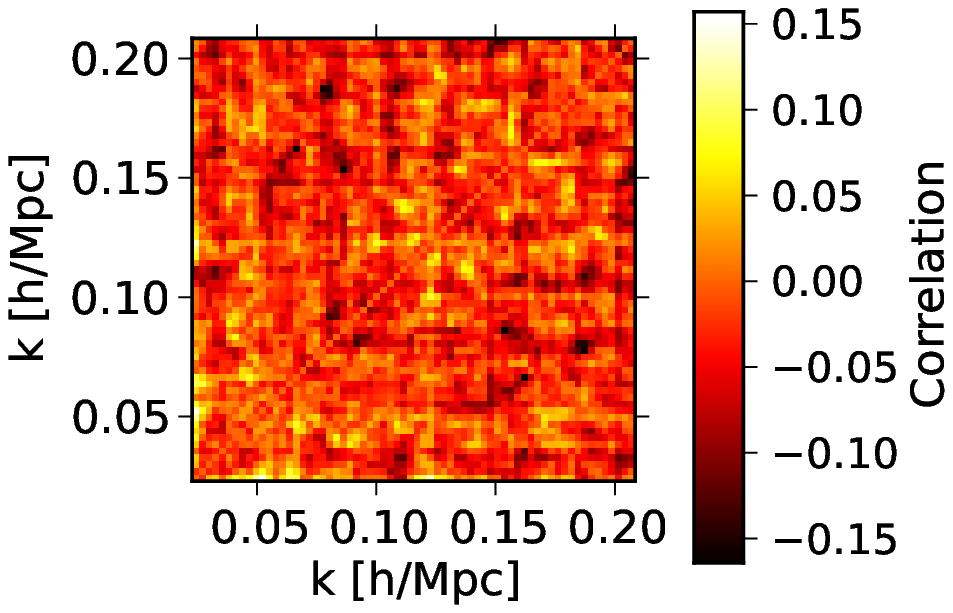} 
  \caption{Differences between the $\pw=40000$ and the $\pw=0$ correlation matrices.} \label{fig:difcor_mock} 
\end{figure}

The choice of $\pw$ has, however, a large effect on the errors. Figures \ref{fig:cor_mock} and \ref{fig:difcor_mock} show, respectively, the correlation matrix for $\pw=40000$ (upper panel) and $\pw=0$ (lower panel) and their difference. At $k\sim0.1\hoM$, for $\pw=40000$ we measure that the correlation is systematically, although not significantly, lower than for $\pw=0$. This is expected since, when using FKP, $P(k\sim0.1)\simeq40000$. At larger wave-number the power spectrum amplitude is smaller than at $k\sim0.1\hoM$ and the variance and the correlation are smaller for small values of $\pw$. This justifies our choice of using $\pw=40000$ in our analysis, since the amplitude of the power spectrum is of this order in the range of scales we are interested in. If the analysis were centred on the small scale power spectrum or correlation function, a smaller value of $\pw$ would be preferable.

\subsection{Impact on the cosmological parameters}\label{ap:test_cosm}

Here we test how the differences in the shape of the LRG power spectrum and the covariance matrix influence the cosmological parameters measured assuming the wCDM cosmology. We combine the LSS information with the WMAP7 data.

Contrary to our expectations based on the results shown in Appendix \ref{ap:test_pk}, the
cosmological parameters are much more sensible to $\pw$ than to $\wi$. For a
fixed $\pw$ the covariance matrix is the same and only the shape of the power
spectrum changes. As the shape of the power spectrum is not affected by the
completeness correction, also the cosmological constraints are insensitive to
it. The fibre collision correction instead changes the bias, over which we
marginalise analytically, without affecting the shot noise amplitude.
Differences in the relative amplitude of the latter term can be absorbed at
least partially by the mode coupling amplitude $A_{\mathrm{MC}}$, which is
systematically, although not significantly, larger when the loss of galaxies
due to fibre collisions is corrected for. Because of this, cosmological
constraints remain almost unchanged for different $\wi$. Changes of $\pw$
instead influence the covariance but only marginally the power spectrum. The
differences in the former influence the cosmological parameters, which differ
by $0.5-1\sigma$, for $\pw=40000$ and 0. For example I obtain the dark
energy equation of state parameter to be $w_{\mathrm{DE}}= -1.02\pm0.13$ for
$\pw=40000$ and $w_{\mathrm{DE}}=-1.10\pm 0.14$ for $\pw=0$.

Given the precision that is possible to achieve with the data used in this article, the differences just highlighted are not distinguishable from the uncertainties in the parameters. But in future, given the big improvements expected, these effects might become important and will need further analysis.

\bsp

\label{lastpage}


\begin{thebibliography}{99}

\bibitem[Abazajian et al.(2009)]{abazajian_DR7} Abazajian, K.~N., et al., 2009, ApJS, 182,
  543

\bibitem[\protect\citeauthoryear{Abbott et al.}{2005}]{des} Abbott T. et al., 2005, preprint (arXiv:astro-ph/0510346)

\bibitem[\protect\citeauthoryear{Ade et al.} {2011}]{plank_mission} Ade P.~A.~R. et al., 2011, preprint (arXiv:1101.2022)

\bibitem[\protect\citeauthoryear{Amanullah et al.} {2010}]{amanullah_10_SN} Amanullah R. et al., 2010, ApJ,  716, 712 

\bibitem[\protect\citeauthoryear{Albrecht et al.} {2006}]{albrecht_06_DETF} 
Albrecht A. et al., 2006, preprint (arXiv:astro-ph/0609591)

\bibitem[\protect\citeauthoryear{Angulo et al.}{2008}]{Angulo_08}Angulo R., Baugh C. M.,
  Frenk C. S., Lacey C. G., 2008, MNRAS, 383, 755

\bibitem[\protect\citeauthoryear{Astier et al.} {2006}]{astier_06_SN} Astier P. et al., 2006, A\&A,  447, 31 

\bibitem[\protect\citeauthoryear{Balaguera-Antol\'inez et al.} {2010}]{balaguera_10}
  Balaguera-Antol\'inez A., S\'anchez A.~G.,  B{\"o}hringer H., Collins C., Guzzo L., Phleps
  S., 2011, MNRAS, in press
\bibitem[\protect\citeauthoryear{Bengochea} {2011}]{bengochea_11} 
Bengochea G.~R., 2011, PhLB,  696, 5


\bibitem[\protect\citeauthoryear{Berlind \& Weinberg} {2002}]{berlind_02} Berlind A.~A.,
  Weinberg D.~H., 2002, ApJ,  575, 587 

\bibitem[\protect\citeauthoryear{Bernardeau et al.}{2002}]{Bernardeau_02} Bernardeau F.,
  Colombi S., Gazta{\~n}aga, E., Scoccimarro R., 2002, PhR, 367, 1

\bibitem[\protect\citeauthoryear{Bernardeau, Crocce \& Scoccimarro}{2008}] {bernardeau_08}
  Bernardeau F., Crocce M., Scoccimarro R., 2008, PhRvD, 78, 10352

\bibitem[\protect\citeauthoryear{Blake et al.} {2011}]{blake_11} Blake 
C. et al., 2011, arXiv,  arXiv:1105.2862 

\bibitem[\protect\citeauthoryear{Brown et al.} {2009}]{brown_09_QUaD} Brown M.~L. et al., 2009, ApJ,  705, 978 

\bibitem[\protect\citeauthoryear{Cabr{\'e} \& Gazta{\~n}aga}{2009a}]{cabre_09a} Cabr\'e A., Gazta\~naga E., 2009, MNRAS, 393, 1183

\bibitem[\protect\citeauthoryear{Cabr{\'e} \& Gazta{\~n}aga} {2009b}]{cabre_09b} Cabr{\'e} A., Gazta{\~n}aga E., 2009, MNRAS,  396, 1119 

\bibitem[\protect\citeauthoryear{Carnero et al.} {2011}]{carnero_11} Carnero A., Sanchez E., Crocce M., Cabre A., Gaztanaga E., 2011, preprint (arXiv:1104.5426)
\bibitem[\protect\citeauthoryear{Chevallier 
\& Polarski} {2001}]{chevallier_01} Chevallier M., Polarski D., 2001, IJMPD,  10, 213

\bibitem[\protect\citeauthoryear{Chuang, Wang \& Hemantha} {2010}]{chuang_10} Chuang C.-H., Wang Y., Hemantha M.~D.~P., 2010, preprint (arXiv:1008.4822)

\bibitem[\protect\citeauthoryear{Cole et al.}{2005}]{cole_05_2dF} Cole S. et al., 2005, MNRAS, 362, 505

\bibitem[\protect\citeauthoryear{Coles \& Jones} {1991}]{coles_91} Coles P., Jones B.,
  1991, MNRAS,  248, 1
  
\bibitem[\protect\citeauthoryear{Colless et al.} {2003}]{colless_03_2dF} Colless M. et al., 2003, preprint (  
arXiv:astro-ph/030658)

\bibitem[\protect\citeauthoryear{Cooray \& Sheth}{2002}]{cooray_HM} Cooray A., Sheth R.,
  2002, PhR, 372, 1

\bibitem[\protect\citeauthoryear{Christensen \& Meyer}{2000}]{christensen_00_MCMC}
  Christensen N., Meyer R., preprint (arXiv:astro-ph/0006401)

\bibitem[\protect\citeauthoryear{Crocce \& Scoccimarro}{2006a}]{crocce_RPT1} Crocce M.,
  Scoccimarro R., 2006, Phys. Rev. D, 73, 063519

\bibitem[\protect\citeauthoryear{Crocce \& Scoccimarro}{2006b}]{crocce_RPT2} Crocce M.,
  Scoccimarro R., 2006, Phys. Rev. D, 73, 063520

\bibitem[\protect\citeauthoryear{Crocce \& Scoccimarro}{2008}]{crocce_nonlinBAO} Crocce
  M., Scoccimarro R., 2008, Phys. Rev. D, 77, 023533

\bibitem[\protect\citeauthoryear{Davis \& Geller} {1976}]{davis_76} Davis M., Geller
  M.~J., 1976, ApJ,  208, 13 

\bibitem[\protect\citeauthoryear{Davis et al.}{1985}]{davis_85} Davis M., Efstathiou G.,
  Frenk C. S., White S. D. M.,   1985, ApJ, 292, 371

\bibitem[\protect\citeauthoryear{Drinkwater et al.} {2010}]{drinkwater_10_wigglez} Drinkwater M.~J. et al., 2010, MNRAS,  401, 1429 

\bibitem[\protect\citeauthoryear{Dunkley et al.} {2009}]{Dunkley_09_WMAP} Dunkley J. et al., 2009, ApJS,  180, 306 

\bibitem[\protect\citeauthoryear{Efstathiou et~al.}{2002}]{efstathiou02} Efstathiou G. et
  al., 2002, MNRAS, 330, L29

\bibitem[\protect\citeauthoryear{Eisenstein \& Hu}{1998}]{Eisenstein_98} Eisenstein D. J.,
  Hu W., 1998, ApJ, 496, 605

\bibitem[\protect\citeauthoryear{Eisenstein \& Hu}{1999}]{EH99} Eisenstein D. J., Hu W., 1998, ApJ. 511, 5

\bibitem[\protect\citeauthoryear{Eisenstein et al.}{2001}]{eisenstein_01} Eisenstein D. J. et al., 2001, AJ, 122, 2267 

\bibitem[\protect\citeauthoryear{Eisenstein et al.} {2005}]{Eisenstein_05} Eisenstein
  D.~J. et al., 2005, ApJ,  633, 560 

\bibitem[\protect\citeauthoryear{Eisenstein et al.} {2011}]{eisenstein_11_sdss3} Eisenstein D.~J. et al., 2011, preprint (arXiv:1101.1529)

\bibitem[\protect\citeauthoryear{Elia et al.} {2010}]{Elia_10} Elia A., Kulkarni S., Porciani C., Pietroni M., Matarrese S., 2010, preprint (arXiv:1012.4833) 

\bibitem[\protect\citeauthoryear{Fang, Hu \& Lewis}{2008}]{fang_08} Fang W., Hu W., Lewis
  A., 2008, PhRvD,  78, 087303 

\bibitem[\protect\citeauthoryear{Feldman, Kaiser \& Peacock} {1994}]{fkp} Feldman H.~A.,
  Kaiser N., Peacock J.~A., 1994, ApJ,  426, 23 

\bibitem[\protect\citeauthoryear{Frieman et al.} {2008}]{frieman_08_SN} Frieman J.~A. et al., 2008, AJ,  135, 338 

\bibitem[\protect\citeauthoryear{Frigo \& Johnson}{2005}]{Frigo_05} Frigo M., Johnson S.
  G., 2005, Proceedings of the IEEE, 93, 216

\bibitem[\protect\citeauthoryear{Gazta{\~n}aga, Cabr{\'e} \& Hui} {2009}]{gaztanaga09} Gazta{\~n}aga E., Cabr{\'e} A., Hui L., 2009, MNRAS,  399, 1663 

\bibitem[\protect\citeauthoryear{Gelman \& Rubin}{1992}]{gelman_92} Gelman A., Rubin D.
  B., 1992, Stat. Sci., 7, 457

\bibitem[\protect\citeauthoryear{Gilks, Richardson \& Spiegelhalter}{1996}]{gilks_MCMC}
  Gilks W. R., Richardson S., Spiegelhalter, D. J.,  1996, MarkovChain Monte Carlo in
  Practice. Chapman and Hall, London, UK

\bibitem[\protect\citeauthoryear{Guy et al.} {2007}]{guy_07_salt2} Guy J. et al., 2007, A\&A,  466, 11 
\bibitem[\protect\citeauthoryear{Heitmann et al.} {2010}]{heitmann_10} 
Heitmann K., White M., Wagner C., Habib S., Higdon D., 2010, ApJ,  715, 104

\bibitem[\protect\citeauthoryear{Hicken et al.} {2009}]{hicken_09_SN} Hicken M.,
  Wood-Vasey W.~M., Blondin S., Challis P., Jha S., Kelly P.~L., Rest A., Kirshner R.~P.,
  2009, ApJ,  700, 1097 
  
\bibitem[\protect\citeauthoryear{Hill et al}{2004}]{hetdex} Hill G. J., Gebhardt K., Komatsu E. \& MacQueen P. J., 2004, The New Cosmology: Conference on Strings and Cosmology, 743, 224 

\bibitem[\protect\citeauthoryear{Hinshaw et al.} {2003}]{hinshaw_wmpa1_aps} Hinshaw G. et al., 2003, ApJS,  148, 135 

\bibitem[\protect\citeauthoryear{Holsclaw et al.} {2010}]{holsclaw_10} Holsclaw T., Alam U., Sans{\'o} B., Lee H., Heitmann K., Habib S., Higdon D., 2010, PhRvD,  82, 103502 

\bibitem[\protect\citeauthoryear{Hu \& Jain} {2004}]{hu_04} Hu W., Jain B., 2004, PhRvD,  70, 043009 

\bibitem[\protect\citeauthoryear{Huff et al.}{2007}]{Huff_07} Huff E., Schulz A. E., White
  M., Schlegel D. J., Warren M. S., 2007, APh, 26, 351

\bibitem[\protect\citeauthoryear{Huterer \& Starkman} {2003}]{huterer_03} Huterer D., Starkman G., 2003, PhRvL,  90, 031301 

\bibitem[\protect\citeauthoryear{Huterer \& Turner} {2001}]{huterer_01} Huterer D., Turner M.~S., 2001, PhRvD,  64, 123527 

\bibitem[\protect\citeauthoryear{Jarosik et al.} {2010}]{jarosik_11_WMAP} Jarosik N. et al., 2010, preprint (arXiv:1001.4744)

\bibitem[\protect\citeauthoryear{Jha, Riess \& Kirshner} {2007}]{jha_07_SN} Jha S., Riess
  A.~G., Kirshner R.~P., 2007, ApJ,  659, 122 

\bibitem[\protect\citeauthoryear{Jennings, Baugh \& Pascoli} {2011}]{Jennings_2010} Jennings E., Baugh C.~M., Pascoli S., 2011, MNRAS,  410, 2081 

\bibitem[\protect\citeauthoryear{Jeong \& Komatsu}{2006}]{Jeong_06} Jeong D., Komatsu E.,
  2006, ApJ, 651, 619

\bibitem[\protect\citeauthoryear{Jeong \& Komatsu}{2009}]{Jeong_09} Jeong D., Komatsu E.,
  2009, ApJ, 691, 569

\bibitem[\protect\citeauthoryear{Jones et al.} {2006}]{jones_06_Boo} Jones W.~C. et al., 2006, ApJ,  647, 823 

\bibitem[\protect\citeauthoryear{Kaiser et al.}{2002}]{panstarrs} Kaiser N. et al., 2002, SPIE, 4836, 154

\bibitem[\protect\citeauthoryear{Kazin et al.} {2010}]{kazin_10a} Kazin E.~A. et al., 2010, ApJ,  710, 1444 

\bibitem[\protect\citeauthoryear{Kessler et al.} {2009}]{kessler_09_SN} Kessler R. et al., 2009, ApJS,  185, 32 

\bibitem[\protect\citeauthoryear{Komatsu et al.}{2009}]{komatsu_wmap09} Komatsu E. et al., 2009, ApJS, 180, 330

\bibitem[\protect\citeauthoryear{Komatsu et al.} {2011}]{komatsu_10} Komatsu E. et al., 2011,  ApJS,  192, 18

\bibitem[\protect\citeauthoryear{Kowalski et al.}{2008}]{kowalski_unionSN1a} Kowalski M. et al, 2008, ApJ, 686, 749

\bibitem[\protect\citeauthoryear{Kuo et al.} {2007}]{kuo_07} Kuo C.~L. et al., 2007, ApJ,  664, 687 

\bibitem[\protect\citeauthoryear{Larson et al.} {2011}]{larson_11_WMAP} Larson D. et al., 2011, ApJS,  192, 16 

\bibitem[\protect\citeauthoryear{Laureijs} {2009}]{laureijis_09_euclid} Laureijs R., 2009, preprint (arXiv:0912.0914)

\bibitem[\protect\citeauthoryear{Lewis \& Bridle}{2002}]{lewis_02} Lewis A., Bridle A.,
  2002, Phys. Rev. D, 66, 103511

\bibitem[\protect\citeauthoryear{Lewis, Challinor \& Lasenby} {2000}]{lewis_00} Lewis A.,
  Challinor A., Lasenby A., 2000, ApJ,  538, 473 

\bibitem[\protect\citeauthoryear{Linder} {2003}]{linder_03} Linder E.~V., 2003, PhRvL, 90,
  091301 

\bibitem[\protect\citeauthoryear{McDonald}{2006}]{mcdonald_renorm} McDonald P., Phys. Rev.
  D, 2006, 74, 103512

\bibitem[\protect\citeauthoryear{McDonald}{2007}]{mcdonald_renormbias} McDonald P., Phys.
  Rev. D, 2007, 75, 043514

\bibitem[\protect\citeauthoryear{MacTavish et al.} {2006}]{mactavish_06_Boo} MacTavish
  C.~J. et al., 2006, ApJ,  647, 799 

\bibitem[\protect\citeauthoryear{Masjedi}{2006}]{masjedi_06} Masjedi M. et al., 2006, ApJ,  644, 54 

\bibitem[\protect\citeauthoryear{Matarrese \& Pietroni}{2007}]{matarrese_rengroup1}
  Matarrese S., Pietroni M., 2007, JCAP, 06, 26

\bibitem[\protect\citeauthoryear{Matarrese \& Pietroni}{2008}]{matarrese_rengroup2}
  Matarrese S., Pietroni M., 2008, Mod.  Phys. Lett. A, 23, 25

\bibitem[\protect\citeauthoryear{Matsubara}{2004}]{Matsubara_04} Matsubara T., 2004, ApJ, 615, 573

\bibitem[\protect\citeauthoryear{Matsubara}{2008a}]{mats_LPT1} Matsubara T., 2008, Phys.
  Rev. D, 77, 063530

\bibitem[\protect\citeauthoryear{Matsubara}{2008b}]{mats_LPT2} Matsubara T., 2008, Phys.
  Rev. D, 78, 083519 

\bibitem[\protect\citeauthoryear{Montesano, S\'anchez \& Phleps}{2010}]{montesano_10a}
  Montesano F., S\'anchez A.~G., Phleps S., 2010, MNRAS, 408, 2397

\bibitem[\protect\citeauthoryear{Montroy et al.} {2006}]{montroy_06_Boo} Montroy T.~E. et al., 2006, ApJ,  647, 813 

\bibitem[\protect\citeauthoryear{Moresco et al.} {2010}]{moresco_11} Moresco M., Jimenez
  R., Cimatti A., Pozzetti L., 2010, preprint (arXiv:1010.0831)

\bibitem[\protect\citeauthoryear{Norberg et al.} {2001}]{norberg_01} Norberg P. et al., 2001, MNRAS,  328, 64 

\bibitem[\protect\citeauthoryear{Norberg et al.} {2002}]{norberg_02} Norberg P. et al., 2002, MNRAS,  332, 827 
\bibitem[\protect\citeauthoryear{Peebles \& Yu} {1970}]{peebles_70} Peebles P.~J.~E., Yu J.~T., 1970, ApJ,  162, 815 

\bibitem[\protect\citeauthoryear{Peebles \& Ratra} {2003}]{Peebles2003} Peebles P.~J., Ratra B., 2003, Rev. Mod. Phys.,  75, 559 


\bibitem[\protect\citeauthoryear{Percival et~al.}{2002}]{percival02} Percival W.~J. et al.,2002, MNRAS, 337, 1068

\bibitem[\protect\citeauthoryear{Percival, Verde \& Peacock} {2004}]{pvp} Percival W.~J.,
  Verde L., Peacock J.~A., 2004, MNRAS,  347, 645 

\bibitem[\protect\citeauthoryear{Percival et~al.}{2007}]{percival07} Percival W.J., Cole
  S., Eisenstein D. J., Nichol R. C., Peacock J. A., Pope A. C., Szalay A. S., 2007,
  MNRAS, 381, 1053

\bibitem[\protect\citeauthoryear{Percival et al.}{2010}]{percival_10_SDSS} Percival W. J. et al., 2010, MNRAS, 401, 2148

\bibitem[\protect\citeauthoryear{Perlmutter et al.}{1999}]{perlmutter_99} Perlmutter S. et al., 1999,  ApJ, 517, 565

\bibitem[\protect\citeauthoryear{Phleps et al.}{2006}]{phleps_06} Phleps S., Peacock J.~A., Meisenheimer K., Wolf C., 2006, A\&A,  457, 145 

\bibitem[\protect\citeauthoryear{Piacentini et al.} {2006}]{piacentini_06_Boo} Piacentini
  F. et al., 2006, ApJ,  647, 833 

\bibitem[\protect\citeauthoryear{Pietroni}{2008}]{pietroni_flowtime} Pietroni M., 2008,
  JCAP, 10, 36

\bibitem[\protect\citeauthoryear{Reichardt et al.} {2009}]{reichardt_09} Reichardt C.~L. et al., 2009, ApJ,  694, 1200 

\bibitem[\protect\citeauthoryear{Reid \& Spergel}{2009}]{reid_09a} Reid B.~A., Spergel
  D.~N., 2009, ApJ,  698, 143

\bibitem[\protect\citeauthoryear{Reid \& White} {2011}]{reid_11_rsdist} Reid B.~A., White M., 2011, preprint (arXiv:1105.4165)

\bibitem[\protect\citeauthoryear{Reid et al.}{2009}]{reid_09b} Reid B. A., Spergel D. N.,
  Bode P., 2009, ApJ, 702, 249f

\bibitem[\protect\citeauthoryear{Reid et al.}{2010}]{reid_10_SDSS} Reid B. A. et al., 2010, MNRAS, 404, 60

\bibitem[\protect\citeauthoryear{Riess et al.}{1998}]{riess_98} Riess A. G. et al., 1998, AJ, 116, 1009
  
\bibitem[\protect\citeauthoryear{Riess et al.} {2004}]{riess_04} Riess A.~G. et al., 2004, ApJ,  607, 665 

\bibitem[\protect\citeauthoryear{Riess et al.} {2007}]{riess_07_SN} Riess A.~G. et al., 2007, ApJ,  659, 98 

\bibitem[\protect\citeauthoryear{Riess et al.} {2009}]{reiss_09_H} Riess A.~G. et al., 2009, ApJ,  699, 539 

\bibitem[\protect\citeauthoryear{Riess et al} {2011}]{reiss_11_H} Riess A.~G., et al., 2011, ApJ,  730, 119 

\bibitem[\protect\citeauthoryear{Samushia, Percival \& Raccanelli} {2011}]{samushia_11} Samushia L., Percival W.~J., Raccanelli A., 2011, preprint(arXiv:1102.1014)

\bibitem[\protect\citeauthoryear{S\'anchez \& Cole }{2008}]{sanchez_cole} S\'anchez A. G., Cole S., 2008, MNRAS, 385, 830

\bibitem[\protect\citeauthoryear{S\'anchez et al.}{2006}]{Sanchez_06} S\'anchez A. G.,
  Baugh C. M., Percival W. J., Peacock J. A., Padilla N. D., Cole S., Frenk C. S., Norberg
  P., 2006, MNRAS, 366, 189 

\bibitem[\protect\citeauthoryear{S\'anchez, Baugh \& Angulo}{2008}]{Sanchez_08} S\'anchez A. G.,
  Baugh C. M., Angulo, R.,  2008, MNRAS, 390, 1470

\bibitem[\protect\citeauthoryear{S\'anchez et al.}{2009}]{Sanchez_09} S\'anchez A. G.,
  Crocce M., Cabre A., Baugh C. M., Gaztanaga E., 2009, MNRAS., 400. 1643

\bibitem[\protect\citeauthoryear{Schlegel, White \& Eisenstein}{2009}]{schlegel_BOSS} Schlegel D., White M., Eisenstein D.,  2009, preprint( arXiv:0902.4680)

\bibitem[\protect\citeauthoryear{Scoccimarro}{2004}]{scoccimarro_04} Scoccimarro R., 2004, Phys. Rev. D, 70, 083007

\bibitem[\protect\citeauthoryear{Serra et al.} {2009}]{serra_09} Serra P., Cooray A., Holz D.~E., Melchiorri A., Pandolfi S., Sarkar D., 2009, Phys. Rev. D,  80, 121302 

\bibitem[\protect\citeauthoryear{Shoji, Jeong \& Komatsu}{2009}]{shoji_09} Shoji M., Jeong D., Komatsu E., 2009, ApJ, 693, 1404

\bibitem[\protect\citeauthoryear{Sievers et al.} {2009}]{sievers_09_cbi} Sievers J.~L. et al., 2009, preprint (arXiv:0901.4540)

\bibitem[\protect\citeauthoryear{Smith et al.} {2003}]{smith_03_halofit} Smith R.~E. et al., 2003, MNRAS,  341, 1311 

\bibitem[\protect\citeauthoryear{Smith, Scoccimarro \& Sheth} {2008}]{smith_08} Smith R.~E., Scoccimarro R., Sheth R.~K., 2008, PhRvD,  77, 043525 

\bibitem[\protect\citeauthoryear{Smith, Hern\'andez-Monteagudo \&
  Seljak}{2009}]{smith09_SZ} Smith R., Hern\'andez-Monteagudo C., Seljak U., Phys.  Rev.
  D, 2009, 80, 063528

\bibitem[\protect\citeauthoryear{Spergel et al.}{2003}]{Spergel_03} Spergel D. N. et al., 2003, ApJS, 148, 175

\bibitem[\protect\citeauthoryear{Spergel et al.}{2007}]{Spergel_07} Spergel D.  N. et al., 2007, ApJS, 170, 377

\bibitem[\protect\citeauthoryear{Sugiyama}{1995}]{Sugiyama_95} Sugiyama N., 1995, ApJS, 100, 28
\bibitem[\protect\citeauthoryear{Sunyaev \& Zeldovich} {1970}]{sunyaev_70} Sunyaev R.~A., Zeldovich Y.~B., 1970, Ap\&SS,  7, 3 

\bibitem[\protect\citeauthoryear{Taruya \& Hiramatsu}{2008}]{taruya_closure} Taruya A.,
  Hiramatsu T., 2008, ApJ, 674, 617

\bibitem[\protect\citeauthoryear{Taruya et al.} {2009}]{taruya_09} Taruya A., Nishimichi
  T., Saito S., Hiramatsu T., 2009, PhRvD,  80, 123503
  
\bibitem[\protect\citeauthoryear{Taruya, Nishimichi \& Saito} {2010}]{taruya_10} Taruya A., Nishimichi T., Saito S., 2010, PhRvD,  82, 063522 

\bibitem[\protect\citeauthoryear{Tegmark, Silk \& Blanchard} {1994}]{tegmark_94} Tegmark
  M., Silk J., Blanchard A., 1994, ApJ,  420, 484 

\bibitem[\protect\citeauthoryear{Tegmark et~al.}{2004}]{tegmark04}Tegmark M. et al., 2004, ApJ 606, 702

\bibitem[\protect\citeauthoryear{Tinker et al.} {2011}]{tinker_11} Tinker J.~L. et al., 2011, preprint (arXiv:1104.1635)

\bibitem[\protect\citeauthoryear{Tsujikawa} {2010}]{tsujikawa_11} Tsujikawa S., 2010, LNP,  800, 99 

\bibitem[\protect\citeauthoryear{Vikhlinin et al} {2009}]{vikhlinin_09} 
Vikhlinin A. et al., 2009, ApJ,  692, 1060 

\bibitem[\protect\citeauthoryear{Wood-Vasey et al.} {2007}]{woodvasey_07_SN} Wood-Vasey
  W.~M. et al., 2007, ApJ,  666, 694 

\bibitem[\protect\citeauthoryear{Xu et al.} {2010}]{xu_10} Xu X. et al., 2010, ApJ,  718, 1224 

\bibitem[\protect\citeauthoryear{Zehavi et al.} {2002}]{zehavi_02} Zehavi I. et al., 2002, ApJ,  571, 172 

\end{thebibliography}
\end{document}